\documentclass[imslayout,preprint]{imsart} 
\def\journal@name{} 
\usepackage[margin=1.5in]{geometry}

\usepackage{mathtools}
\usepackage[colorlinks,citecolor=blue,urlcolor=black,linkcolor=blue,pdfborder={0 0 0}]{hyperref}
\usepackage{url}
\usepackage{comment}
\usepackage{graphicx}
\usepackage{enumitem}
\usepackage{framed}
\usepackage{dcolumn}
\usepackage{chngcntr}
\usepackage{subcaption}
\usepackage[sort]{natbib}
\usepackage[capitalize]{cleveref}  
\usepackage{datetime}
\usepackage{algorithm}
\usepackage{algpseudocode}
\usepackage[normalem]{ulem}

\crefname{nlem}{Lemma}{Lemmas}
\crefname{nprop}{Proposition}{Propositions}
\crefname{ncor}{Corollary}{Corollaries}
\crefname{nthm}{Theorem}{Theorems}
\crefname{nnon}{Conjecture}{Conjectures}
\crefname{assumption}{Assumption}{Assumptions}
\crefname{exa}{Example}{Examples}

\RequirePackage[usenames,dvipsnames]{color}

\usepackage{jhh-misc2}
\usepackage{mathrsfs}
\usepackage{autonum}

\newcommand{\cf}[1]{\psi_{#1}} 
\newcommand{\ii}{i} 

\newcommand{\indicatorfn}{\mathds{1}}
\newcommand{\id}{\operatorname{id}}

\newcommand{\littleo}{o}

\newcommand{\empcov}{\hat\Sigma}

\newcommand{\optsym}{\circ}

\newcommand{\obsspace}{\mathbb X}
\newcommand{\numobs}{N}
\newcommand{\data}{x}
\newcommand{\datarv}{X}
\newcommand{\dataarg}[1]{x_{1:#1}}
\newcommand{\datarvarg}[1]{X_{1:#1}}

\newcommand{\alldatarv}{\datarvarg{\infty}}

\newcommand{\obs}[1]{x_{#1}}
\newcommand{\obsrv}[1]{X_{#1}}
\newcommand{\datameanarg}[1]{\bar\data_{#1}}
\newcommand{\datarvmeanarg}[1]{\bar\datarv_{#1}}

\newcommand{\postdist}[1]{\Pi_{#1}}
\newcommand{\postdistfull}[2]{\Pi(#1 \given #2)}
\newcommand{\ppostdist}[2]{\Pi_{#1}^{#2}}
\newcommand{\postdensity}[1]{\pi_{#1}}
\newcommand{\postdensityfull}[2]{\pi(#1 \given #2)}

\newcommand{\priordist}{\postdist{0}}
\newcommand{\priordensity}{\postdensity{0}}

\newcommand{\likdist}[1]{P_{#1}}
\newcommand{\likfun}[1]{p_{#1}}
\newcommand{\lik}[2]{\likfun{#2}(#1)}
\newcommand{\loglik}[2]{\ell_{#2}(#1)}
\newcommand{\loglikfun}[1]{\ell_{#1}}

\newcommand{\gradloglik}[2]{\dot\ell_{#2}(#1)}
\newcommand{\gradloglikfun}[1]{\dot\ell_{#1}}
\newcommand{\hessloglik}[2]{\ddot\ell_{#2}(#1)}
\newcommand{\hessloglikfun}[1]{\ddot\ell_{#1}}

\newcommand{\marginallikfull}[1]{p(#1)}

\newcommand{\param}{\theta}
\newcommand{\paramspace}{\Theta}
\newcommand{\trueparam}{\param_{\dagger}}
\newcommand{\optparam}{\param_{\optsym}}
\newcommand{\paramsample}{\vartheta}

\newcommand{\bbparamsample}{\paramsample^{\bbsym}}
\newcommand{\bbparamsamplecopy}{\paramsample^{\bbsym\prime}}

\newcommand{\mle}[1]{\hat\param_{#1}}

\newcommand{\Ehessloglik}[1]{J_{#1}}
\newcommand{\Vargradloglik}[1]{I_{#1}}

\newcommand{\obsdist}{P_{\optsym}}
\newcommand{\empdist}{\Pr_{\numobs}}

\newcommand{\bbsym}{*} 
\newcommand{\bsnumobs}{M}

\newcommand{\bscount}[1]{K_{#1}}
\newcommand{\bscounts}{K_{1:\numobs}}
\newcommand{\bsobs}[1]{\obs{#1}^{\bbsym}}

\newcommand{\bsdata}{\data^{\bbsym}}
\newcommand{\bsdatarv}{\datarv^{\bbsym}}
\newcommand{\bsdatasample}[1]{\data^{\bbsym}_{(#1)}}

\newcommand{\bsdatarvarg}[1]{\bsdatarv_{1:#1}}

\newcommand{\bbpostdist}[1]{\postdist{#1}^{\bbsym}}
\newcommand{\bbpostdistfull}[2]{\Pi^{\bbsym}(#1 \given #2)}
\newcommand{\bbpostdensity}[1]{\postdensity{#1}^{\bbsym}}
\newcommand{\bbpostdensityfull}[2]{\postdensity{}^{\bbsym}(#1 \given #2)}

\newcommand{\bsdatarvmeanarg}[1]{\bar\datarv^{\bbsym}_{#1}}
\newcommand{\bbempdist}{\Pr_{\numobs}^{\bbsym}}

\newcommand{\bsscale}{c}
\newcommand{\optMasympt}{\bsnumobs_{\infty,\text{opt}}}
\newcommand{\optMasymptest}{\hat\bsnumobs_{\infty,\text{opt}}}
\newcommand{\optMfs}{\bsnumobs_{\text{fs,opt}}}
\newcommand{\optMfsest}{\hat\bsnumobs_{\text{fs,opt}}}
\newcommand{\optMest}{\hat\bsnumobs_{\text{opt}}}
\newcommand{\modelmismatch}{\mathcal{I}}
\newcommand{\nan}{\textsf{NA}}
\newcommand{\concconst}{C_{\numobs}}

\def\norm#1{\left\|{#1}\right\|} 
\newcommand{\twonorm}[1]{\norm{#1}_2} 
\def\staticnorm#1{\|{#1}\|} 
\newcommand{\statictwonorm}[1]{\staticnorm{#1}_2} 

\graphicspath{{figures/}}

\allowdisplaybreaks[2]

\begin{document}

\begin{frontmatter}

\title{Robust Inference and Model Criticism \\ Using Bagged Posteriors}
\runtitle{~Robust Inference Using Bagged Posteriors}
\runauthor{J.\ H.\ Huggins and J.\ W.\ Miller~}

\begin{aug}

\author[A]{\fnms{Jonathan H.} \snm{Huggins}\ead[label=e1]{huggins@bu.edu}} %
\and
\author[B]{\fnms{Jeffrey W.} \snm{Miller}\ead[label=e2]{jwmiller@hsph.harvard.edu}}
\address[A]{Department of Mathematics \& Statistics, Boston University, \printead{e1}}
\address[B]{Department of Biostatistics, Harvard University, \printead{e2}}

\end{aug}

\begin{abstract}
Standard Bayesian inference is known to be sensitive to model misspecification, leading to unreliable uncertainty quantification and poor predictive performance. 
However, finding generally applicable and computationally feasible methods for robust Bayesian inference under misspecification has proven to be a 
difficult challenge.
An intriguing, easy-to-use, and widely applicable approach is to use bagging on the Bayesian posterior (``BayesBag''); that is, 
to use the average of posterior distributions conditioned on bootstrapped datasets.
In this paper, we develop the asymptotic theory of BayesBag,
propose a model--data mismatch index for model criticism using BayesBag,
and empirically validate our theory and methodology on synthetic and real-world data in linear regression,
sparse logistic regression, and a hierarchical mixed effects model. 
We find that in the presence of significant misspecification, BayesBag yields more reproducible inferences and has better predictive accuracy
than the standard Bayesian posterior;
on the other hand, when the model is correctly specified, BayesBag produces superior or equally good results.
Overall, our results demonstrate that BayesBag combines the attractive modeling features of standard Bayesian inference with 
the distributional robustness properties of frequentist methods,
providing benefits over both Bayes alone and the bootstrap alone.
\end{abstract}


\end{frontmatter}

\section{Introduction} \label{sec:introduction}

Bayesian inference is premised on the data being generated from the assumed model.
In practice, however, it is widely recognized that models are (sometimes gross) approximations to reality~\citep{Box:1979,Box:1980,Cox:1990,Lehmann:1990}.
Moreover, even when the model is nearly correct, the optimal parameter (in terms of log loss or Kullback--Leibler divergence) may be extremely unlikely under the prior distribution, which can
bias the posterior distribution and lead to poor predictive performance.
Thus, in order to effectively use Bayesian methods, it is crucial to be able to both diagnose and correct for mismatch between the model and the data
during the model building process~\citep{Gelman:2013,Blei:2014}.
The task of diagnosing mismatch/misspecification is often termed ``model criticism'' or ``model assessment''~\citep{Gelman:2011,Vehtari:2012}.
The model building process eventually concludes either because a model is deemed satisfactory or because of limited resources,
such as the data analyst's time, knowledge of the phenomena under study, or computational capability.
If some model--data mismatch does remain, the analyst may need to rely on a robust %
estimation method to ensure correct uncertainty quantification and good predictive accuracy.

This article develops the theory and practice of \emph{BayesBag}, a simple and widely applicable approach to robust Bayesian inference that also provides diagnostics for model criticism. 
Originally suggested by \citet{Waddell:2002} and \citet{Douady:2003} in the context of phylogenetic inference and then independently proposed by~\citet{Buhlmann:2014} (where the name was coined), the idea of BayesBag is to apply bagging~\citep{Breiman:1996} to the Bayesian posterior. 
The \emph{bagged posterior} $\bbpostdensityfull{\param}{\data}$ is defined by taking bootstrapped copies $\bsdata \defined (\bsobs{1},\ldots,\bsobs{\bsnumobs})$ 
of the original dataset $\data \defined (\obs{1},\ldots,\obs{\numobs})$ and averaging over the posteriors obtained by treating each bootstrap dataset as the observed data:
\[
\bbpostdensityfull{\param}{\data} \defined \frac{1}{\numobs^{\bsnumobs}} \sum_{\bsdata} \postdensityfull{\param}{\bsdata}, \label{eq:bayesbag-definition}
\]
where $\postdensityfull{\param}{\bsdata} \propto \priordensity(\param)\prod_{m=1}^{\bsnumobs}\lik{\bsobs{m}}{\param}$ 
is the standard posterior density given data $\bsdata$ and the sum is over all $N^M$ possible bootstrap datasets of $\bsnumobs$ samples drawn 
with replacement from the original dataset.
In practice, we can approximate $\bbpostdensityfull{\param}{\data}$ by generating $B$ bootstrap datasets $\bsdatasample{1},\dots,\bsdatasample{B}$,
where $\bsdatasample{b}$ consists of $\bsnumobs$ samples drawn with replacement from $\data$, yielding the approximation
\[
\bbpostdensityfull{\param}{\data} \approx \frac{1}{B} \sum_{b=1}^{B} \postdensityfull{\param}{\bsdatasample{b}}.  \label{eq:bayesbag-approximation}
\]
BayesBag is easy to use since the bagged posterior is simply an average over standard Bayesian posteriors, which means no additional algorithmic tools are needed beyond 
those required for posterior inference in the original model. 
While using the approximation in \cref{eq:bayesbag-approximation} does require more computational resources since 
one must approximate $B$ posteriors, %
where typically $B \approx$ 50--100, 
each posterior can be approximated in parallel, which is ideal for modern cluster-based high-performance computing environments. 
Surprisingly, despite these attractive features, there has been little practical or theoretical investigation of BayesBag.
In the only previous work of which we are aware, \citet{Buhlmann:2014} (which is a short discussion paper) presented only a few simulation results in a simple 
Gaussian location model, while \citet{Waddell:2002} and \citet{Douady:2003} undertook limited investigations in the setting of phylogenetic tree inference in papers focused primarily on 
speeding up model selection (in the former) and comparing Bayesian inference and the bootstrap (in the latter).

In this paper, we show that parameter estimation and prediction with the bagged posterior has appealing statistical properties 
in the presence of model misspecification, while also being easy to use and computationally tractable on a range of practical problems. 
The bagged posterior integrates the attractive features of Bayesian inference -- such as flexible hierarchical modeling, automatic integration over nuisance parameters, and the use of prior 
information -- with the distributional robustness of frequentist methods, nonparametrically accounting for sampling variability and model misspecification. 
Moreover, rather than just providing robustness to misspecification, our BayesBag methodology can simultaneously diagnose the degree of misspecification, 
or more generally, the degree of model--data mismatch.
In a companion article, we explore the benefits of using BayesBag for model selection \citep{Huggins:2019:BayesBagII}. 

\paragraph*{Overview}
The organization and main contributions of the paper are as follows.
In \cref{sec:overview}, %
we first sketch out the theoretical motivation for the bagged posterior.
We then introduce our methodology in detail and describe our \emph{model--data mismatch index} for performing model criticism.
This mismatch index is based on how much the standard and bagged posterior variances differ, compared to what would be expected if the model were correctly specified.

In \cref{sec:parameter-estimation}, we present our asymptotic theory for the bagged posterior. 
We prove that the bagged posterior is asymptotically normal (that is, it satisfies a Bernstein--Von Mises theorem)
and that bagged posterior credible intervals for the optimal parameter are asymptotically conservative
when the bootstrapped datasets are the same size as the original dataset.
Moreover, we show that if the size of the bootstrapped datasets is appropriately selected, then
the credible intervals have asymptotically correct frequentist coverage. 
In short, we show that when used for parameter inference, the bagged posterior can improve upon the standard posterior by accounting for sampling variance, 
as in traditional bootstrapping~\citep{Efron:1979}.

In \cref{sec:simulations,sec:application}, we validate our theory and methodology through simulations and a real-data application.
Our first simulation study shows that, in the case of linear regression, the mismatch index is useful for (a) diagnosing misspecification in the likelihood, and
(b) detecting poorly chosen priors that either make the true parameter extremely unlikely or lead to poorly identified model parameters. 
Remarkably, we find that BayesBag often produces superior results compared to standard Bayesian inference even when the likelihood model is correct. 
In our second simulation study, we consider a hierarchical mixed effects model and compare BayesBag to alternative robust approaches. 
Finally, we apply BayesBag {and the mismatch index} to real-world data with a sparse logistic regression model.
Overall, our empirical results demonstrate that in the presence of significant misspecification, the bagged posterior produces more stable inferences
and has better predictive accuracy than the standard posterior; on the other hand, when the model is correctly specified, 
the bagged posterior produces equally good or better results.
We conclude in \cref{sec:discussion} with a more detailed discussion of related work and possible extensions.

\section{Methodology and motivation} \label{sec:overview}

In this section, we  %
introduce our BayesBag methodology and provide an overview of the statistical theory that justifies our methodology.

\subsection{Theoretical motivation} \label{sec:parameter-theory-sketch}

To elucidate the behavior of the bagged posterior, we begin by deriving its mean and covariance
in the case where $\param \in \reals^{D}$.
Given data $\data$, let $\bsdatarv$ be a random bootstrap dataset and let $\bbparamsample\given\bsdatarv\dist \postdensityfull{\param}{\bsdatarv}$ 
be distributed according to the standard posterior given data $\bsdatarv$. (We denote random variables with capital Latin letters, e.g., $\datarv$ rather than $\data$, or ``curly'' Greek letters, e.g., $\paramsample$ rather than $\param$.)
Marginalizing out $\bsdatarv$, we have $\bbparamsample\given\data \dist \bbpostdensityfull{\param}{\data}$.
Let $\paramsample \given \data \dist \postdensityfull{\param}{\data}$ and define 
$\mu(\data) \defined \EE(\paramsample\given\data) = \int \param\,\postdensityfull{\param}{\data}\dee\param$
to be the standard posterior mean given $\data$.
By the law of total expectation, the mean of the bagged posterior is 
\[ \label{eq:bayesbag-posterior-mean}
\EE(\bbparamsample\given\data) = \EE\big\{\EE(\bbparamsample\given\bsdatarv)\given\data\big\} = \EE\{\mu(\bsdatarv)\mid\data\}
= \frac{1}{\numobs^{\bsnumobs}} \sum_{\bsdata} \mu(\bsdata).
\]
By the law of total covariance, the covariance matrix of the bagged posterior is
\[ \label{eq:bayesbag-posterior-covariance}
\cov(\bbparamsample\given\data) &= \EE\big\{\cov(\bbparamsample\given\bsdatarv)\given\data\big\} + \cov\big\{\EE(\bbparamsample\given\bsdatarv)\given\data\big\} \\
 &= \EE\{\Sigma(\bsdatarv)\mid \data\} + \cov\{\mu(\bsdatarv)\mid \data\} ,
\]
where $\Sigma(\data) \defined \cov(\paramsample\given\data) = \int \{\param-\mu(\data)\}\{\param-\mu(\data)\}^{\top} \postdensityfull{\param}{\data}\dee\param$
is the standard posterior covariance.
In this decomposition of $\cov(\bbparamsample\given\data)$,
the first term approximates the mean of the posterior covariance matrix under the sampling distribution, and the
second term approximates the covariance of the posterior mean under the sampling distribution.
Thus, the first term reflects Bayesian model-based uncertainty averaged with respect to frequentist sampling variability,
and the second term reflects frequentist sampling-based uncertainty of the Bayesian model-based estimate.
In \cref{app:jeffrey-conditionalization}, we show that the bagged posterior can also be interpreted in terms of Jeffrey conditionalization,
which provides an alternative view of \cref{eq:bayesbag-posterior-mean,eq:bayesbag-posterior-covariance}.
The following example makes these concepts more concrete.

\bexa[BayesBag for the Gaussian location model] \label{exa:multivariate-gaussian-location}
Consider the Gaussian location model in which observations $\obs{n}$ are i.i.d.\ $\distNorm(\param,V)$ with known covariance matrix $V$, 
and assume a conjugate prior on the mean: $\param\dist\distNorm(0, V_{0})$.
Given data $\data=(\obs{1},\ldots,\obs{\numobs})$, the posterior distribution is $\param\given\data \dist \distNorm(R_{\numobs}\datameanarg{\numobs}, V_{\numobs})$, where 
$\datameanarg{\numobs} \defined \numobs^{-1}\sum_{n=1}^{\numobs}\obs{n}$,
$R_{\numobs} \defined (V_{0}^{-1}V/\numobs + I)^{-1}$, and 
$V_{\numobs} \defined (V_{0}^{-1} + \numobs V^{-1})^{-1}$.
For intuition, one can think of $R_{\numobs} \approx I$ since $\|R_{\numobs} - I\| = O(\numobs^{-1})$.
The bagged posterior mean and covariance are 
\[
\EE(\bbparamsample\given\data)
&= \EE(R_{\bsnumobs}\bsdatarvmeanarg{\bsnumobs}\given\data) 
= R_{\bsnumobs}\datameanarg{\numobs} \label{eq:bayesbag-posterior-mean-normal-location} \\
\cov(\bbparamsample\given\data)
&= \EE(V_{\bsnumobs}\given\data) + \cov(R_{\bsnumobs}\bsdatarvmeanarg{\bsnumobs}\given\data) 
= V_{\bsnumobs} + \bsnumobs^{-1}R_{\bsnumobs}\empcov_{\numobs}R_{\bsnumobs}, \label{eq:bayesbag-posterior-covariance-normal-location}
\]
where $\empcov_{\numobs} \defined \numobs^{-1}\sum_{n=1}^{\numobs} (\obs{n}-\datameanarg{\numobs})(\obs{n}-\datameanarg{\numobs})^{\top}$
is the sample covariance.
In particular, when $\bsnumobs = \numobs$, these expressions simplify to $\EE(\bbparamsample\given\data)  = \EE(\paramsample\given\data)$ and 
$\cov(\bbparamsample\given\data)  = \cov(\paramsample\given\data) +  \numobs^{-1}R_{\numobs}\empcov_{\numobs}R_{\numobs}$.
Unlike the standard posterior, which simply assumes the data have covariance $V$, the bagged posterior accounts for the true covariance of the data through the inclusion 
of the term involving $\empcov_{\numobs}$. 
Thus, we see that the bagged posterior covariance is the sum of
the Bayesian model-based uncertainty plus the frequentist sampling uncertainty. 
\eexa

The decomposition of the bagged posterior covariance in \cref{eq:bayesbag-posterior-covariance} also provides insight into when
we can expect the bagged posterior to be advantageous compared to the standard posterior.
Again, the case of the Gaussian location model is instructive. 

\bexa[Uncertainty calibration in the Gaussian location model] \label{exa:multivariate-gaussian-location-cont} 
Suppose the data are $\obsrv{1},\ldots,\obsrv{\numobs}$ i.i.d.\ $\dist\obsdist$,
and denote the mean and covariance of a single observation by $\mu_{\optsym} \defined \EE(\obsrv{1})$ and $\Sigma_{\optsym} \defined \cov(\obsrv{1})$, respectively. Assume $\mu_{\optsym}$ and $\Sigma_{\optsym}$ are finite.
Then the optimal parameter (in terms of log loss) is $\optparam = \mu_{\optsym}$ and 
$\param\given\datarvarg{\numobs}$ converges in distribution to a point mass at $\mu_{\optsym}$
(see \cref{sec:parameter-estimation} for details).
For quantifying uncertainty about $\optparam$, the posterior is appropriately calibrated if the posterior covariance is equal to 
\[
\EE[(R_{\numobs}\datarvmeanarg{\numobs} - \mu_{\optsym})(R_{\numobs}\datarvmeanarg{\numobs} - \mu_{\optsym})^{\top}] 
&= \numobs^{-1}R_{\numobs}\Sigma_{\optsym}R_{\numobs} + (R_{\numobs} - I)\mu_{\optsym}\mu_{\optsym}^{\top}(R_{\numobs} - I) \\
&= \numobs^{-1}\Sigma_{\optsym} + O(N^{-2}).
\]
If the model is correctly specified, then $\Sigma_{\optsym} = V$.
Hence, in this case the posterior is correctly calibrated, as expected.
If we choose $\bsnumobs = 2\numobs$, then BayesBag is also correctly calibrated since 
the covariance of the bagged posterior is approximately $\numobs^{-1}V = \numobs^{-1}\Sigma_{\optsym}$.
On the other hand, if we use the default choice of $\bsnumobs = \numobs$,
then the covariance of the bagged posterior is approximately $2\numobs^{-1}V = 2\numobs^{-1}\Sigma_{\optsym}$,
overestimating the true uncertainty by only a factor of 2. 

If the model is misspecified, then
the posterior covariance underestimates the true uncertainty unless
$\Sigma_{\optsym} \preceq \numobs V_{\numobs} = V R_{\numobs}$ (that is, unless $V R_{\numobs} - \Sigma_{\optsym}$ is positive semidefinite).
More generally, when $\numobs$ is sufficiently large, the posterior covariance will underestimate (respectively, overestimate) the true uncertainty
about $\param^{\top}v$ for some $v \in \reals^{D}$ if any eigenvalue of $V - \Sigma_{\optsym}$ is negative (respectively, positive). 
Meanwhile, the bagged posterior covariance with $\bsnumobs = \numobs$ is 
\[
V_{\numobs} + \numobs^{-1}R_{\numobs}\empcov_{\numobs}R_{\numobs} \approx  N^{-1}(V + \Sigma_{\optsym}),
\]
so it provides an (asymptotically) conservative uncertainty estimate. 
In the worst-case scenario of $V \ll \Sigma_{\optsym}$, the standard posterior dramatically underestimates the true uncertainty about $\optparam$ while
the bagged posterior is correctly calibrated. 
\eexa

Returning to the generic decomposition given by \cref{eq:bayesbag-posterior-covariance}, we 
show in \cref{sec:parameter-estimation} that for any sufficiently smooth finite-dimensional parametric model,
the covariance of the bagged posterior behaves (qualitatively) like it does in 
\cref{exa:multivariate-gaussian-location,exa:multivariate-gaussian-location-cont}.
In particular, if $\obsrv{1},\ldots,\obsrv{\numobs}$ i.i.d.\ $\dist\obsdist$, then as $N \to \infty$,
\[ \label{eq:bayesbag-asymptotic-covariance}
\cov(\bbparamsample\given\data) \approx \bsnumobs^{-1}\Ehessloglik{\optsym}^{-1} + \bsnumobs^{-1}\Ehessloglik{\optsym}^{-1}\Vargradloglik{\optsym}\Ehessloglik{\optsym}^{-1}, 
\]
where $\Ehessloglik{\optsym}^{-1}$ is the ``model'' covariance (analogous to $V$) and $\Ehessloglik{\optsym}^{-1}\Vargradloglik{\optsym}\Ehessloglik{\optsym}^{-1}$ 
is the ``sandwich'' covariance (analogous to $\Sigma_{\optsym}$).
Behavior analogous to that of the Gaussian location model case holds: 
(1) the (rescaled) sandwich covariance $\numobs^{-1}\Ehessloglik{\optsym}^{-1}\Vargradloglik{\optsym}\Ehessloglik{\optsym}^{-1}$ correctly quantifies the (asymptotic) 
uncertainty about the optimal parameter $\optparam$;
(2) when the model is correctly specified, $\Ehessloglik{\optsym}^{-1} = \Ehessloglik{\optsym}^{-1}\Vargradloglik{\optsym}\Ehessloglik{\optsym}^{-1}$; and 
(3) when the model is misspecified, typically $\Ehessloglik{\optsym}^{-1} \ne \Ehessloglik{\optsym}^{-1}\Vargradloglik{\optsym}\Ehessloglik{\optsym}^{-1}$.

\subsection{BayesBag in practice}

Recall from \cref{eq:bayesbag-approximation} that we can approximate the bagged posterior by averaging standard posteriors conditioned on each 
of $B$ bootstrap datasets $\bsdatasample{1},\dots,\bsdatasample{B}$, where $\bsdatasample{b} \defined (\bsobs{(b)1},\ldots,\bsobs{(b)\bsnumobs})$ 
consists of $\bsnumobs$ samples drawn with replacement from the original dataset $\data = (\obs{1},\ldots,\obs{\numobs})$.
For each $b$, expectations with respect to $\postdensityfull{\param}{\bsdatasample{b}}$ can be computed by whatever method is most
appropriate -- for example, a closed-form solution, Markov chain Monte Carlo, or quadrature. 
In this section, we discuss the choice of the bootstrap size $\bsnumobs$, we show how these considerations naturally 
lead us to use BayesBag for model criticism, and 
we discuss how to choose the number of bootstrap datasets $B$.

\subsubsection{Choosing the bootstrap size for the bagged posterior} \label{sec:M-opt-selection}

A crucial question for the practical application of BayesBag is how to select the bootstrap dataset size $\bsnumobs$.
For a real-valued function of interest $f$, let $v_{\bsnumobs}$ and $v_{\numobs}^{\bbsym}$ denote, respectively, the standard and bagged posterior variances of $f(\param)$.
Since we can redefine $\param$ to be $f(\param)$, without any loss of generality, for the moment we assume that $\param \in \reals$
is the quantity of interest
and $f = \id$ is the identity function, $f(\theta) = \theta$.
Let $\sigma^{2}_{\optsym} \defined  \Ehessloglik{\optsym}^{-1}$ denote the model-based asymptotic variance
and let $s^{2}_{\optsym} \defined \Ehessloglik{\optsym}^{-1}\Vargradloglik{\optsym}\Ehessloglik{\optsym}^{-1}$ denote the sampling-based (sandwich) variance. 

Conceptually, the situation is as follows.
Bootstrapping with $M=N$ typically increases the variance relative to the standard posterior, and as $\bsnumobs$ grows the variance decreases. 
We are trying to balance these two tendencies in order to match the frequentist sandwich (co)variance.
In particular, the bagged posterior variance needs to be approximately $s^{2}_{\optsym}/\numobs$ in order to be well-calibrated.
If follows from \cref{eq:bayesbag-asymptotic-covariance} that the bagged posterior variance when using a bootstrap sample size $\bsnumobs$ is 
$v_{\bsnumobs}^{\bbsym} \approx (\sigma^{2}_{\optsym} + s_{\optsym}^{2})/\bsnumobs$. 
Setting $(\sigma_{\optsym}^{2} + s_{\optsym}^{2})/\bsnumobs = s_{\optsym}^{2}/\numobs$ and solving for $\bsnumobs$
shows that we should choose
\[
\bsnumobs = \optMasympt(\id) \defined (1 + \sigma_{\optsym}^{2} / s_{\optsym}^{2}) \numobs. \label{eq:optimal-asymptotic-M}
\]
Thus, if $s_{\optsym}^{2} = \sigma_{\optsym}^{2}$ (i.e., the model variance is correctly specified),
then we should choose $\bsnumobs = 2\numobs$;
this is in agreement with \cref{exa:multivariate-gaussian-location-cont}.
If $s_{\optsym}^{2} > \sigma_{\optsym}^{2}$, then the sampling-based term is larger, and we should choose $\bsnumobs \in [\numobs, 2\numobs)$.
If $s_{\optsym}^{2} < \sigma_{\optsym}^{2}$, then $\bsnumobs > 2\numobs$ is preferred.

A conservative default would be to choose $\bsnumobs = \numobs$ since
this protects against having an over-confident posterior in the presence of misspecification
and only over-inflates the posterior variance by a factor of $2$ when the model is correct.
Alternatively, we can estimate $\optMasympt$ using \cref{eq:optimal-asymptotic-M} by plugging in an estimate of $\sigma_{\optsym}^{2}/s_{\optsym}^{2}$.
To obtain such an estimate, we use the fact that the posterior variance satisfies $v_{\numobs} \approx \sigma_{\optsym}^{2}/\numobs$ and 
the bagged posterior variance satisfies $v_{\numobs}^{\bbsym} \approx (\sigma^{2}_{\optsym} + s_{\optsym}^{2})/\numobs$.
Combining these two equations and solving, we find that $\sigma_{\optsym}^{2}/s_{\optsym}^{2} \approx v_{\numobs}/(v_{\numobs}^{\bbsym} - v_{\numobs})$. 
Plugging this into \cref{eq:optimal-asymptotic-M} yields the estimator
\[
\optMasymptest(\id) \defined \frac{v_{\numobs}^{\bbsym}}{v_{\numobs}^{\bbsym} - v_{\numobs}}\numobs. \label{eq:optimal-asymptotic-M-est}
\]

Returning to the case of general $f$, we can define $\optMasympt(f)$ and $\optMasymptest(f)$ analogously to the special case of $f = \id$. 
If $\optMasymptest(f)$ differs significantly from $\numobs$, then we suggest recomputing the bagged posterior with $\bsnumobs = \optMasymptest(f)$. 
Note that since $v_{\numobs}$ and $v_{\numobs}^{\bbsym}$ are nonnegative, it follows that $\optMasymptest(f) \in (-\infty, 0) \cup [\numobs, \infty]$. 
If $\optMasymptest(f)  < 0$, then $v_{\numobs} > v_{\numobs}^{\bbsym}$, which indicates that using the asymptotically optimal estimate of bootstrap size is not appropriate.

For a set of functions of interest $\mcF$,
we suggest taking the most conservative sample size: $\optMasymptest(\mcF) \defined \inf_{f \in \mcF}\optMasymptest(f)$. 
In general, $\mcF$ can be chosen to reflect the quantity or quantities of interest to the ultimate statistical analysis. 
When $\param \in \reals^{D}$, two natural choices for the function class are $\mcF_{1} \defined \{ \param \mapsto w^{\top}\param : \twonorm{w} = 1 \}$ 
and $\mcF_{\text{proj}} = \{ \param \mapsto \param_{d} : d = 1,\dots,D\}$.
In our experiments we use the latter and therefore adopt the shorthand notation $\optMasymptest \defined \optMasymptest(\mcF_{\text{proj}})$. 

Using the finite-sample covariance expression in \cref{eq:bayesbag-posterior-covariance-normal-location} for the bagged posterior under the Gaussian location model,
we can also define a finite-sample version of $\optMasympt$, denoted $\optMfs$.
To construct an estimator for $\optMfs$, let $v_{0}$ denote the prior variance and define the estimators 
$\hat\sigma_{\optsym}^{2} \defined \numobs v_{0} v_{\numobs}/(v_{0} - v_{\numobs})$
and
\[
\hs_{\optsym}^{2} &\defined \frac{v_{0}^{2}}{(v_{0} - v_{\numobs})^{2}}(v^{\bbsym}_{\numobs} - v_{\numobs})\numobs.
\]
The estimator for $\optMfs$ is given by
\[
\optMfsest(f)
&\defined
   \frac{\numobs}{2} + \frac{\numobs\hat\sigma_{\optsym}^{2}}{2\hs_{\optsym}^{2}}  - \frac{\hat\sigma_{\optsym}^{2}}{v_{0}}
  +   \left\{ \left(\frac{\numobs}{2} + \frac{\numobs\hat\sigma_{\optsym}^{2}}{2\hs_{\optsym}^{2}}\right)^{2} - \frac{\numobs\hat\sigma_{\optsym}^{2}}{v_{0}}\right\}^{1/2} 
\label{eq:optimal-fs-M}
\]
when the right hand side of \cref{eq:optimal-fs-M} is well-defined and positive; otherwise, we set $\optMfsest = \numobs$. 
See \cref{app:finite-sample-optimal-bootstrap-size} for the derivation of \cref{eq:optimal-fs-M}.

\begin{rmk}
Fundamentally, we construct $\optMasymptest$ and $\optMfsest$ using an estimator for the sandwich variance $s_{\optsym}^{2}$,
which may be difficult to accurately estimate.
However, note that we can always default to the conservative choice $\bsnumobs = \numobs$
when it is hard to estimate the optimal choice of $M$.
Further, the optimal bootstrap sample size estimators will tend to be effective when $f(\param)$ is roughly Gaussian-distributed,
and since we only need to estimate the sandwich variance for the univariate quantity $f(\param)$, it is plausible to find 
roughly Gaussian behavior even with relatively small samples sizes and even if $\param$ is high-dimensional. 
Thus, the applicability of  $\optMasymptest$ and $\optMfsest$ is greater than it might first appear.
\end{rmk}

\subsubsection{Model criticism with BayesBag}

We can also use $\optMasymptest(f)$ or $\optMfsest(f)$ for model criticism, which is the task of diagnosing any mismatch between the assumed model 
and the observed data. 
Let $\optMest(f)$ denote either $\optMasymptest(f)$ or $\optMfsest(f)$.
When the model is correctly calibrated, we expect $\optMest(f) \approx 2 \numobs$. 
When the standard posterior is overconfident (respectively, under-confident), we expect $\optMest(f) < 2 \numobs$ (respectively, $\optMest(f) > 2 \numobs$).
These observations suggest we could defined a measure of model fit in terms of $\numobs$ and $\optMest(f)$.
Toward that end, we propose the \emph{model--data mismatch index}, which we define as
\[
\modelmismatch(f) \defined \begin{cases}
2\numobs/\optMest(f) - 1 & \text{if $\optMest(f) \in [\numobs,\infty)$} \\
\nan & \text{otherwise.}
\end{cases} \label{eq:simple-mismatch-index}
\] 
If we are using $\optMfsest(f)$, then we also set $\modelmismatch(f) = \nan$ when the right hand side of \cref{eq:optimal-fs-M} is not well-defined or is negative. 
For a function class $\mcF$, we define $\modelmismatch(\mcF)$ by replacing $\optMasymptest(f)$ with $\optMasymptest(\mcF)$ in \cref{eq:simple-mismatch-index}.
Also, let $\modelmismatch \defined \modelmismatch(\mcF_{\text{proj}})$. 

The mismatch index provides a simple, intuitive, and theoretically well-grounded method for measuring the fit of the model to the data. 
Its interpretation is as follows: $\modelmismatch(f) \approx 0$ indicates no evidence of mismatch since then $\optMest(f) \approx 2 \numobs$;
$\modelmismatch(f) > 0$ (respectively, $\modelmismatch(f) < 0$) indicates the standard posterior is overconfident (respectively, under-confident)
since then $\optMest(f) < 2 \numobs$ (respectively, $\optMest(f) > 2 \numobs$);
$\modelmismatch(f)= \nan$ indicates either that the assumptions required to use $\optMest(f)$ do not hold 
(e.g., due to multimodality in the posterior or small sample size) or there is a severe model--data mismatch. 
To more fully understand why we set $\modelmismatch(f) = \nan$ when $\optMest(f) < \numobs$, we consider the cases of $\optMasymptest(f)$ and $\optMfsest(f)$ separately.
By construction, $\optMasymptest \in [\numobs, \infty)$ unless $v^{\bbsym}_{\numobs} < v_{\numobs}$; in which case $\optMasymptest(f) < 0$, which is nonsensical.
On the other hand, $\optMfsest(f) < \numobs$ when $\hs_{\optsym}^{2}/\numobs > 0.5v_{0}\{1 + v_{0}/(2\numobs \hsigma_{\optsym}^{2} + v_{0})\}$;
in other words, when the (estimated) optimal posterior variance is large relative to the prior variance, which indicates that the assumptions used to construct $\optMfsest(f)$ 
do not hold (since posterior variance should generally be smaller than prior variance).
In either case, $\optMest(f) < \numobs$ indicates that either 
(1) there is severe model--data mismatch or (2) the posterior is multimodal or otherwise far from a Gaussian approximation. 
Hence, we choose to set $\modelmismatch(f) = \nan$ when $\optMest(f) < \numobs$.

\subsubsection{Choosing the number of bootstrap datasets for BayesBag}  \label{sec:choosing-B}

If we wish to use \cref{eq:bayesbag-approximation}  to approximate the bagged posterior,
then %
we must select the number of bootstrap datasets $B$. 
Assume that we can approximate $\postdensityfull{\param}{\bsdatasample{b}}$ to high accuracy.
Then evaluating the accuracy of the BayesBag approximation given by \cref{eq:bayesbag-approximation} reduces to the well-studied problem of estimating the accuracy of
a simple Monte Carlo approximation~\citep[e.g.,][]{Koehler:2009}. 
In practice, we have found it sufficient to take $B = 50$ or $100$ since the quantities we wish to estimate seem to be fairly low-variance.
Thus, we suggest starting with $B = 50$, estimating the Monte Carlo error of any quantities of interest such as parameter 
means and variances, and then increasing $B$ if the estimated error is unacceptably large. 
On the other hand, in some scenarios it may be desirable to reduce computational expense by balancing the number of bootstrap samples $B$
versus the accuracy of the approximation to 
$\postdensityfull{\param}{\bsdatasample{b}}$ (e.g., in terms of the length of Markov chain Monte Carlo runs). 
See \cref{app:computation} for a discussion of this computational trade-off.

\section{Theory of parameter inference with BayesBag} \label{sec:parameter-estimation}

This section formalizes and generalizes the results sketched out in \cref{sec:parameter-theory-sketch}.
Readers who are not interested in these technical details can safely proceed to \cref{sec:simulations}. 
Our main result (\cref{thm:bb-bvm}) is a Bernstein--Von Mises theorem for the bagged posterior 
under sufficiently regular finite-dimensional models.
In particular, we show that while the standard Bayesian posterior may be arbitrarily under- or over-confident when the model is misspecified,
the bagged posterior avoids over-confident uncertainty quantification by accounting for sampling variability.
Since \cref{thm:bb-bvm} is asymptotic in nature, 
it ignores the potentially significant finite-sample benefits of both the bootstrap and the posterior, neither of which requires
the normality assumptions of our asymptotic analysis. 
Nevertheless, the theorem offers valuable statistical justification for BayesBag in general and 
for the use of $\optMasymptest$ and $\modelmismatch$ in particular.
\cref{prop:bb-bbvm-gaussian-location} is a simpler version of the same result for the univariate Gaussian location model,
for which the statement and proof of the theorem are much more transparent.

\subsection{Background} \label{sec:background}

We now restate more precisely the setting that was informally introduced in \cref{sec:introduction,sec:overview}. 
Consider a model $\{\likdist{\param} : \param \in \paramspace\}$ for independent and identically distributed (\iid)\ data 
$\obs{1},\dots,\obs{\numobs}$, where $\obs{n}\in\obsspace$ and $\paramspace\subset\reals^{D}$.
Denote $\dataarg{\numobs} = (\obs{1},\dots,\obs{\numobs})$ and suppose $\likfun{\param}$ is the density of $\likdist{\param}$ with respect to some reference measure.
The standard Bayesian posterior distribution given $\dataarg{\numobs}$ is 
\[ 
\postdistfull{\dee\param}{\dataarg{\numobs}} 
\defined \frac{\prod_{n=1}^{\numobs}\lik{\obs{n}}{\param}}{\marginallikfull{\dataarg{\numobs}}}\priordist(\dee\param), \label{eq:iid-data-posterior}
\]
where $\priordist(\dee\param)$ is the prior distribution and
$\marginallikfull{\dataarg{\numobs}} \defined \int\{\prod_{n=1}^{\numobs}\lik{\obs{n}}{\param}\}\priordist(\dee\param)$ is the marginal likelihood. 
When convenient, we will use the shorthand notation $\postdist{\numobs} \defined \postdistfull{\cdot}{\dataarg{\numobs}}$.

Assume the observed data $\obsrv{1},\dots,\obsrv{\numobs}$ is actually generated \iid\ from some unknown distribution $\obsdist$.
Suppose there is a unique parameter $\optparam$ that 
minimizes the Kullback--Leibler divergence from $\obsdist$ to the model, or equivalently, $\optparam = \argmax_{\param \in \paramspace}\EE\{\log \lik{\obsrv{1}}{\param}\}$.
Denote the log-likelihood by $\loglikfun{\param} \defined \log \likfun{\param}$, its gradient by $\gradloglikfun{\param} \defined \grad_{\param}\loglikfun{\param}$,
and its Hessian by $\hessloglikfun{\param} \defined \grad_{\param}^2\loglikfun{\param}$.
Furthermore, define the information matrices $\Ehessloglik{\param} \defined -\EE\{\hessloglik{\obsrv{1}}{\param}\}$ 
and $\Vargradloglik{\param} \defined \cov\{\gradloglik{\obsrv{1}}{\param}\}$.

Let $\mle{\numobs} \defined \argmax_{\param} \prod_{n=1}^{\numobs}\lik{\obsrv{n}}{\param}$ denote the maximum likelihood estimator.
Under regularity conditions, $\mle{\numobs}$ is asymptotically normal in the sense that 
\[ \label{eq:mle-asympotics}
\numobs^{1/2}(\mle{\numobs} - \optparam) \convD \distNorm(0,\Ehessloglik{\optparam}^{-1}\Vargradloglik{\optparam}\Ehessloglik{\optparam}^{-1}),
\]
where $\Ehessloglik{\optparam}^{-1}\Vargradloglik{\optparam}\Ehessloglik{\optparam}^{-1}$ is known as the sandwich covariance~\citep{White:1982}.
Under mild conditions, the Bernstein--Von Mises theorem (\citealp[Ch.~10]{vanderVaart:1998} and \citealp{Kleijn:2012}) guarantees that for $\paramsample \dist \postdist{\numobs}$,
\[ \label{eq:bayes-asymptotics}
\numobs^{1/2}(\paramsample - \mle{\numobs}) \given \datarvarg{\numobs} \convD \distNorm(0, \Ehessloglik{\optparam}^{-1}).
\]
Hence, the standard posterior is correctly calibrated, asymptotically, if the covariance matrices of the Gaussian distributions in \cref{eq:mle-asympotics,eq:bayes-asymptotics}
coincide -- that is, if $\Ehessloglik{\optparam}^{-1}\Vargradloglik{\optparam}\Ehessloglik{\optparam}^{-1} =  \Ehessloglik{\optparam}^{-1}$,
which is implied by $\Vargradloglik{\optparam} = \Ehessloglik{\optparam}$.
In particular, if $\Vargradloglik{\optparam} = \Ehessloglik{\optparam}$, then Bayesian credible sets are (asymptotically) 
valid confidence sets in the frequentist sense: sets of posterior probability $1 - \alpha$ contain the true parameter with $\obsdist^{\infty}$-probability $1 - \alpha$, under mild conditions.

If the model is well-specified, that is, if $\obsdist = \likdist{\trueparam}$ for some parameter $\trueparam \in \paramspace$
(and thus $\optparam = \trueparam$ by the uniqueness assumption), then $\Vargradloglik{\optparam} = \Ehessloglik{\optparam}$ under very mild conditions.
On the other hand, if the model is misspecified -- that is, if $\obsdist \ne \likdist{\param}$ for all $\param \in \paramspace$ --
then although \cref{eq:bayes-asymptotics} still holds, typically $\Vargradloglik{\optparam} \ne \Ehessloglik{\optparam}$. 
If $\Vargradloglik{\optparam} \ne \Ehessloglik{\optparam}$, then the standard posterior is not correctly calibrated, and in fact, asymptotic Bayesian credible sets may be arbitrarily 
over- or under-confident. 

Although \cref{eq:mle-asympotics,eq:bayes-asymptotics} are only asymptotic, we should take little comfort that the non-asymptotic situation
will somehow be better.
We have already seen in \cref{exa:multivariate-gaussian-location-cont} that, no matter the sample size, the standard posterior for the Gaussian location model 
can be badly miscalibrated.
The prior may help to down-weight \emph{a priori} unlikely hypotheses, but it cannot account for misspecification in the likelihoods 
among ``reasonable'' parameter values. 

\subsection{BayesBag for parameter inference}

Let $\bsdatarvarg{\bsnumobs}$ denote a bootstrapped copy of $\datarvarg{\numobs}$ with $\bsnumobs$ observations;
that is, each observation $\obsrv{n}$ is replicated $\bscount{n}$ times in $\bsdatarvarg{\bsnumobs}$,
where $\bscounts \dist \distMulti(\bsnumobs, 1/\numobs)$ is a multinomial-distributed count vector of length $\numobs$.
We define the \emph{bagged posterior} $\bbpostdistfull{\cdot}{\datarvarg{\numobs}}$ by setting
\[
\bbpostdistfull{A}{\datarvarg{\numobs}} \defined \EE\{\postdistfull{A}{\bsdatarvarg{\bsnumobs}} \given \datarvarg{\numobs} \}
\]
for all measurable $A \subseteq \paramspace$.
Note that this is equivalent to the informal definition in \cref{eq:bayesbag-definition}.
In order words, BayesBag uses bootstrapping to average posteriors over approximate realizations of data from the true data-generating distribution. 
Depending on $\bsnumobs$, the bagged posterior may be more diffuse or less diffuse than the standard posterior. 

To avoid notational clutter, we suppress the dependence of $\bbpostdistfull{\cdot}{\datarvarg{\numobs}}$ on $\bsnumobs$.
We will typically use the shorthand notation $\bbpostdist{\numobs} \defined \bbpostdistfull{\cdot}{\datarvarg{\numobs}}$ and
we let $\bbparamsample\given\datarvarg{\numobs} \dist \bbpostdist{\numobs}$ denote a random variable distributed according to the bagged posterior.
We assume $\paramspace$ is an open subset of $\reals^{D}$ and we write $\dee\param$ to denote Lebesgue measure on $\paramspace$.
Further, we assume $\postdist{\numobs}$ and $\bbpostdist{\numobs}$ have densities $\postdensity{\numobs}$ and $\bbpostdensity{\numobs}$, respectively, 
with respect to Lebesgue measure.  Note that $\bbpostdensity{\numobs}$ exists if $\postdensity{\numobs}$ exists.

\subsection{Asymptotic normality of BayesBag for the Gaussian location model} \label{sec:bb-gaussian-location}

Before developing a general Bernstein--Von Mises theorem for BayesBag, as a warmup we prove \cref{prop:bb-bbvm-gaussian-location}, 
a simpler version of the result 
in the case of the Gaussian location model from \cref{exa:multivariate-gaussian-location,exa:multivariate-gaussian-location-cont}.
Although it is a special case of \cref{thm:bb-bvm},
the statement and proof of \cref{prop:bb-bbvm-gaussian-location} are much easier to follow
and it still captures the essence of the general result.
For maximal clarity, we consider the case of univariate data.

\bnthm \label{prop:bb-bbvm-gaussian-location}
Let $\obsrv{1},\obsrv{2},\ldots\in\reals\;\iid$ such that $\EE(|\obsrv{1}|^{3}) < \infty$.
Let $\bbparamsample\given\datarvarg{\numobs} \dist \bbpostdist{\numobs}$ and 
suppose $\bsscale \defined \lim_{\numobs \to \infty} \bsnumobs/\numobs \in (0,\infty)$ for $\bsnumobs = \bsnumobs(\numobs)$.
Then for almost every $(\obsrv{1},\obsrv{2},\ldots)$,
\[
\numobs^{1/2}\big\{\bbparamsample - \EE(\bbparamsample\given\datarvarg{\numobs})\big\}  \given \datarvarg{\numobs} \convD \distNorm(0, \;V/\bsscale + \var(\obsrv{1})/\bsscale).  \label{eq:bb-bbvm-gaussian-location}
\]
\enthm
In other words, with probability 1, the bagged posterior converges weakly to $\distNorm(0, \;V/\bsscale + \var(\obsrv{1})/\bsscale)$
after centering at its mean and scaling by $\numobs^{1/2}$.
The proof of \cref{prop:bb-bbvm-gaussian-location} is in \cref{sec:proof-of-bb-bbvm-gaussian-location}.

\subsection{Bernstein--Von Mises theorem for BayesBag}
\label{sec:bvm}

We now turn to the main result of this section: a general Bernstein--Von Mises theorem for BayesBag. 
Recall that if the standard posterior were asymptotically correctly calibrated, it would have asymptotic covariance 
$\Ehessloglik{\optparam}^{-1}\Vargradloglik{\optparam}\Ehessloglik{\optparam}^{-1}$ (the sandwich covariance), whereas in fact it has asymptotic covariance
$\Ehessloglik{\optparam}^{-1}$. %
Letting $\bsscale \defined \lim_{\numobs \to \infty} \bsnumobs/\numobs$ as in \cref{prop:bb-bbvm-gaussian-location},
we show that the asymptotic covariance of the bagged posterior is 
$\Ehessloglik{\optparam}^{-1}/\bsscale + \Ehessloglik{\optparam}^{-1}\Vargradloglik{\optparam}\Ehessloglik{\optparam}^{-1}/\bsscale$,
which mirrors the form of \cref{eq:bb-bbvm-gaussian-location} but is much more general.
Our technical assumptions %
are essentially the same as those used by \citet{Kleijn:2012} to prove the Bernstein--Von Mises theorem under misspecification for the standard posterior. 

For a measure $\nu$ and function $f$, we will make use of the shorthand $\nu(f) \defined \int f \dee \nu$.  
Let $\alldatarv$ denote the infinite sequence of data $(\obsrv{1},\obsrv{2},\dots)$.

\bnthm \label{thm:bb-bvm}
Suppose $\obsrv{1},\obsrv{2},\dots\;\iid\dist\obsdist$ and assume that:
\begin{enumerate}[label=(\roman*)]
\item $\param \mapsto \loglik{\obsrv{1}}{\param}$ is differentiable at $\optparam$ in probability;
\item there is an open neighborhood $U$ of $\optparam$
and a function $m_{\optparam} : \obsspace \to \reals$ such that $\obsdist(m_{\optparam}^{3}) < \infty$ and for all $\param,\param' \in U$,
$|\loglikfun{\param} - \loglikfun{\param'}| \le m_{\optparam}\twonorm{\param - \param'}$ a.s.$[\obsdist]$;
\item $-\obsdist(\loglikfun{\param} - \loglikfun{\optparam}) = \frac{1}{2}(\param - \optparam)^{\top}\Ehessloglik{\optparam}(\param - \optparam) + \littleo(\twonorm{\param - \optparam}^{2})$ as $\param \to \optparam$;
\item $\Ehessloglik{\optparam}$ is an invertible matrix;
\item letting $\bbparamsample \dist \bbpostdist{\numobs}$,  it holds that conditionally on $\alldatarv$, for almost every $\alldatarv$, for every sequence of constants $\concconst\to\infty$,
\[
\EE\Big\{\postdistfull{\twonorm{\bbparamsample - \optparam} > \concconst/\bsnumobs^{1/2}}{\bsdatarvarg{\bsnumobs}} \;\Big\vert\; \datarvarg{\numobs}\Big\} \to 0;
\]
and
\item  $\bsscale \in (0,\infty)$. %
\end{enumerate}
Then, letting $\bbparamsample \dist \bbpostdist{\numobs}$, we have that conditionally on $\alldatarv$, for almost every $\alldatarv$,
\[
\numobs^{1/2}(\bbparamsample - \optparam) -\Delta_{\numobs} \given \datarvarg{\numobs} \convD \distNorm(0, \Ehessloglik{\optparam}^{-1}/\bsscale + \Ehessloglik{\optparam}^{-1}\Vargradloglik{\optparam}\Ehessloglik{\optparam}^{-1}/\bsscale),
\]
where $\Delta_{\numobs} \defined \numobs^{1/2}\Ehessloglik{\optparam}^{-1}(\empdist - \obsdist)\gradloglikfun{\optparam}$
and $\empdist \defined \numobs^{-1}\sum_{n=1}^{\numobs}\delta_{\obsrv{n}}$.
\enthm

The proof of \cref{thm:bb-bvm} is in \cref{sec:proof-of-bb-bvm}.
To interpret this result, it is helpful to compare it to the behavior of the standard posterior. 
Under the conditions of \cref{thm:bb-bvm}, letting $\paramsample \dist \postdist{\numobs}$,
then almost surely $\numobs^{1/2}(\paramsample - \optparam) - \Delta_{\numobs} \convD \distNorm(0, \Ehessloglik{\optparam}^{-1})$
by \citet[Theorem 2.1 and Lemma 2.1]{Kleijn:2012}.
Thus, the bagged posterior and the standard posterior for $\numobs^{1/2}(\param - \optparam)$  have the same asymptotic mean, $\Delta_{\numobs}$, 
but the bagged posterior has asymptotic covariance $\Ehessloglik{\optparam}^{-1}/\bsscale + \Ehessloglik{\optparam}^{-1}\Vargradloglik{\optparam}\Ehessloglik{\optparam}^{-1}/\bsscale$
instead of $\Ehessloglik{\optparam}^{-1}$.

\subsection{Extensions}

There are many possible extensions to \cref{thm:bb-bvm}. 

\paragraph*{Regression models}
\cref{thm:bb-bvm} applies equally well to the regression setting with random regressors where the data take the form $\obsrv{n} = (Y_{n}, Z_{n})$ 
and the models $\lik{y \given z}{\param}$ are conditional.

\paragraph*{Alternative bootstrap methods}
Much of bootstrap theory extends beyond the multinomial distribution for $\bscounts$ to other distributions
such as those where $\bscount{1},\ldots,\bscount{\numobs}$ are weakly correlated random variables with mean $1$ and variance $1$~\citep{vanderVaart:1996}.
Thus, we conjecture that
\cref{thm:bb-bvm} also holds when we use the bagged posterior proportional to $\priordensity(\param)\prod_{n=1}^{\numobs}\lik{\obsrv{n}}{\param}^{\bscount{n}}$, 
where $\bscount{1},\ldots,\bscount{\numobs}$ are independent nonnegative random variables satisfying $\EE(\bscount{n})=1$ and $\var(\bscount{n})=1$ $(n=1,\dots,\numobs)$. 

\paragraph*{Dependent observations}
We have also focused on the case of independent observations $\obsrv{1},\ldots,\obsrv{\numobs}$,
but it is feasible to extend our theory and methodology to more complex models.
One possibility is to draw on the rich existing bootstrap literature for time series and spatial models~\citep[e.g.,][]{Kunsch:1989,Peligrad:1998}.
Alternatively, a model-based bootstrap approach could be used by employing a nonparametric or rich parametric model to approximate $\obsdist$. 
We leave investigation of these extensions for future work.

\section{Simulations} \label{sec:simulations}

In this section, we validate the effectiveness of BayesBag through two simulation studies.
Some implementation details are deferred to \cref{app:experiment-details}. 

\subsection{Linear regression} \label{sec:linreg}

We performed an extensive set of simulations
to assess the performance of BayesBag in the setting of linear regression.
Linear regression is an ideal model for investigating the properties of BayesBag and the usefulness of the mismatch index $\modelmismatch$
and optimal bootstrap sample size estimator $\optMest$, since all computations of posterior quantities can be done in closed form,
yet it is a rich enough model that we can explore many kinds of model--data mismatch. 
The data consist of regressors $Z_{n} \in \reals^{D}$ and observations $Y_{n} \in \reals~(n=1,\dots,\numobs)$,
and the parameter is $\param = (\param_{0}, \dots, \param_{D}) = (\log \sigma^{2}, \beta_{1},\dots,\beta_{D}) \in \reals^{D+1}$.
Assuming conjugate priors, the assumed model is 
\[
\sigma^{2} &\dist \distInvGam(a_{0}, b_{0}) \\
\beta_{d} &\given \sigma^{2} \distiid\distNorm(0, \sigma^{2}/\lambda)  & d &=1,\dots,D, \\
Y_{n} &\given Z_{n}, \beta, \sigma^{2} \distind \distNorm(Z_{n}^{\top}\beta, \sigma^{2}) & n &=1,\dots,\numobs,
\]
where $a_{0}, b_{0}$, and $\lambda$ are hyperparameters that will be specified later. 
We simulated data by generating 
$Z_{n} \distiid G$, $\eps_{n} \distiid \distNorm(0,1)$, and 
\[
Y_{n} = f(Z_{n})^{\top}\beta_{\dagger} + \eps_{n}  \label{eq:simulated-linreg-data}
\]
for $n =1,\dots,N$,
where we used two settings for each of the regressor distribution $G$, %
the regression function $f$, and the coefficient vector $\beta_{\dagger} \in \reals^{D}$.

\bitems
\item \textbf{Regressor distribution $G$.} By default, we used $G = \distNorm(0, I)$ to simulate data; we refer to this as the \textsf{uncorrelated} setting.
Alternatively, we used a \textsf{correlated} setting, where, for $h = 10$, $Z \dist G$ was defined by generating
$\xi\dist\chi^{2}(h)$ and then 
$Z \given \xi \dist\distNorm(0,\Sigma)$ where $\Sigma_{dd'} = \exp\{-(d-d')^{2}/64\} / (\xi_{d}\xi_{d'})$ and
$\xi_{d} = \sqrt{\xi / (h-2)}^{\indicatorfn(d \text{ is odd})}$.
The motivation for the \textsf{correlated} sampling procedure was to generate correlated regressors that have different
tail behaviors while still having the same first two moments, since regressors are typically standardized to have mean 0 and variance 1. 
Note that, marginally, $Z_{1}, Z_{3}, \dots$ are each rescaled $t$-distributed random variables with $h$ degrees of freedom such that $\var(Z_{1}) = 1$,
and $Z_{2}, Z_{4},\dots$ are standard normal. 

\item \textbf{Regression function $f$.} By default, we used a \textsf{linear} regression function $f(z) = z$ to simulate data.
Alternatively, we used the \textsf{nonlinear} function $f(z) = (z_{1}^{3},\ldots,z_{D}^3)^{\top}$.

\item \textbf{Coefficient vector $\beta_{\dagger}$.} By default, we used a \textsf{dense} vector with $\beta_{\dagger d} = 2^{(5-d)/2}$ for $d = 1,\ldots,D$ to simulate data.
Alternatively, we used a \textsf{$k$-sparse} vector with $\beta_{\dagger d} = 1$ if $d \in \{ \lfloor j (D+\tfrac{1}{2})/(k+1)\rfloor \given j = 1,\dots,k\}$ and $\beta_{\dagger d} = 0$ otherwise. 
\eitems

For brevity, we omit the default setting labels when indicating which settings of $G$, $f$, and $\beta_{\dagger}$ were used to generate each dataset.
For example, we abbreviate \textsf{uncorrelated-nonlinear-2-sparse} as \textsf{nonlinear-2-sparse}, and \textsf{correlated-linear-dense} as \textsf{correlated}.
We refer to the dataset with all three defaults as \textsf{default}.

Unless stated otherwise, the data were generated with $D=10$ and $N=50$, and the model
hyperparameters were set to $a_{0} = 2$, $b_{0} = 1$, and $\lambda = 1$.
Each experimental setting was replicated 50 times. 
BayesBag was run with $\bsnumobs = \optMfsest$ and $B=100$ since pilot experiments revealed no noticeable differences when larger values of $B$ were used.

\begin{figure}[tbp]
\begin{center}
\begin{subfigure}[b]{.49\textwidth}
\centering
\includegraphics[height=1.5in]{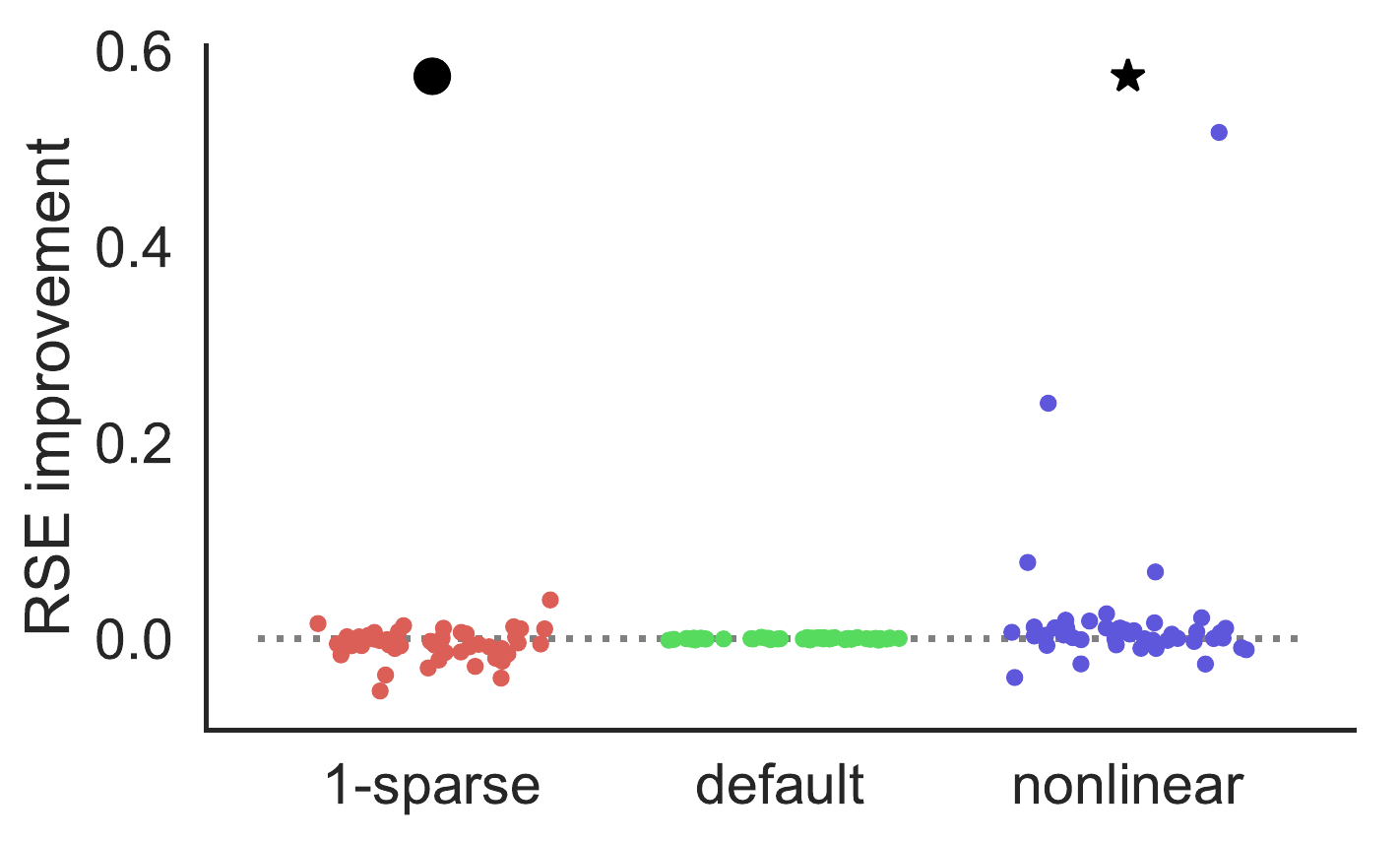}
\caption{$\lambda = 1$}
\end{subfigure}%
\begin{subfigure}[b]{.49\textwidth}
\centering
\includegraphics[height=1.5in]{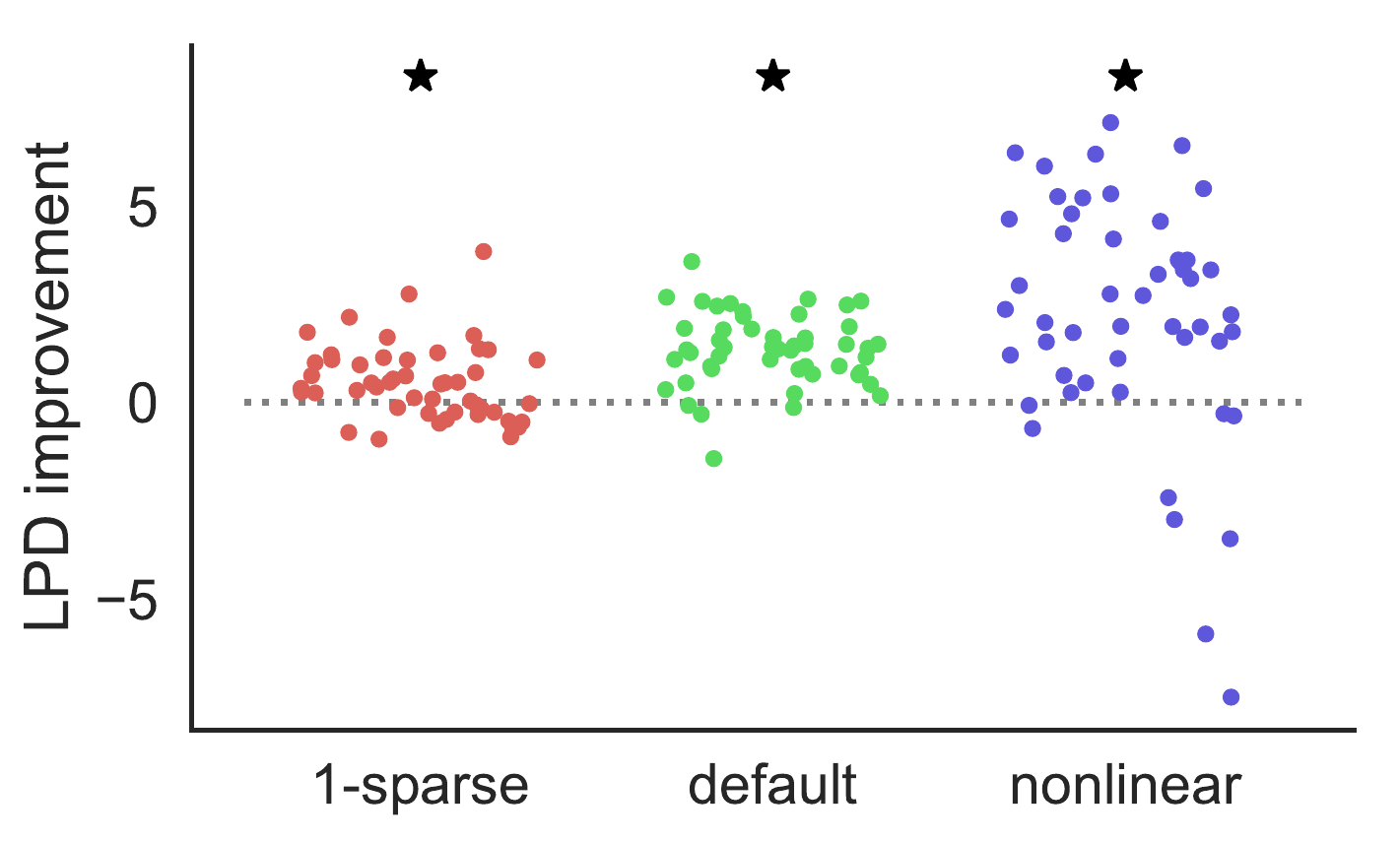}
\caption{$\lambda = 1$}
\end{subfigure} \\
\begin{subfigure}[b]{.49\textwidth}
\centering
\includegraphics[height=1.5in]{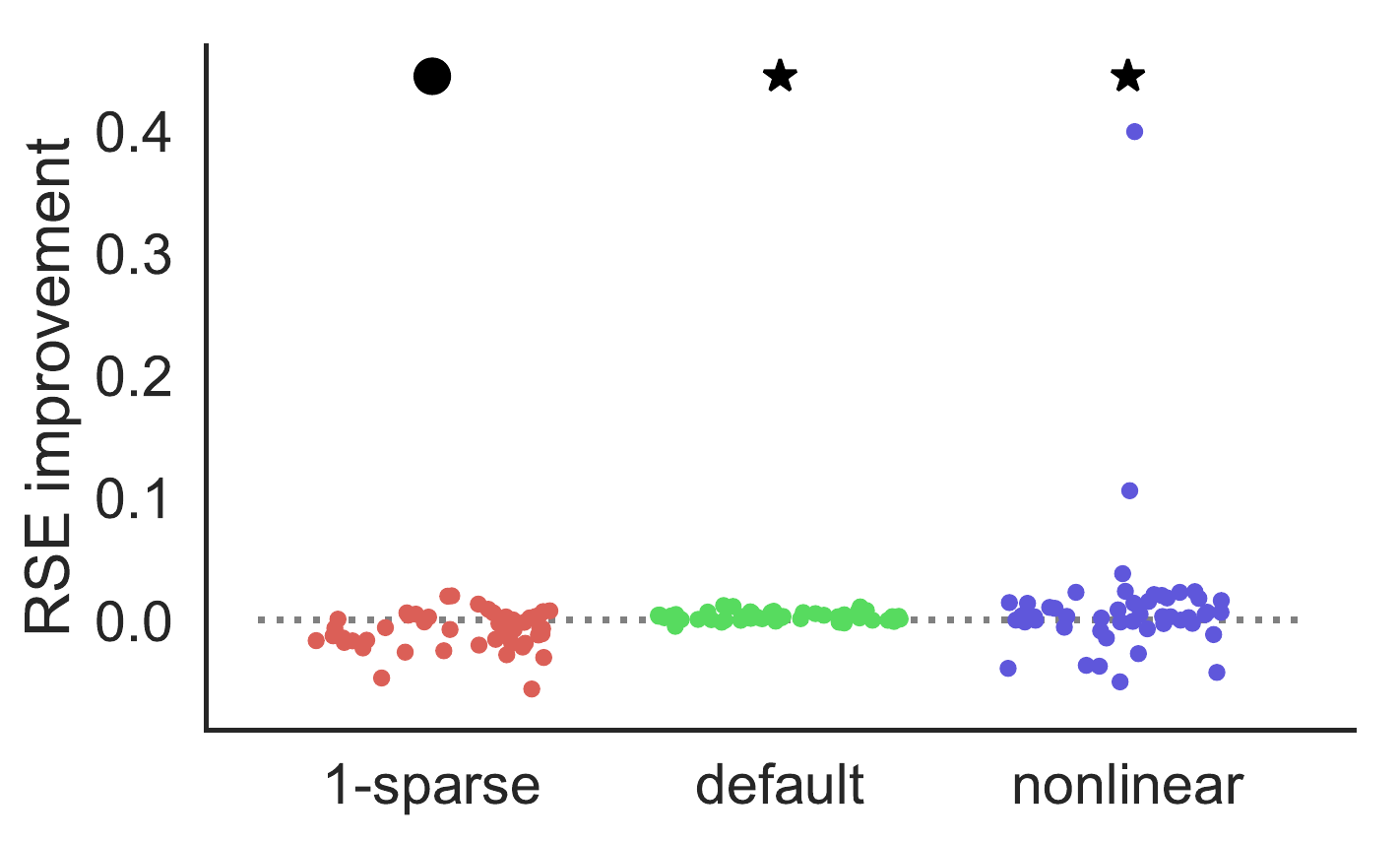}
\caption{$\lambda = 4$}
\end{subfigure}%
\begin{subfigure}[b]{.49\textwidth}
\centering
\includegraphics[height=1.5in]{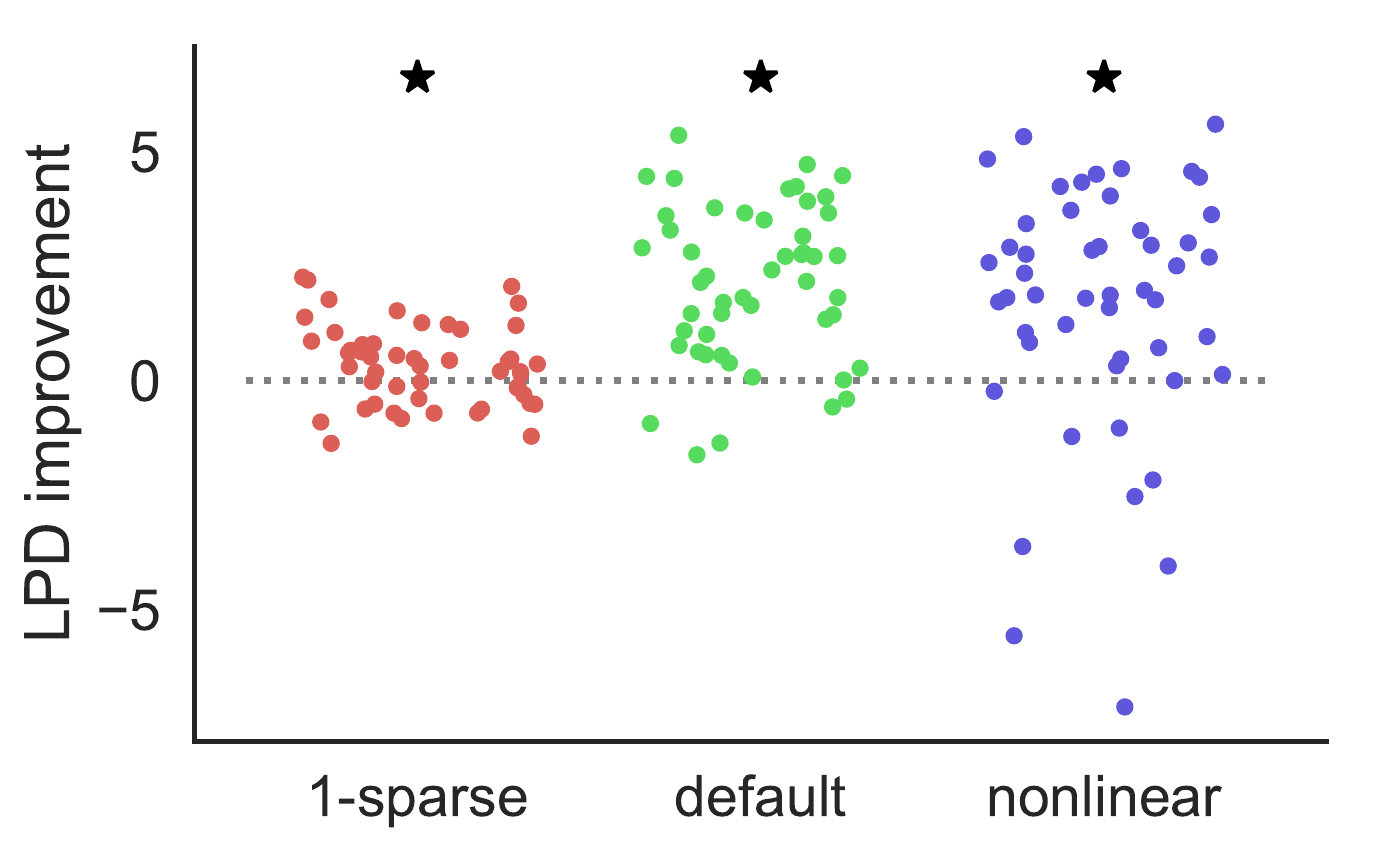}
\caption{$\lambda = 4$}
\end{subfigure} \\
\begin{subfigure}[b]{.49\textwidth}
\centering
\includegraphics[height=1.5in]{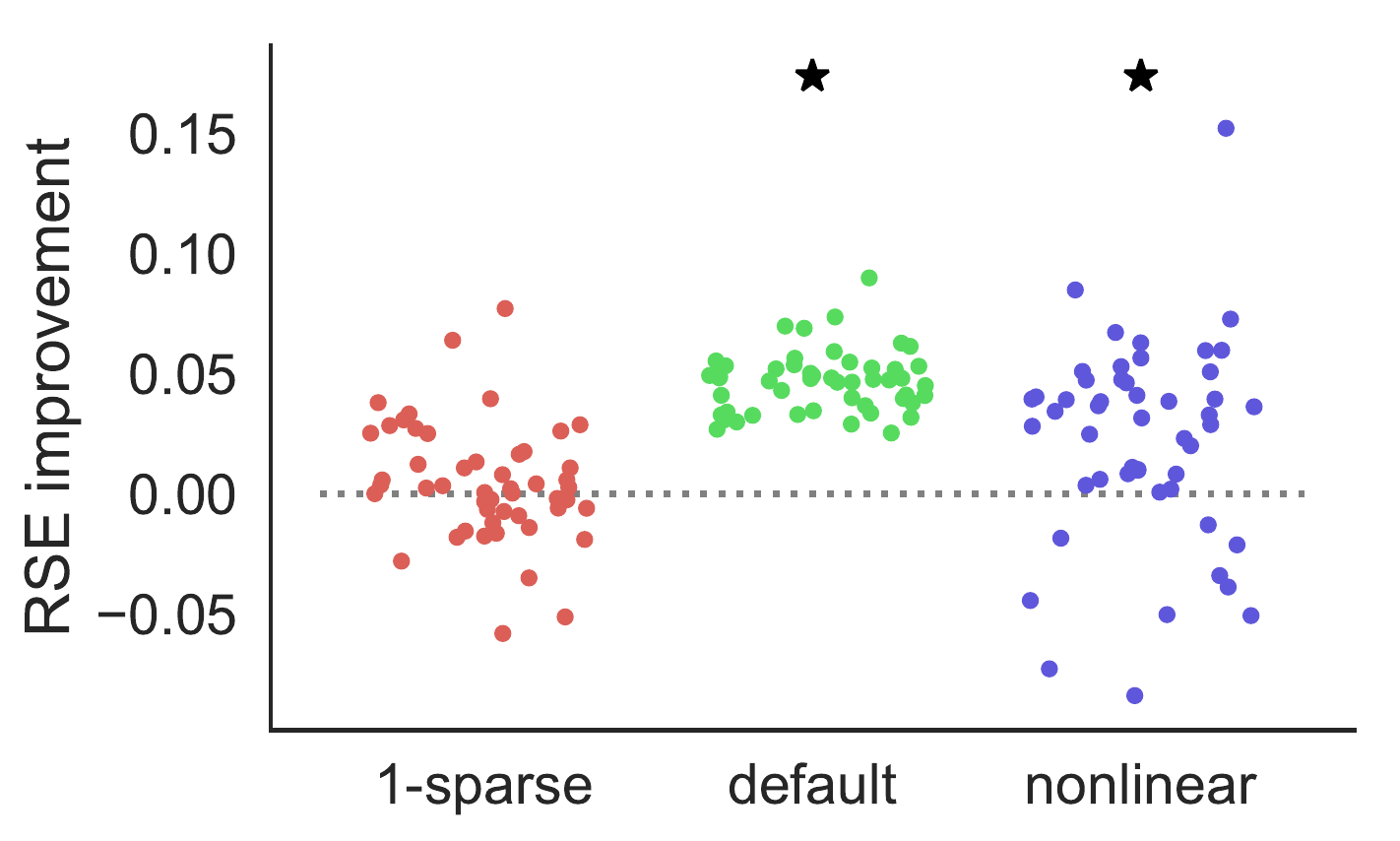}
\caption{$\lambda = 16$}
\end{subfigure}%
\begin{subfigure}[b]{.49\textwidth}
\centering
\includegraphics[height=1.5in]{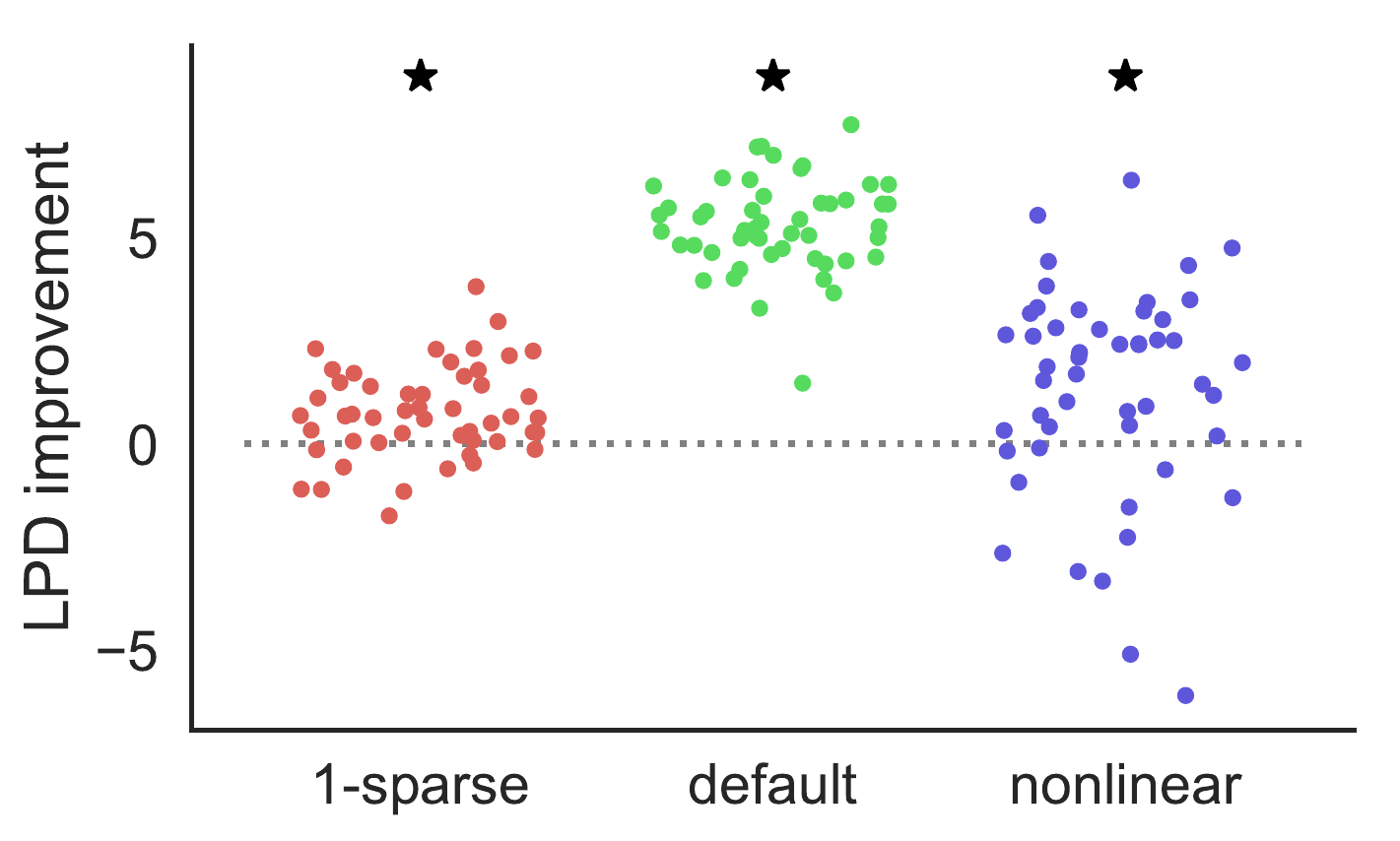}
\caption{$\lambda = 16$}
\end{subfigure}
\caption{Parameter inference performance on \textsf{default}, \textsf{1-sparse}, and \textsf{nonlinear} data for $\lambda \in \{1,4,16\}$. 
A positive improvement value indicates that the bagged posterior outperformed the standard posterior on that dataset. 
Scenarios marked with a $\star$ (respectively, $\bullet$) exhibited a statistically significant difference in the positive (respectively, negative) direction
($p < 0.05$, two-sided Wilcoxon signed-rank test).
RSE = relative squared error of $\beta_{\optsym}$. 
LPD = log posterior density at $\beta_{\optsym}$.
}
\label{fig:linear-regression-sparse-and-nonlinear-parameter-estimation}
\end{center}
\end{figure}

\begin{figure}[tbp]
\begin{center}
\begin{subfigure}[b]{.4\textwidth}
\centering
\includegraphics[height=1.5in]{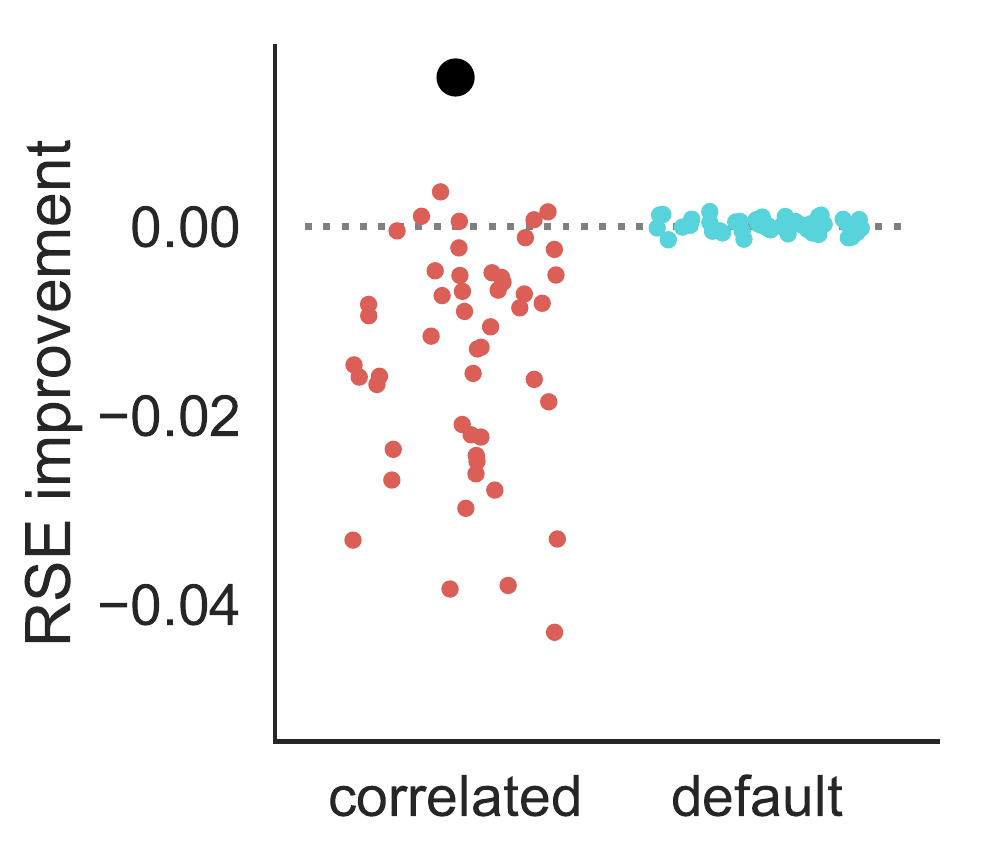}
\caption{$N = 50$}
\end{subfigure}
\begin{subfigure}[b]{.4\textwidth}
\centering
\includegraphics[height=1.5in]{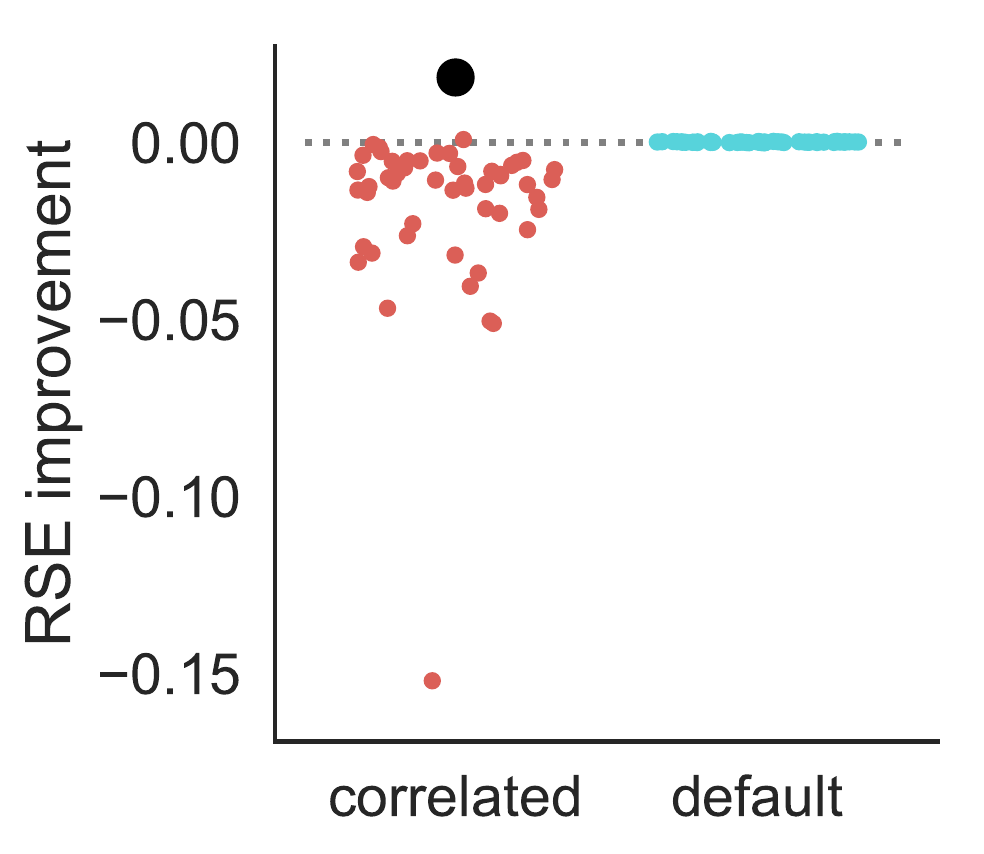}
\caption{$N = 200$}
\end{subfigure} \\
\begin{subfigure}[b]{.4\textwidth}
\centering
\includegraphics[height=1.5in]{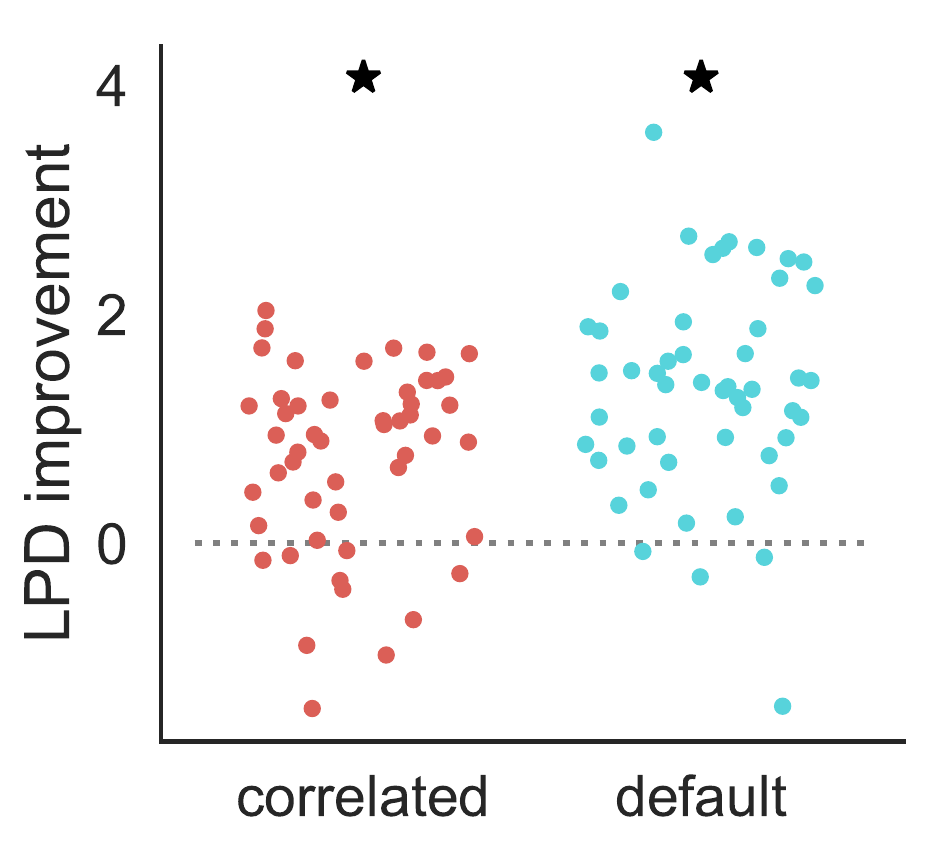}
\caption{$N = 50$}
\end{subfigure}
\begin{subfigure}[b]{.4\textwidth}
\centering
\includegraphics[height=1.5in]{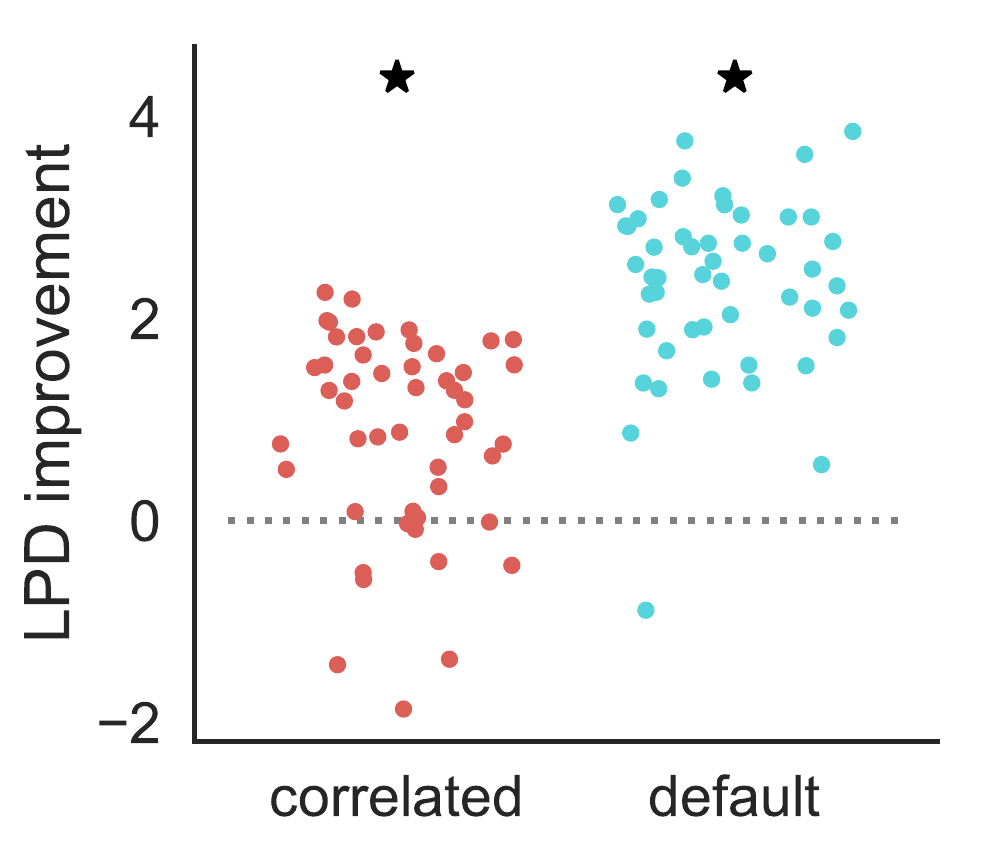}
\caption{$N = 200$}
\end{subfigure}
\caption{Parameter inference performance on \textsf{default} and \textsf{correlated} data for $N \in \{50, 200\}$.
See the caption of \cref{fig:linear-regression-sparse-and-nonlinear-parameter-estimation} for further explanation.
}
\label{fig:linear-regression-correlated-parameter-estimation}
\end{center}
\end{figure}

\subsubsection{Parameter inference} 

We begin by assessing how well the standard posterior and the bagged posterior estimated the optimal coefficient vector 
$\beta_{\optsym} \defined \argmin_{\beta \in \reals^{d}} \EE\{(Y_{1} - Z_{1}^{\top}\beta)^{2}\}$.
For the \textsf{linear} simulation setting (regardless of $G$), $\beta_{\optsym} = \beta_{\dagger}$; 
for the \textsf{uncorrelated} and \textsf{nonlinear} setting, $\beta_{\optsym} = 3\beta_{\dagger}$. 
Let $\hat\beta$ and $\log \postdensity{}(\beta)$ denote, respectively, the (standard or bagged) posterior mean and log posterior density of $\beta$. 
We quantify estimation accuracy by computing the relative squared error (RSE) $\statictwonorm{\hat\beta - \beta_{\optsym}}^{2} / \statictwonorm{\beta_{\optsym}}^{2}$
and the log posterior density (LPD) $\log \postdensity{}(\beta_{\optsym})$. 
For all experiments, we first ran BayesBag with $\bsnumobs = \numobs$ in order to compute $\optMfsest$ and $\modelmismatch$. 
If $\modelmismatch \ne \nan$, we reran BayesBag using $\bsnumobs = \optMfsest$.
If $\modelmismatch = \nan$, we reran BayesBag using $\bsnumobs = 2\numobs$ because we found that $\modelmismatch = \nan$ typically indicated a poorly chosen prior
(either because the true parameter was unlikely or the model was poorly identified), which we could best mitigate by using more data. 

In almost every case, the bagged posterior performed as well as or better than the standard posterior in terms of both relative squared error
and log posterior density. 
\Cref{fig:linear-regression-sparse-and-nonlinear-parameter-estimation} compares performance on \textsf{default}, \textsf{1-sparse}, and \textsf{nonlinear}
data for the varying prior choices $\lambda \in \{1,4,16\}$. 
The benefits of BayesBag were especially large for the excessively strong $\lambda = 16$ prior.
As expected, BayesBag was also particularly effective on the misspecified \textsf{nonlinear} data. 
\Cref{fig:linear-regression-correlated-parameter-estimation} compares performance on \textsf{default} and \textsf{correlated} data for $N \in \{50, 500\}$.
Because the prior on $\beta$ was very weak, there were significant identifiability issues when the data were heavily correlated---particularly 
in the small data regime of $N = 50$.
The only case in which BayesBag performed noticeably worse than the standard posterior 
was in terms of relative squared error on the correlated data (\cref{fig:linear-regression-correlated-parameter-estimation}a--b).
However, BayesBag still performed better in terms of log posterior density (\cref{fig:linear-regression-correlated-parameter-estimation}c--d),
indicating superior calibration of the parameter estimate at the cost of a small additional bias.

\begin{figure}[tbp]
\begin{center}
\begin{subfigure}[b]{.32\textwidth}
\centering
\includegraphics[width=1.6in]{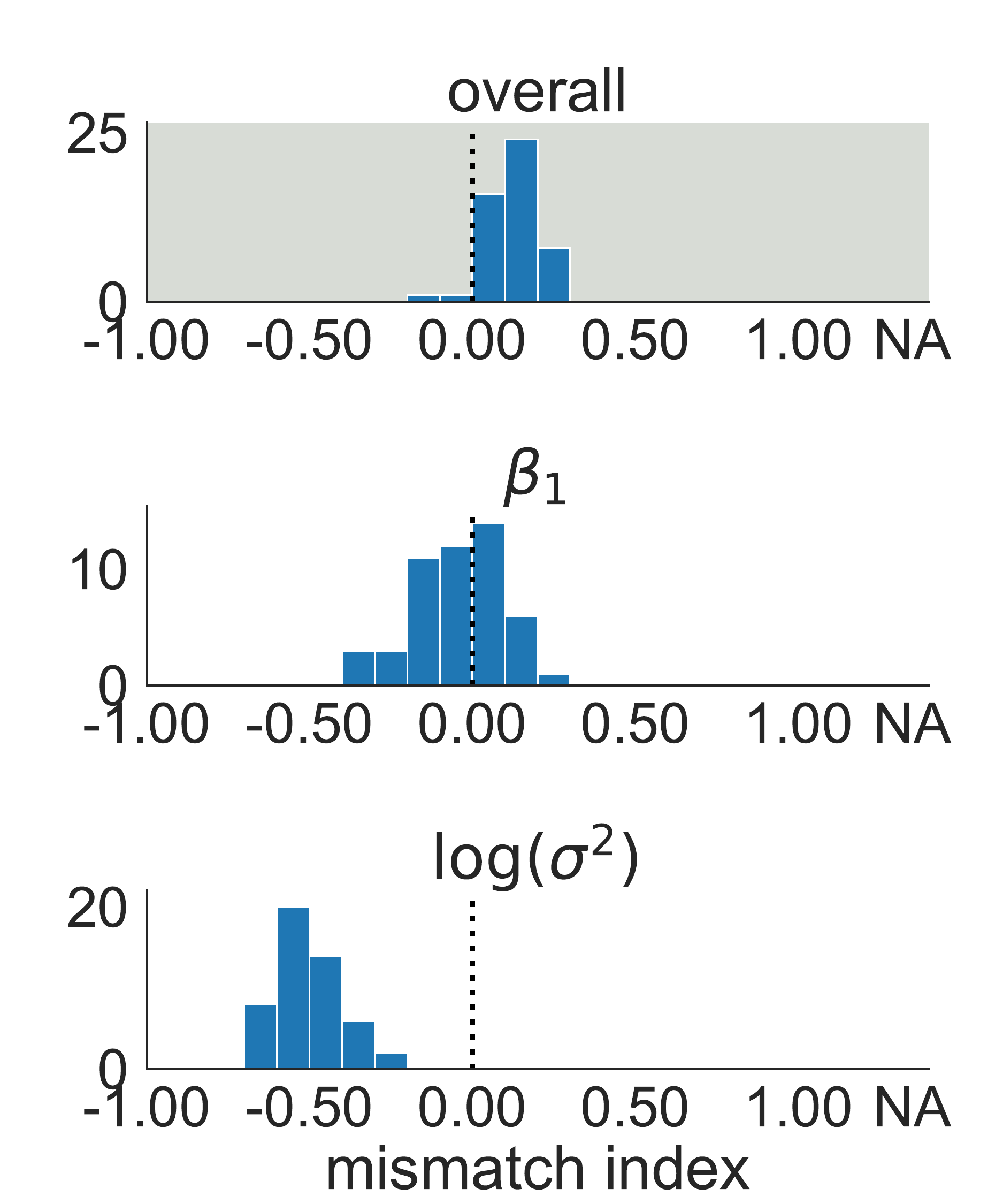}
\caption{\textsf{default}, $\lambda = 1$}
\end{subfigure}
\begin{subfigure}[b]{.32\textwidth}
\centering
\includegraphics[width=1.6in]{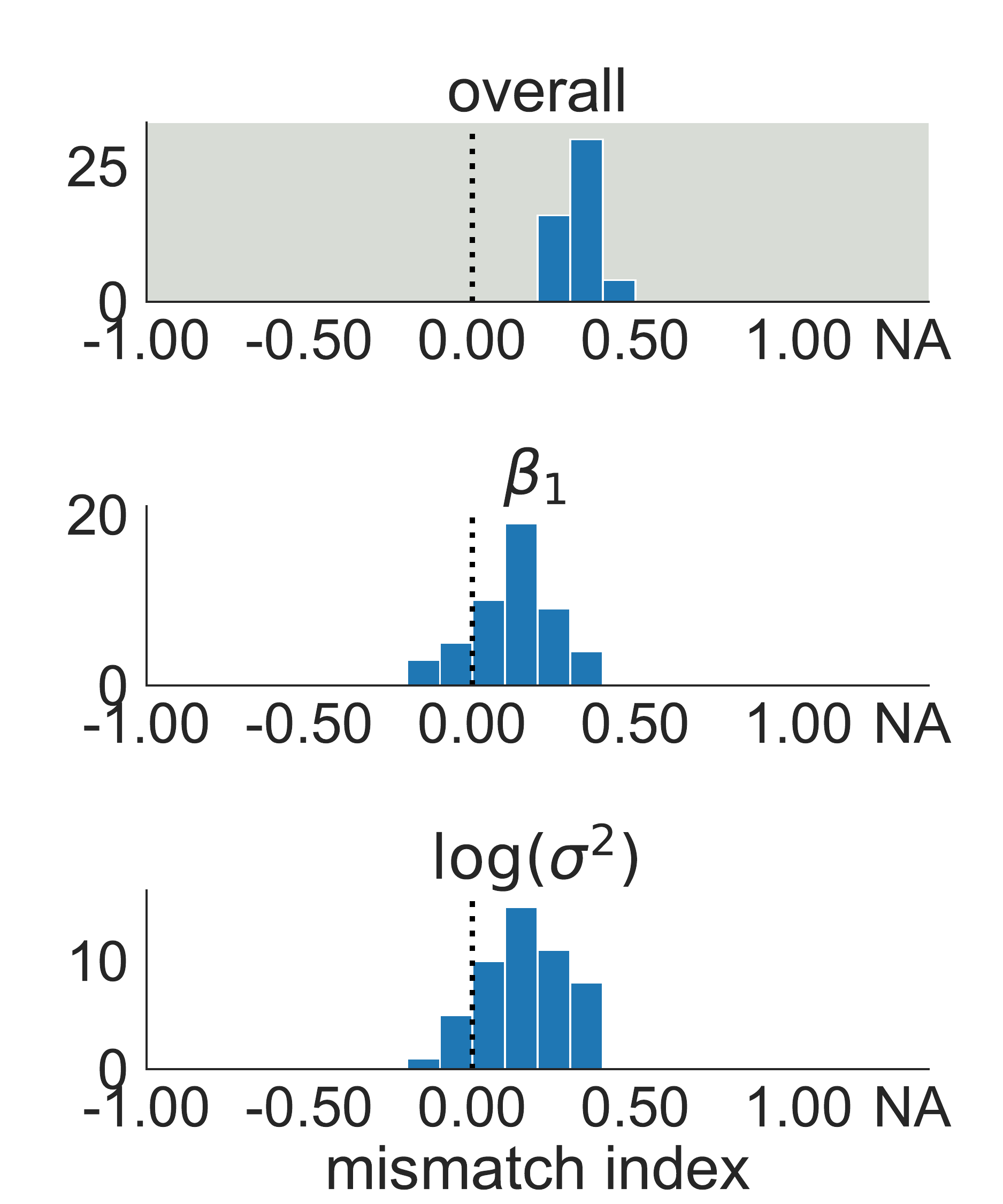}
\caption{\textsf{1-sparse}, $\lambda = 1$}
\end{subfigure} 
\begin{subfigure}[b]{.32\textwidth}
\centering
\includegraphics[width=1.6in]{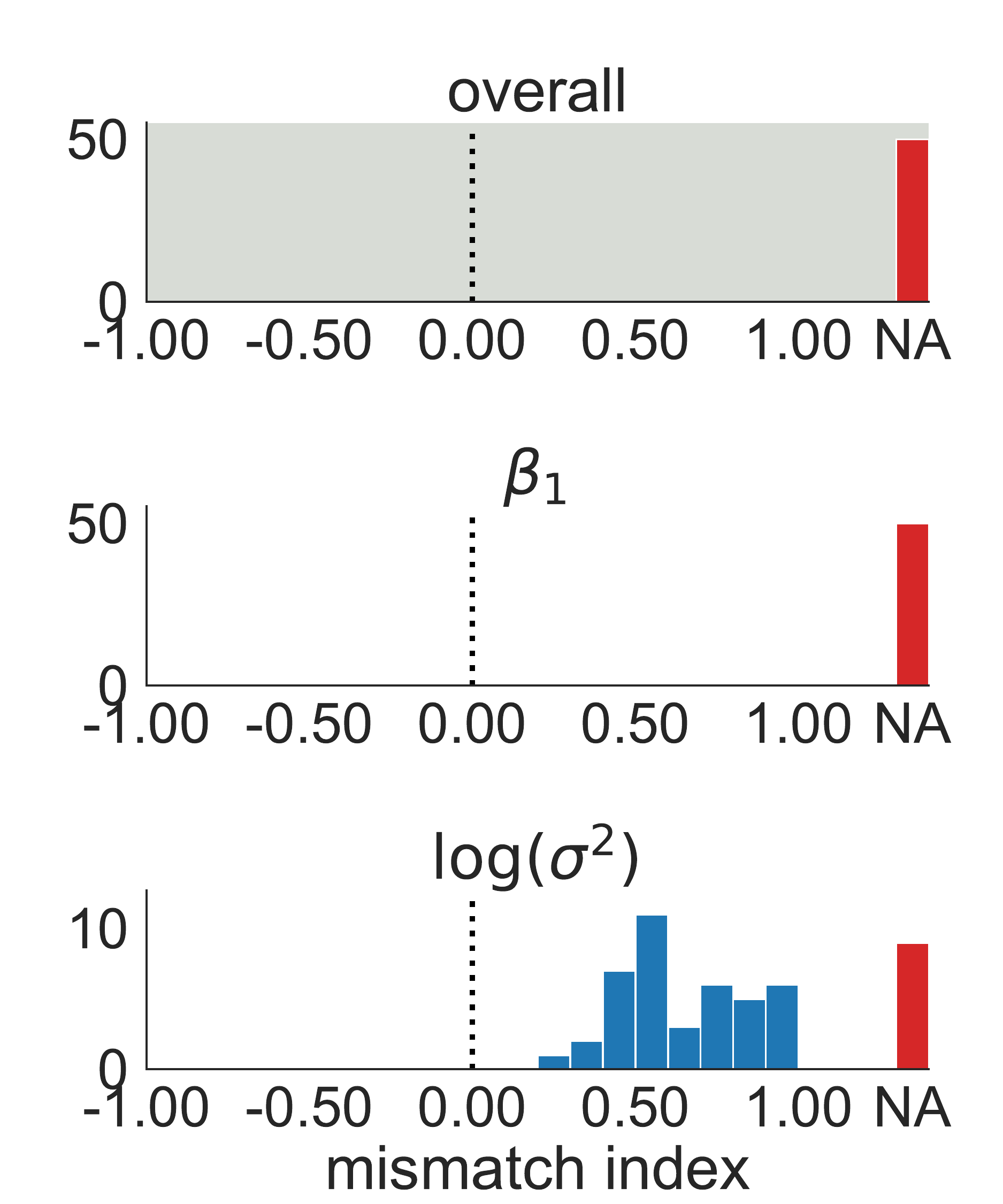}
\caption{\textsf{nonlinear}, $\lambda = 1$}
\end{subfigure}  \\
\begin{subfigure}[b]{.32\textwidth}
\centering
\includegraphics[width=1.6in]{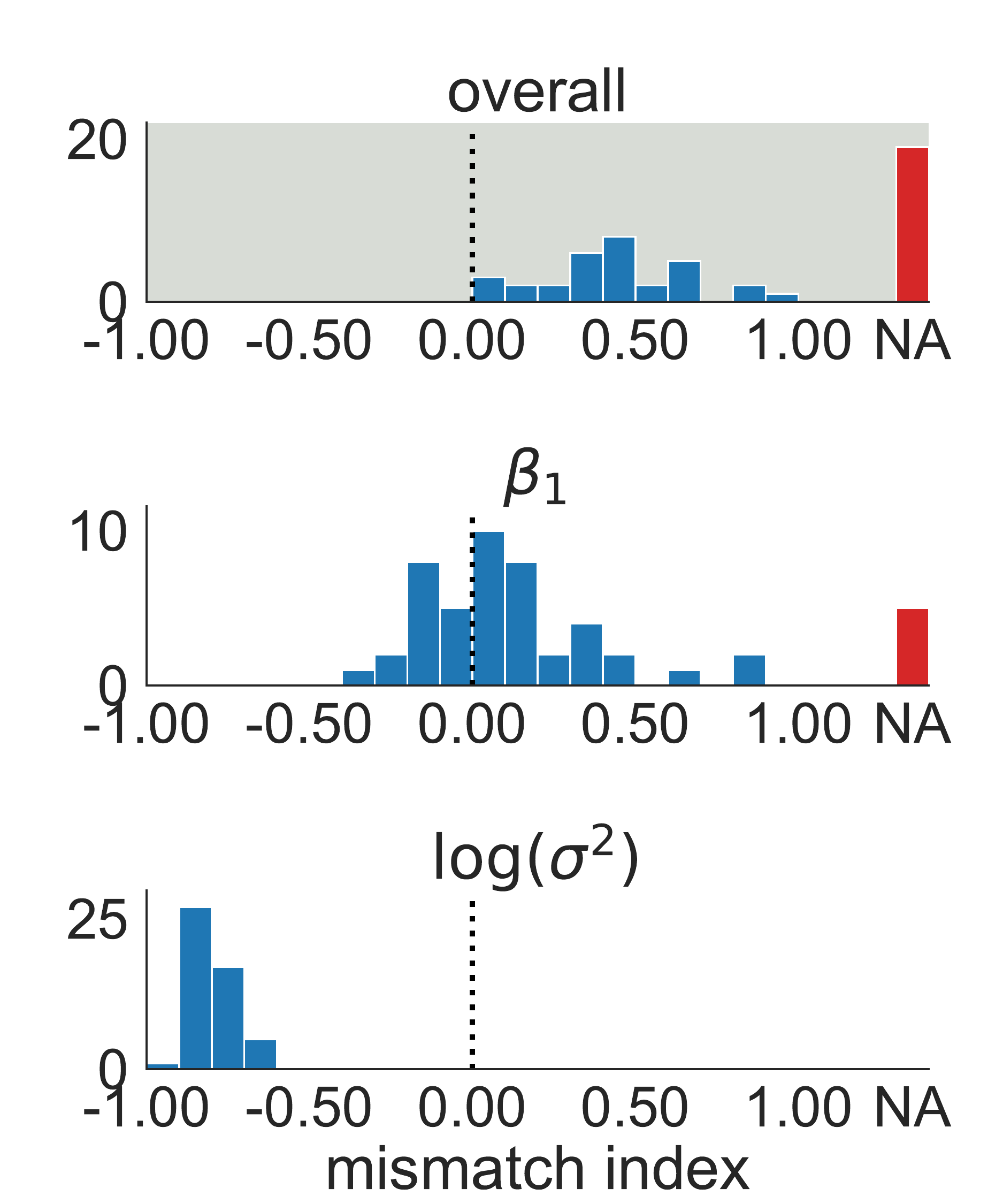}
\caption{\textsf{default}, $\lambda = 4$}
\end{subfigure}
\begin{subfigure}[b]{.32\textwidth}
\centering
\includegraphics[width=1.6in]{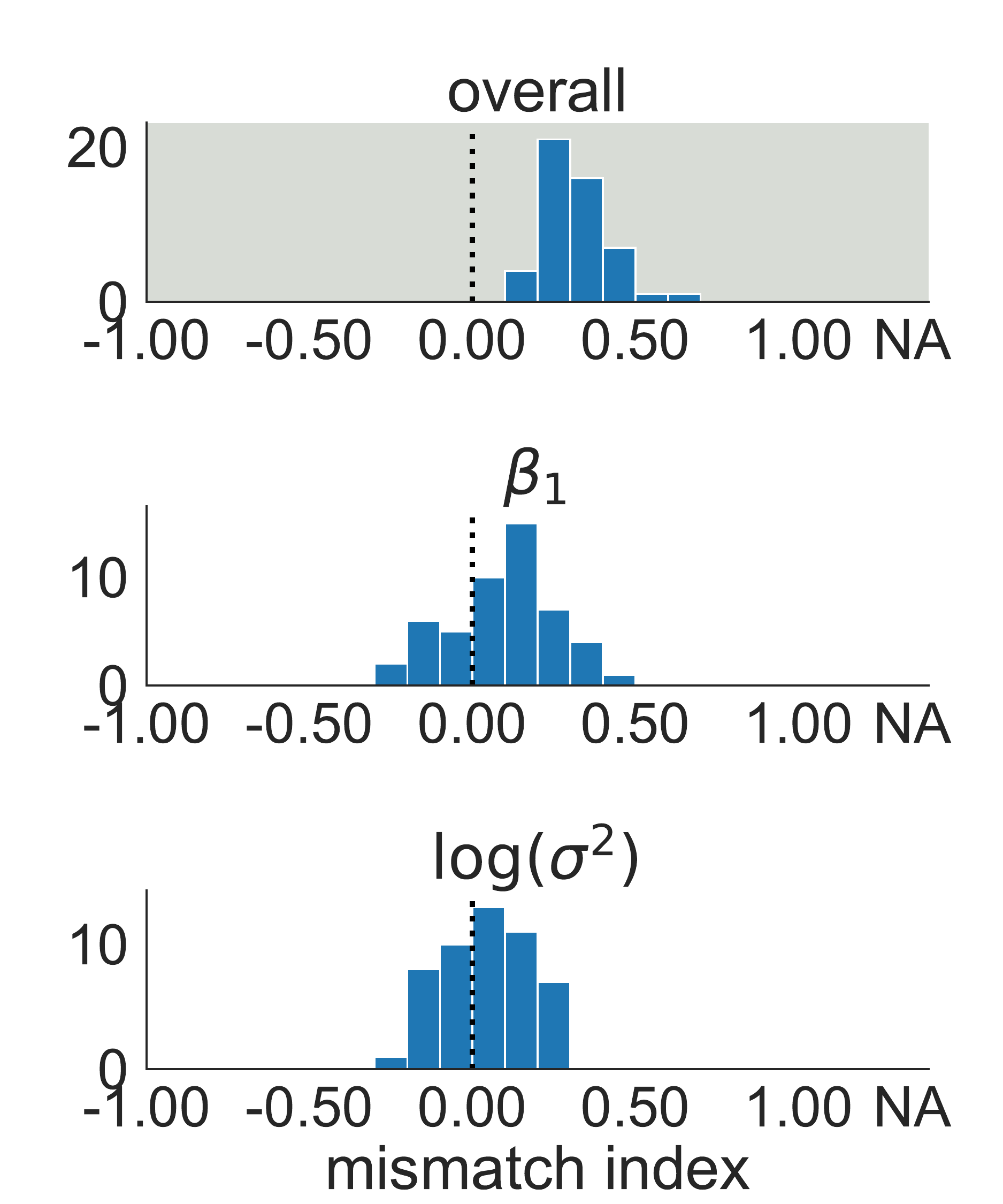}
\caption{\textsf{1-sparse}, $\lambda = 4$}
\end{subfigure} 
\begin{subfigure}[b]{.32\textwidth}
\centering
\includegraphics[width=1.6in]{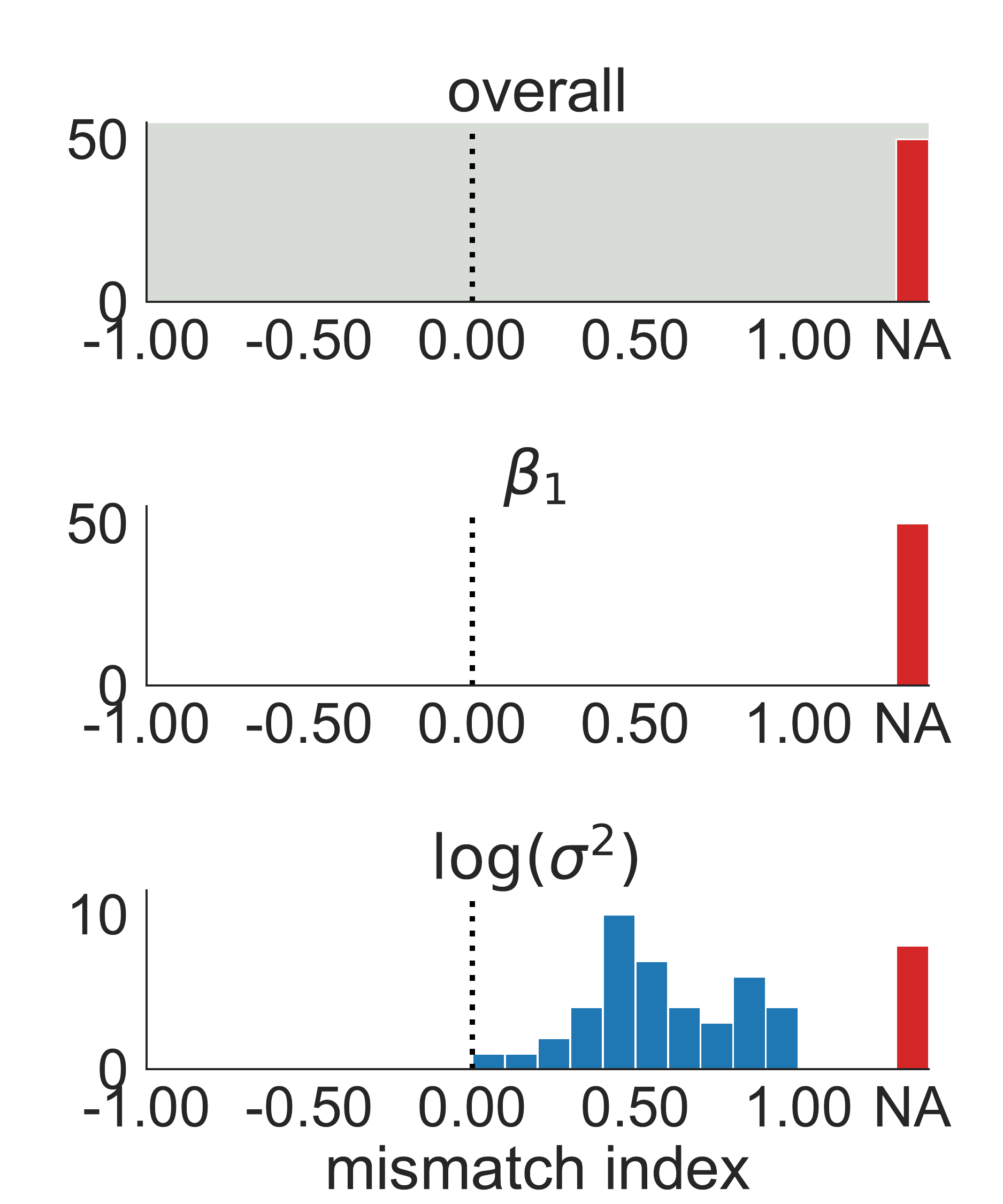}
\caption{\textsf{nonlinear}, $\lambda = 4$}
\end{subfigure}  \\
\begin{subfigure}[b]{.32\textwidth}
\centering
\includegraphics[width=1.6in]{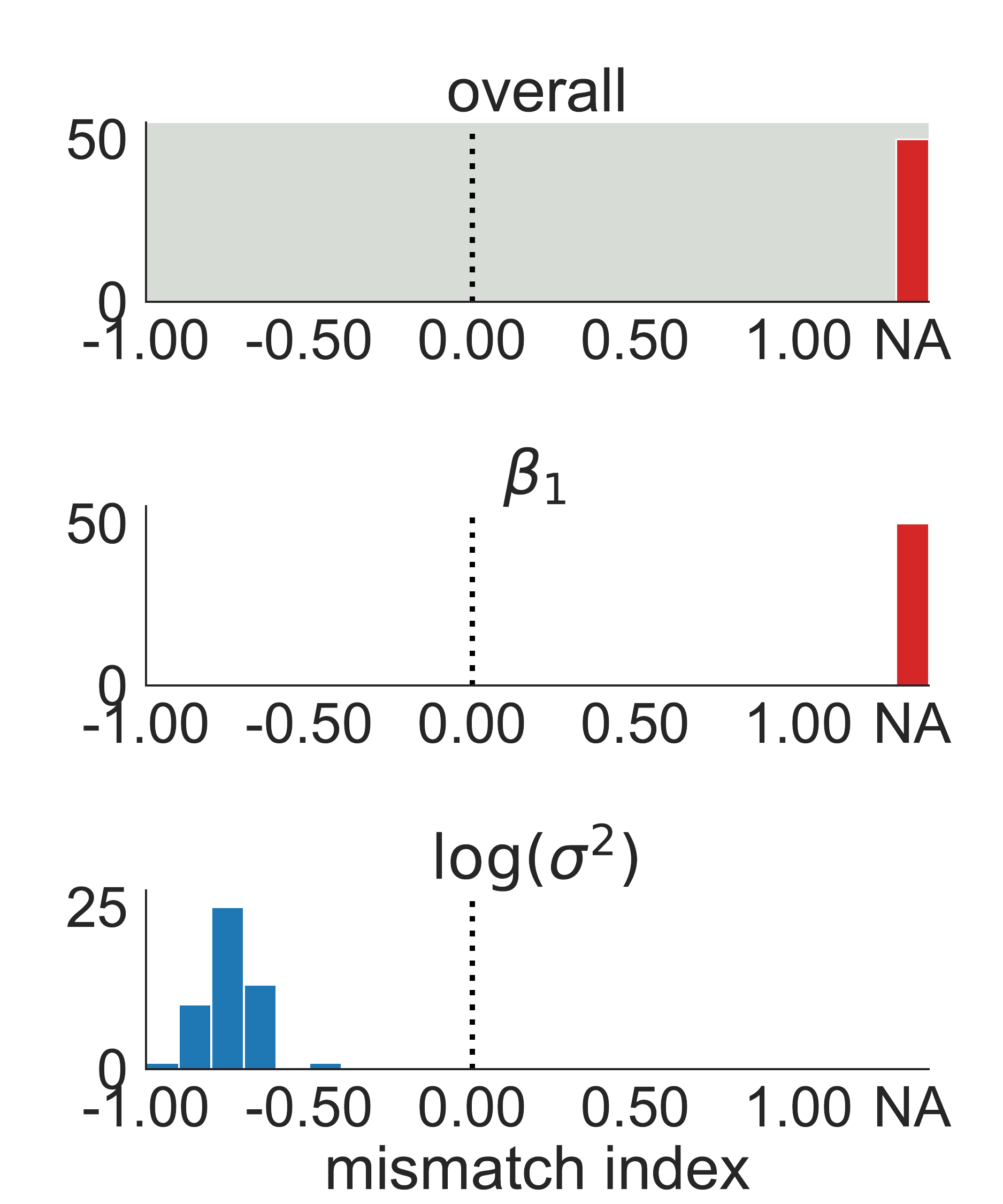}
\caption{\textsf{default}, $\lambda = 16$}
\end{subfigure}
\begin{subfigure}[b]{.32\textwidth}
\centering
\includegraphics[width=1.6in]{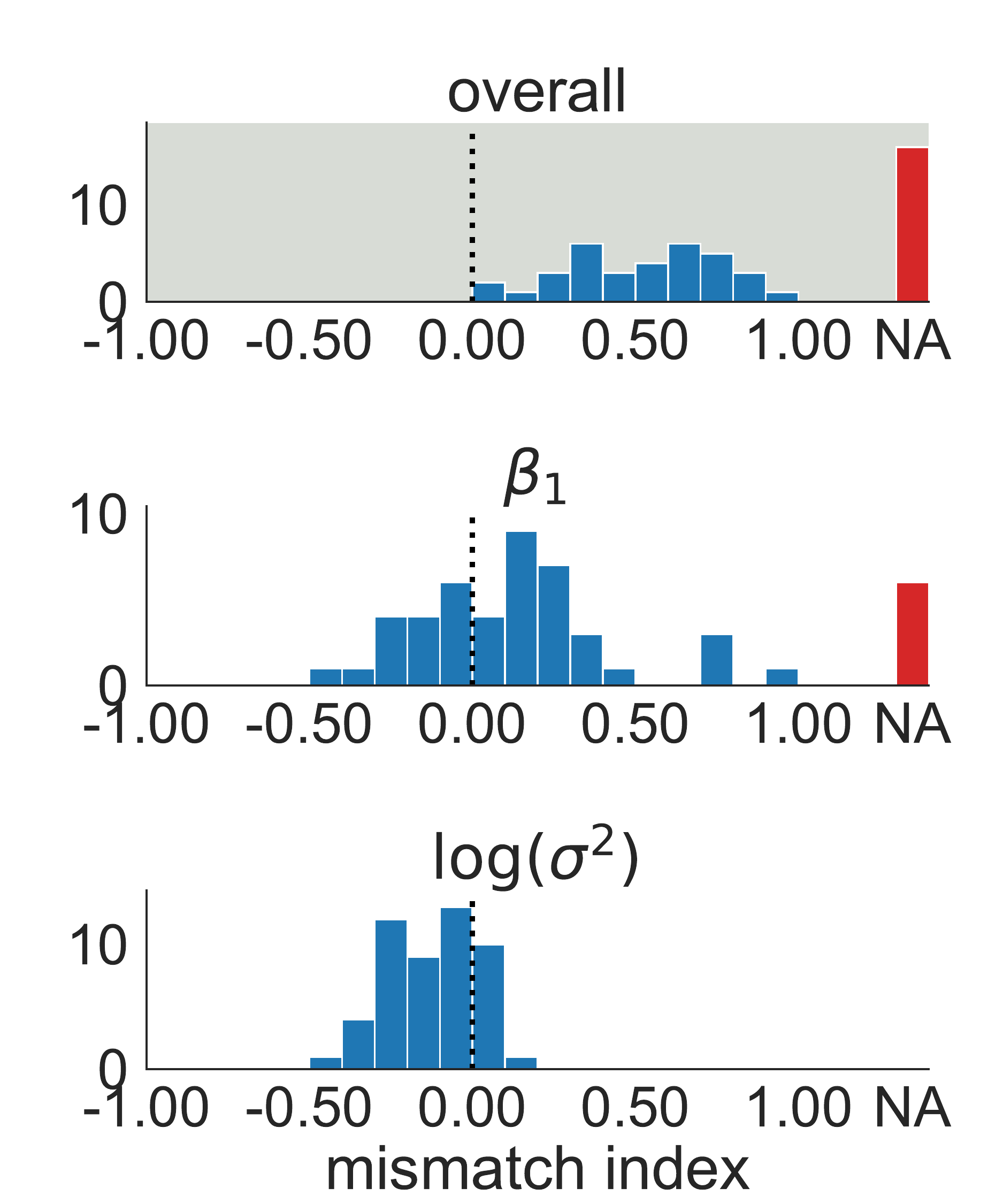}
\caption{\textsf{1-sparse}, $\lambda = 16$}
\end{subfigure} 
\begin{subfigure}[b]{.32\textwidth}
\centering
\includegraphics[width=1.6in]{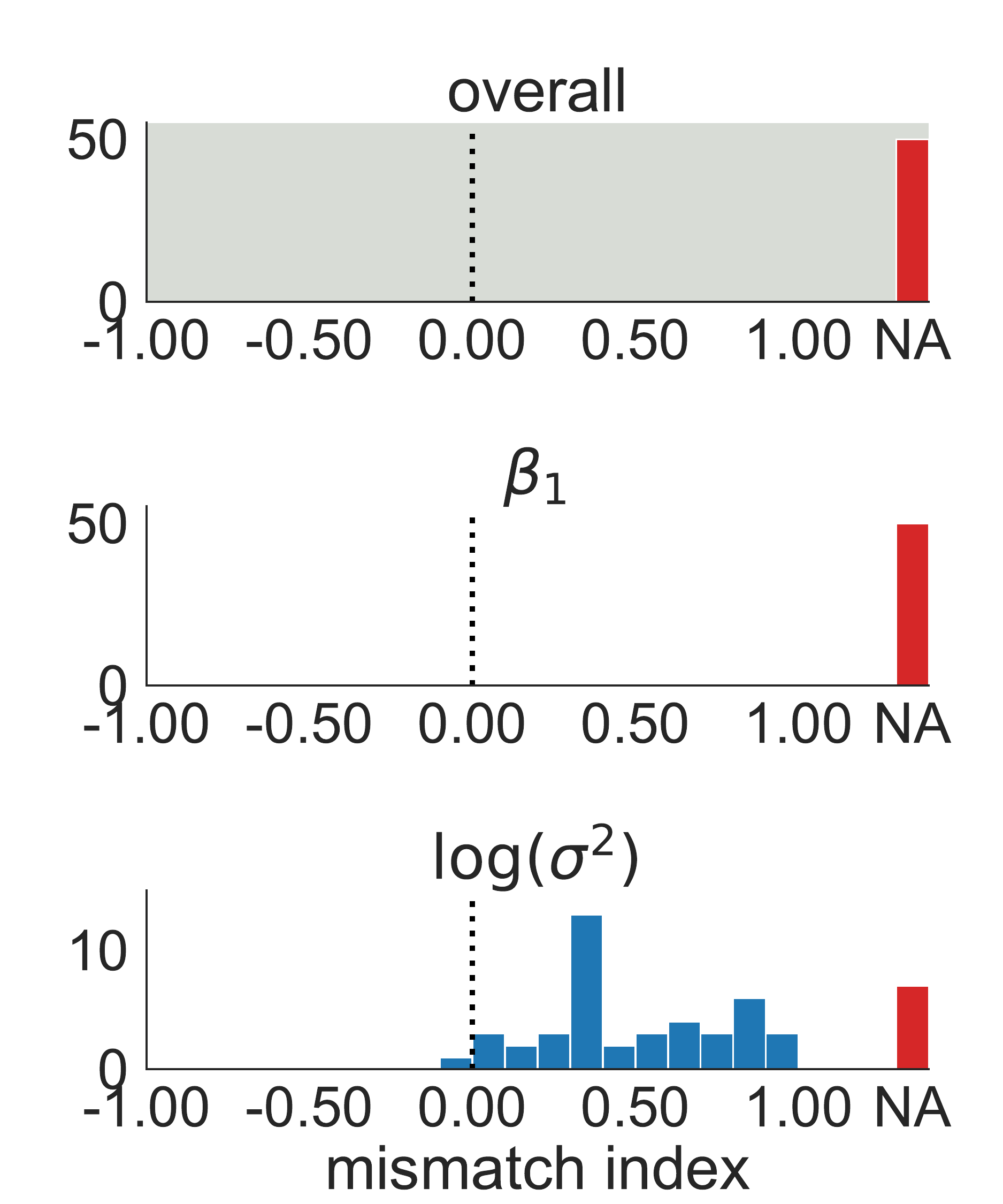}
\caption{\textsf{nonlinear}, $\lambda = 16$}
\end{subfigure}   \\
\caption{Model--data mismatch indices $\modelmismatch$ for selected parameters as well as the overall $\modelmismatch$ value
for \textsf{default}, \textsf{1-sparse}, and \textsf{nonlinear} data for $\lambda \in \{1,4,16\}$.
We only display one component of $\beta$ since the $\modelmismatch$ values followed the same distribution for all components. 
}
\label{fig:linear-regression-sparse-and-nonlinear-model-mismatch}
\end{center}
\end{figure}

\subsubsection{Model criticism using the mismatch index} \label{sec:model-mismatch-simulations}

Next, we reconsider the parameter inference scenarios from the perspective of model criticism.
Our results demonstrate how $\modelmismatch$ can be used to detect model--data mismatch when either 
(a) the likelihood is misspecified or
(b) the likelihood is well-specified but the prior is poorly chosen or some model parameters are poorly identified.

\paragraph*{Dense versus sparse coefficient vector.} 
We first consider the effects of varying prior choices. 
\Cref{fig:linear-regression-sparse-and-nonlinear-model-mismatch} compares the mismatch index with \textsf{default} and \textsf{1-sparse} data for $\lambda \in \{1,4,16\}$. 
For the \textsf{default} data with a \textsf{dense} coefficient vector, the larger $\lambda$ values result in a prior on $\beta$ that is too concentrated near zero, 
leading to larger $\modelmismatch$ values.
The settings of $\lambda = 4$ and $\lambda = 16$ resulted in $\modelmismatch = \nan$ for many of the datasets. 
For the \textsf{1-sparse} data, on the other hand, a larger $\lambda$ is more appropriate since all but one coefficient is zero. 
Thus, $\modelmismatch$ was at most 0.5 for $\lambda \in \{1,4\}$, although the stronger prior resulting from $\lambda = 16$ 
led to $\modelmismatch = \nan$.

\paragraph*{Linear versus nonlinear regression function.} 
Next, we consider misspecified data.
\Cref{fig:linear-regression-sparse-and-nonlinear-model-mismatch} compares the mismatch index with \textsf{default} and \textsf{nonlinear} data for $\lambda \in \{1,4,16\}$. 
Due to the misspecification, $\modelmismatch = \nan$ for all choices of $\lambda$. 

\paragraph*{Correlated versus uncorrelated regressors.} In this case the mismatch index is able to detect that there are poorly identified
parameters when the data is \textsf{correlated}. See \cref{app:additional-figs} for details.

\subsection{Hierarchical mixed effects logistic regression model} \label{sec:RELR}

Next, we considered the canonical setting of mixed effects models, where Bayesian methods can provide superior inferences compared to
maximum likelihood and quasi-likelihood methods \citep{Browne:2006} and, even with significant amounts of data,
can lead to dramatically different inferences \citep[e.g.,][]{Giordano:2018:LRVB}. 
Our objective was to compare the predictive performance of the bagged posterior (BayesBag) to both the standard Bayesian posterior and alternative likelihood-based methods.
Specifically, we considered a 3-level logistic regression model with mixed effects and a balanced design: 
\[
v_{k} &\distiid \distNorm(0, \sigma_{v}^{2}),  &
u_{jk} &\distiid \distNorm(0, \sigma_{u}^{2}),  \\
Y_{ijk} \given Z_{ijk}, u_{jk}, v_{k} &\distind \distBern(p_{ijk}), &
p_{ijk} &= \operatorname{logit}^{-1}(\beta_{0} + Z_{ijk}^{\top}\beta + u_{jk} + v_{k}),
\]
for $k = 1,\dots,K$, $j=1,\dots,J$, $i=1,\dots,I$, $Z_{ijk} \in \reals^{D}$, $\beta_{0} \in \reals$, and $\beta \in \reals^{D}$. 
For example, \citet{Browne:2006} take $Y_{ijk}$ to be a binary indicator of whether a woman received modern prenatal care during a pregnancy,
with $i$ indexing the birth, $j$ indexing the mother, and $k$ indexing the Guatemalan community to which the mother belonged.
Note that in this particular example one would not expect a balanced design, but we used the balanced case for simplicity. 
For the Bayesian model, we used relatively diffuse priors:
\[
\sigma_{v} &\dist \distUnif(0,100), &
\sigma_{u} &\dist \distUnif(0,100), \\
\beta_{d} &\distiid \distNorm(0, 10^{2}), \quad d=0,\dots,D. 
\]
In our experiments, we took $I = 3$, $J = 8$, $K = 100$, and $D = 3$.
We considered a well-specified scenario and a misspecified scenario.  
In the well-specified scenario, we generated covariates $Z_{ijkd} \distiid \distNorm(0, 1)$ and generated responses $Y_{ijk}$ according to the assumed model with 
$\beta_{0} = 0.65$, $\beta = (1,1,1)$, and $\sigma_{v} = \sigma_{u} = 3$.
For the misspecified scenario, we generated data as in the well-specified case except that the random effects had unmodeled 
correlation structure: $(v_{k}, u_{1k}, \dots, u_{Jk})$ was jointly Gaussian with
correlation $\rho$ between each pair of components, where $\rho = 0.99$. %

\begin{figure}[tbp]
\begin{center}
\begin{subfigure}[b]{.49\textwidth}
\centering
\includegraphics[width=\textwidth]{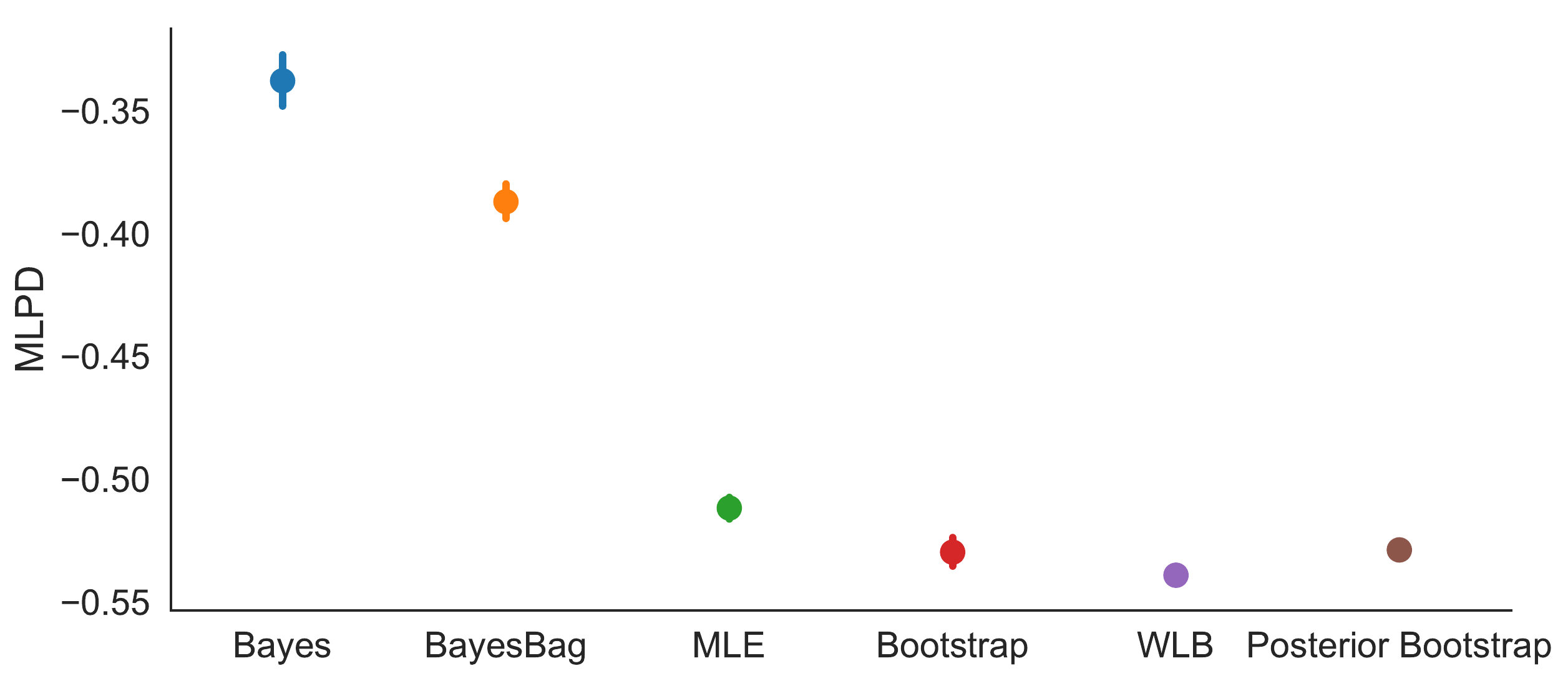}
\caption{well-specified}
\end{subfigure}
\begin{subfigure}[b]{.49\textwidth}
\centering
\includegraphics[width=\textwidth]{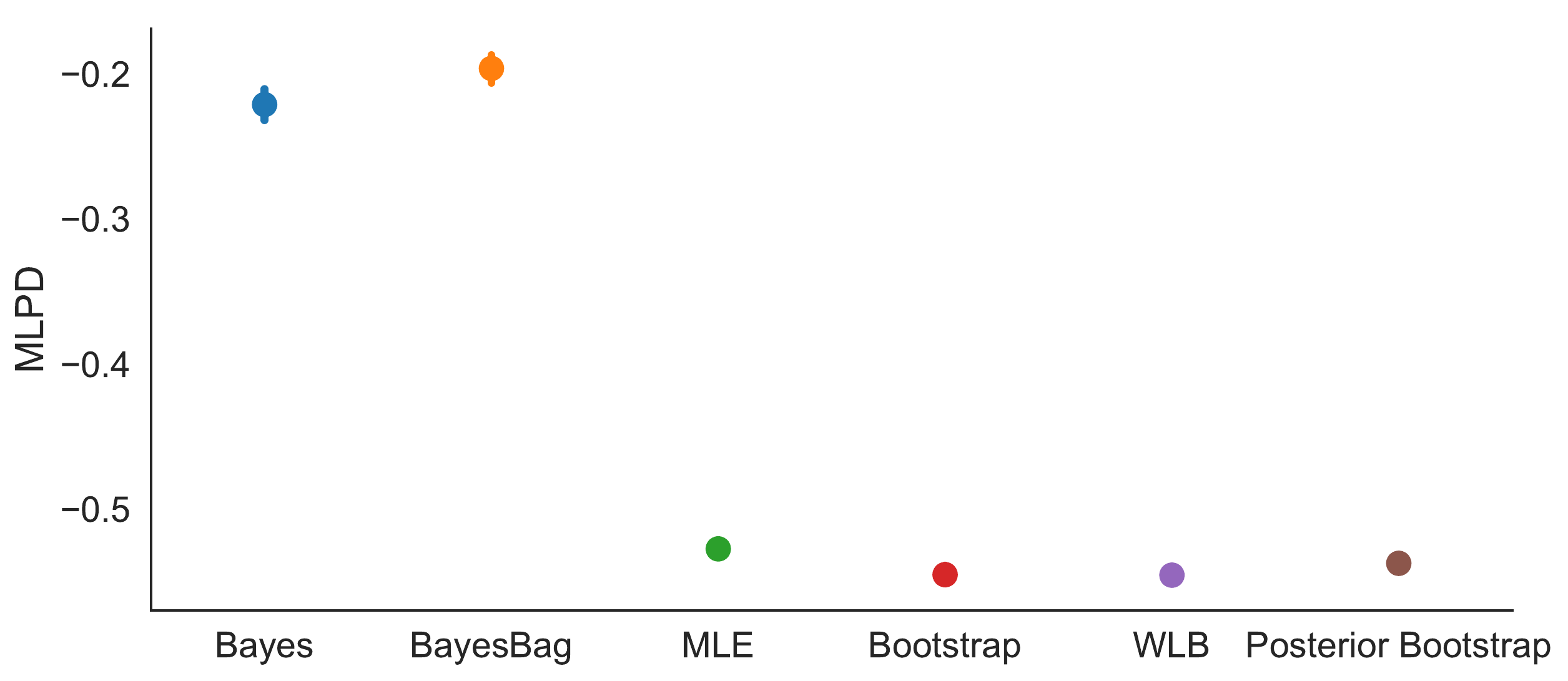}
\caption{misspecified}
\end{subfigure}
\caption{Predictive performance comparison of the standard posterior (Bayes), the bagged posterior (BayesBag), and four other methods based on 
maximum likelihood estimation (MLE) on held-out data for the hierarchical mixed effects logistic regression model.
We show the mean log predictive density (MLPD) of each method on held-out data and 95\% confidence intervals. 
The differences between Bayes, BayesBag, and the MLE-based methods are all statistically significant
($p < 0.0001$, paired $t$-test)} %
\label{fig:RELR-overview}
\end{center}
\end{figure}

We compared the predictive performance of the standard posterior, the bagged posterior, and four methods based on 
maximum likelihood estimation (with the random effects integrated out): the standard MLE, the bootstrapped MLE, 
the weighted likelihood bootstrap \citep{Newton:1994}, and the posterior bootstrap \citep{Lyddon:2018}.
See \cref{sec:discussion} for further discussion of these approaches.
\Cref{fig:RELR-overview} shows the predictive performance of each method relative to the standard Bayesian posterior. 
Both the standard and bagged posteriors outperformed the MLE-based methods.
In the well-specified scenario, standard Bayes was better than BayesBag, as expected.
Meanwhile, in the misspecified scenario, BayesBag had superior predictive performance compared to standard Bayes.

\section{Application to cancer microarray data} \label{sec:application}

We next consider an application to cancer microarray data using a sparse logistic regression model.
We used the model and four cancer microarray datasets from \citet{Piironen:2017}.
Since we do not have access to ground truth parameters, we followed the procedure of \citet{Piironen:2017} and 
computed the mean log predictive density (MLPD) on 50 random train--test splits of each dataset, holding out 20\% as test data on each split. 
The posterior for the regression coefficients $\beta \in \reals^{D}$ is multimodal, so the standard mismatch index $\modelmismatch$ is not applicable. 
Instead, we computed %
$\modelmismatch(f)$ with $f(\theta) = \log(\sum_{d=1}^{D}\beta_{d}^{2})$, which has a unimodal distribution with minimal skew.  
Since $\modelmismatch(f)$ was small for all datasets (less than 0.1), we used $\bsnumobs = 2\numobs$. 
We used $B=50$ bootstrap samples to approximate the bagged posterior. Nearly identical results were obtained with $B=25$, indicating that
it was not necessary to make $B$ larger.
See  \cref{app:experiment-details} for further implementation details.

\begin{figure}[tbp]
\begin{center}
\begin{subfigure}[b]{\textwidth}
\centering
\includegraphics[height=1.3in]{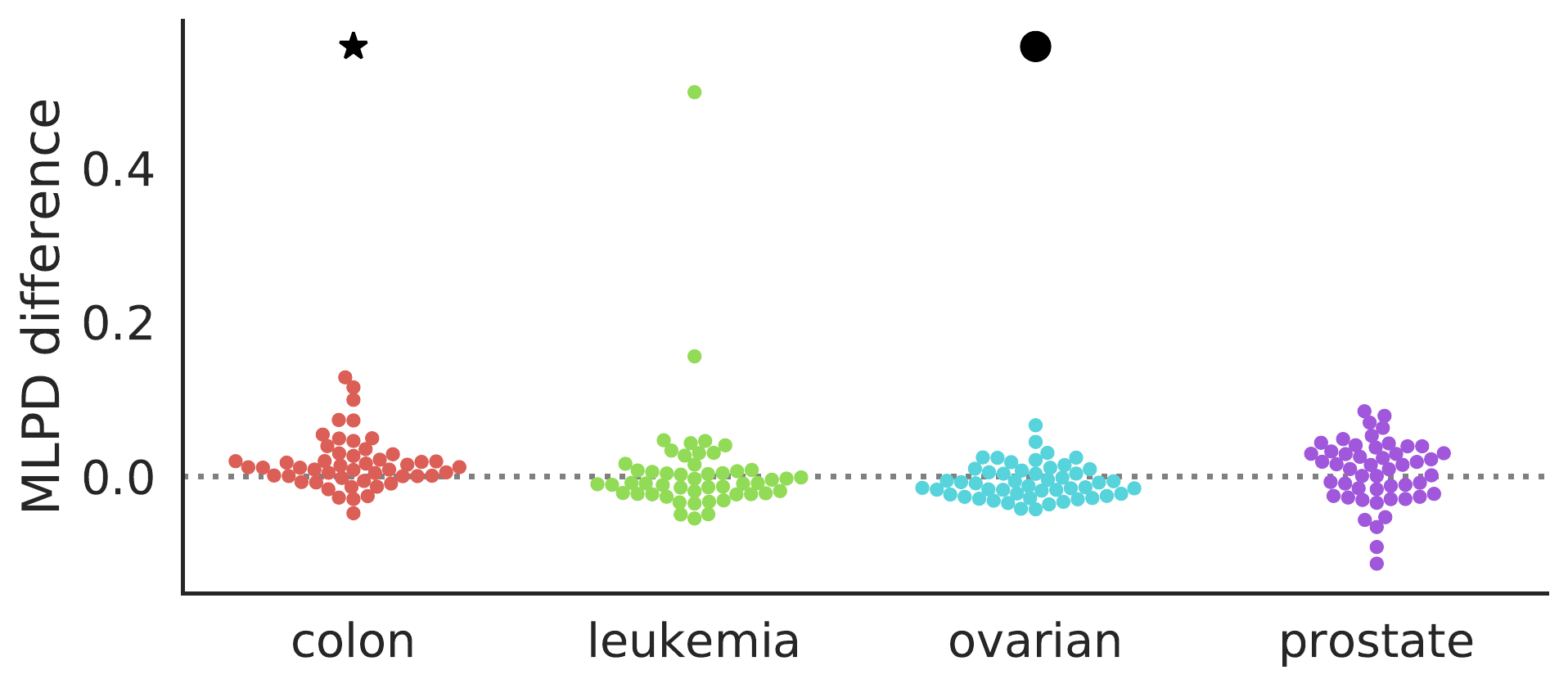}\hspace{5em}
\caption{}
\end{subfigure} \\
\begin{subfigure}[b]{\textwidth}
\centering
\includegraphics[height=1.3in]{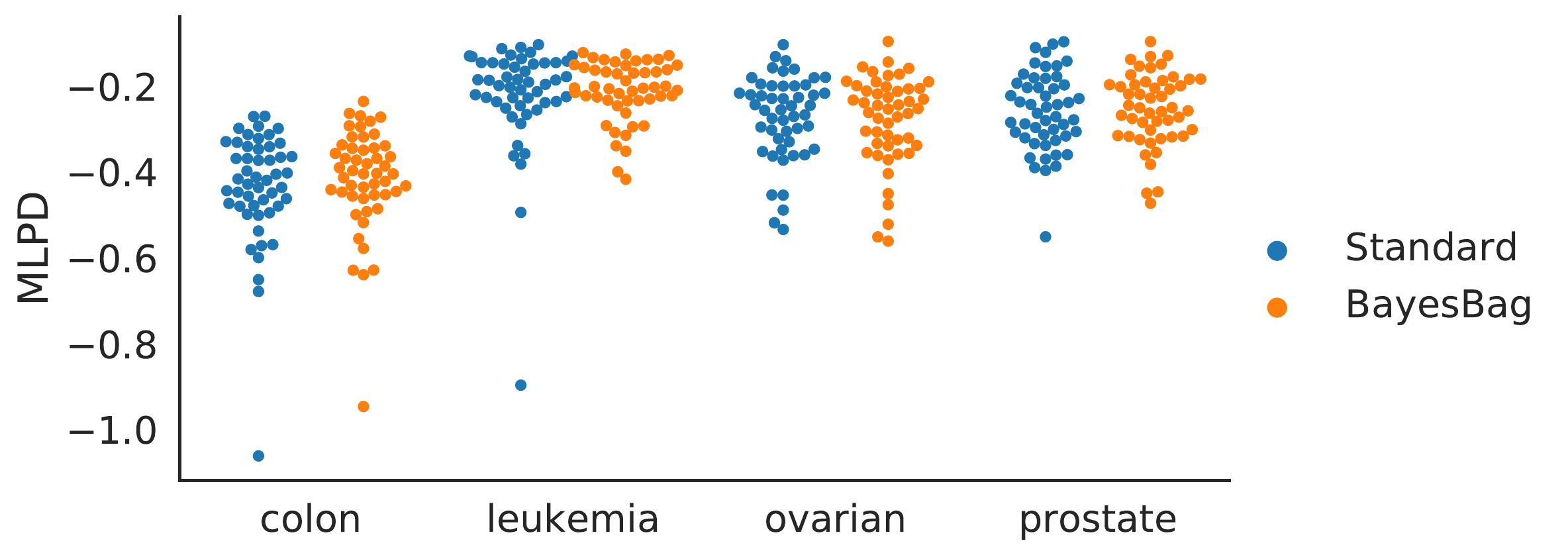}
\caption{}
\end{subfigure}
\caption{Predictive performance on 20\% held-out test data across 50 random splits of the data for sparse logistic regression. 
(a) Differences in mean log predictive density (MLPD) between BayesBag and standard Bayes; a positive difference means BayesBag outperformed standard Bayes.
Datasets marked with a $\star$ (respectively, $\bullet$) exhibited a statistically significant difference in the positive (respectively, negative) direction
($p < 0.05$, two-sided Wilcoxon signed-rank test).
(b) MLPD of BayesBag and standard Bayes. The ratios of variances of the MLPDs for colon, leukemia, ovarian, and prostrate were $1.2$, $3.1$, $0.89$, and $1.1$. 
}
\label{fig:sparse-logistic-regression-prediction}
\end{center}
\end{figure}

\begin{figure}[tbp]
\begin{center}
\centering
\includegraphics[height=2in]{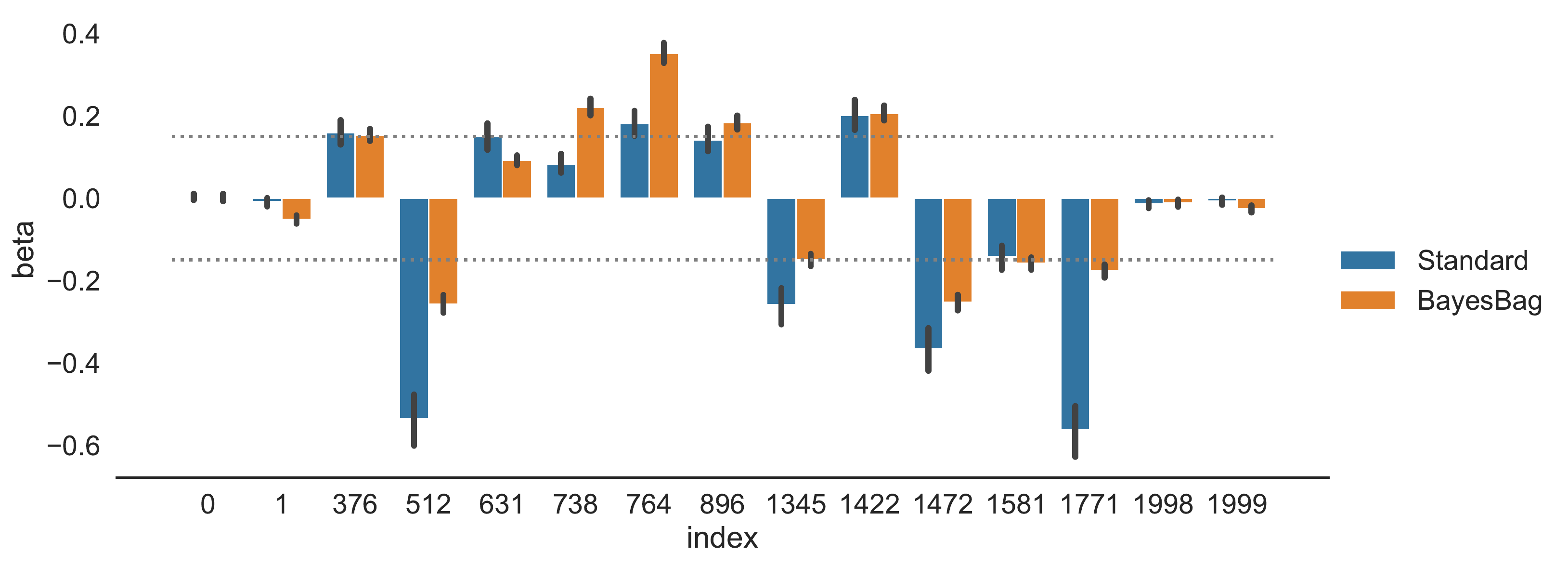}
\caption{For sparse logistic regression on the colon dataset, posterior means of components with significant regression coefficients (absolute mean greater than 0.15) and
representative non-significant coefficients (with component indices 0, 1, $D-2$, and $D-1$, where $D=2000$). 
The standard deviations are shown with black lines. 
}
\label{fig:sparse-logistic-regression-qualitative}
\end{center}
\end{figure}

Compared to the standard posterior, the predictive performance of the bagged posterior 
tends to be (1)~equal or slightly better on average and (2)~more stable---that is, lower-variance---across splits (\cref{fig:sparse-logistic-regression-prediction}).
The only exception was the ovarian dataset, where BayesBag had slightly lower MLPD on average and slightly higher variance MLPDs.
However, the average MLPD difference was only $-0.007$ nats on the ovarian dataset, while it was $0.018$ nats on the colon dataset. 
The average MLPD difference for the leukemia and prostate dataset were $-0.008$ and $0.006$ nats, respectively, although these
differences were not statistically significant. 

\Cref{fig:sparse-logistic-regression-qualitative} shows that in the colon dataset, using BayesBag also leads to qualitatively 
different inferences compared to the standard posterior.
While the bagged posterior means of the regression coefficients tended to be shrunk toward zero relative to the standard posterior means, in some cases (e.g., components 376 and 1422)
there was almost no difference, and in other cases (e.g., components 738 and 764) the bagged posterior means were more than twice the magnitude as the standard posterior means. 
Thus, BayesBag cannot be viewed as performing a simple operation to the standard posterior, such as increasing posterior variances or shrinking regression
coefficients toward zero. 
\Cref{app:additional-figs} provides similar results for the other three datasets.

\section{Discussion} \label{sec:discussion}

We conclude by situating BayesBag in the wider literature on robust Bayesian inference and model criticism.
With this additional context in place, we highlight the strengths of our approach, while also suggesting fruitful directions for future development. 

\subsection{Robust Bayesian inference}

Two common themes emerge when surveying existing methods for robust Bayesian inference. 
First, many methods require choosing a free parameter. %
However, the proposals for choosing free parameters tend to be either (a) heuristic, (b) strongly dependent on being in the asymptotic regime, 
or (c) computationally prohibitive for most real-world problems. 
Second, those methods without a free parameter lose key parts of what makes the Bayesian approach attractive. 
For example, they strongly rely on asymptotic assumptions, make a Gaussian assumption, or do not incorporate a prior distribution. 

The power posterior is perhaps the most widely studied method for making the posterior robust to model misspecification \citep{Grunwald:2012,Holmes:2017,Grunwald:2017,Miller:2018:coarsening,Syring:2018,Lyddon:2019}. 
For a likelihood function $L(\param)$, prior distribution $\priordist$, and any $\zeta \ge 0$, the \emph{$\zeta$-power posterior} is defined as 
$\ppostdist{}{(\zeta)}(\dee\param) \propto L(\param)^{\zeta}\priordist(\dee\param)$.
Hence, $\ppostdist{}{(1)}$ is equal to the standard posterior and $\ppostdist{}{(0)}$ is equal to the prior. 
Typically, $\zeta$ is set to a value between these two extremes, as 
there is significant theoretical support for the use of power posteriors with 
$\zeta \in (0,1)$~\citep{Bhattacharya:2019,Walker:2001,Miller:2018:coarsening,Royall:2003,Grunwald:2012}.
However, there are two significant methodological challenges. 
First, computing the power posterior often requires new computational methods or additional approximations, particularly in latent variable models~\citep{AntonianoVillalobos:2013,Miller:2018:coarsening}.
Second, choosing an appropriate value of $\zeta$ can be difficult.
\citet{Grunwald:2012} proposes SafeBayes, a theoretically sound method which is evaluated empirically in \citet{Grunwald:2017} and \citet{deHeide:2019}. 
However, SafeBayes is computationally prohibitive except with simple models and very small datasets.
In addition, the underlying theory relies on strong assumptions on the model class. 
Many other methods for choosing $\zeta$ have been suggested, but they are either heuristic or rely on strong asymptotic
assumptions such as the accuracy of the plug-in estimator for the sandwich covariance~\citep{Royall:2003,Holmes:2017,Miller:2018:coarsening,Syring:2018,Lyddon:2019}.

More in the spirit of BayesBag are a number of bootstrapped point estimation approaches~\citep{Rubin:1981:BayesianBootstrap,Newton:1994,Chamberlain:2003,Lyddon:2018,Lyddon:2019}.
However, unlike BayesBag, these methods compute a collection of \emph{maximum a posteriori} (MAP) or \emph{maximum likelihood} (ML) estimates. 
The weighted likelihood bootstrap of \citet{Newton:1994} and a generalization proposed by \citet{Lyddon:2019} do not incorporate a prior, and therefore lose 
many of the benefits of Bayesian inference.
The related approach of \citet{Lyddon:2018}, which includes the weighted likelihood bootstrap and standard Bayesian inference as limiting cases, draws the bootstrap samples partially from
the posterior and partially from the empirical distribution. 
Unfortunately, there is no accompanying theory to guide how much the empirical distribution and posterior distribution should be weighted relative to each other -- nor rigorous 
robustness guarantees. 
Moreover, bootstrapped point estimation methods can behave poorly when the MAP and ML estimates are not well-behaved -- for example, 
due to the likelihood being peaked (or even tending to infinity) 
in a region of low posterior probability. 

\citet{Muller:2013} suggests replacing the standard posterior with a Gaussian distribution with covariance proportional to a plug-in estimate of the sandwich covariance.
However, the applicability of such an approach seems rather limited,
since it requires that the dataset be large enough that (1) the sandwich covariance can be well-estimated, and (2) the posterior uncertainty can be represented as approximately Gaussian.
If both these conditions hold, then Bayesian inference may be adding minimal value anyway. 

\subsection{Misspecification and decision theory}

When the model is well-specified, Bayesian inference is the optimal procedure for updating beliefs in light of new data, no matter the loss function~\citep{Robert:2007,BS:2000}.
When the model is misspecified, our analysis of BayesBag shows (near) optimality under log loss, but not necessarily under other loss functions.
When some other loss function is ultimately of interest, there is no reason to assume BayesBag will provide high-quality inferences, although 
the log loss does serve as a reasonable and universally-applicable default choice.
When the model is misspecified and a particular loss function is of interest, generalized belief updating (that is, using a Gibbs posterior) may be more 
appropriate \citep{Bissiri:2016,Syring:2017,Syring:2019}.
It is conceptually straightforward to combine our BayesBag methodology with generalized belief updates to obtain better-calibrated 
inferences that are (near) optimal for the loss function of interest. 

\subsection{Model criticism}

Methods for Bayesian model criticism typically involve ``predictive checks.'' 
For example, prior and posterior predictive checks compare the observed data to data generated from the prior and posterior predictive distributions, respectively~\citep{Guttman:1967,Box:1980,Rubin:1984,Gelman:1996,Vehtari:2012}.
The version most closely related to BayesBag is the \emph{population} predictive check~\citep{Ranganath:2019}, which is based on a fusion of Bayesian and frequentist thinking.
The population predictive check aims to avoid the data-reuse of posterior predictive checks by comparing data generated from the posterior predictive to data generated from
the true distribution.
As a computable approximation to the ideal population predictive check, \citet{Ranganath:2019} suggest bootstrapping datasets and computing the predictive check 
on the data not included in the bootstrap sample.
However, predictive checks and the mismatch index aim to diagnose different aspects of misspecification: predictive checks concern the fit of the observed data
to the likelihood model while the mismatch index addresses calibration of the posterior distribution.
The methods more directly analogous to the mismatch index are various frequentist tests for model misspecification based on the idea of determining whether 
$\Vargradloglik{\optparam} = \Ehessloglik{\optparam}$~\citep{White:1982,Presnell:2004,Zhou:2012}.

\subsection{The benefits of BayesBag}

In view of previous work, the BayesBag approach has a number of attractive features that make it flexible, easy-to-use, and widely applicable. 
From a methodological perspective,  BayesBag is general-purpose.
It relies only on carrying out standard posterior inference, it is applicable to a wide range of models, and it can make full use of 
modern probabilistic programming tools: the only other requirement is the design of a bootstrapping scheme. 
Although this paper focuses on using BayesBag with independent observations, future work can draw on the large literature devoted to adapting the bootstrap to more
complex models such as those involving time-series and spatial data.
BayesBag is also general-purpose in the sense that it is useful no matter whether the ultimate goal of Bayesian inference is parameter estimation, prediction, or model selection;
see \citet{Huggins:2019:BayesBagII} for model selection with BayesBag. 

Another appeal of BayesBag as a methodology is that the only hyperparameter -- the bootstrap dataset size $\bsnumobs$ -- is straightforward to set
because it has a natural, theoretically well-justified choice of $\bsnumobs = \numobs$ that results in conservative inferences.
Meanwhile, when the posterior quantities of interest are sufficiently close to being Gaussian-distributed, the bootstrap dataset size can also be selected 
in a data-driven way using $\optMasymptest$ or $\optMfsest$, making inferences less conservative when the data supports it. 
(See \cref{sec:M-opt-selection} for further discussion of these points.)

In terms of computation, when using the approximation in \cref{eq:bayesbag-approximation}, there is an additional cost due to the need to compute 
the posterior for each bootstrapped dataset.
However, it is trivial to compute the bootstrapped posteriors in parallel.
As described in \cref{sec:choosing-B}, validating that the number of bootstrap datasets $B$ is sufficiently large only requires
computing simple Monte Carlo error bounds. 
Moreover, defaulting to $B=50$ or $100$ appears to be an empirically sound choice across a range of problems. 
Nonetheless, speeding up BayesBag with more specialized computational methods could be worthwhile in some applications.
For example, in \cref{app:computation}, we suggest one simple approach to speeding up Markov chain Monte Carlo (MCMC) runs when using BayesBag.
Pierre Jacob has proposed using more advanced unbiased MCMC techniques for potentially even greater 
computational efficiency.\footnote{\small\url{https://statisfaction.wordpress.com/2019/10/02/bayesbag-and-how-to-approximate-it/}}

A final benefit of BayesBag is that it incorporates robustness features of frequentist methods into Bayesian inference without sacrificing 
the core benefits of the Bayesian approach such as flexible modeling, straightforward integration over nuisance parameters, and the use of prior information. 
This synthesis of Bayesian and frequentist approaches compares favorably to existing methods, which, as we described above, either sacrifice some useful part
of standard Bayesian inference or introduce tuning parameters that are difficult to choose. 
Indeed, BayesBag can actually diagnose how much robustness is necessary via the the model--data mismatch index.
An exciting direction for future work is to better understand the finite-sample properties of BayesBag and the mismatch index.

\subsection*{Acknowledgments}

Thanks to Pierre Jacob for bringing P.~B\"uhlmann's BayesBag paper to our attention.
Thanks also to Ryan Giordano and Pierre Jacob for helpful feedback on an earlier draft of this paper,
and to Peter Gr\"unwald, Natalia Bochkina, Mathieu Gerber, and Anthony Lee for helpful discussions.

\bibliographystyle{imsart-nameyear}
\bibliography{library,../bayesbag}

\newpage

\appendix 

\counterwithin{figure}{section}
\renewcommand{\thefigure}{\Alph{section}.\arabic{figure}}
\counterwithin{table}{section}
\renewcommand{\thetable}{\Alph{section}.\arabic{table}}

\section{Interpretation of the bagged posterior in terms of Jeffrey conditionalization}\label{app:jeffrey-conditionalization}

The bagged posterior has an insightful interpretation in terms of Jeffrey conditionalization. 
This interpretation elegantly unifies the Bayesian and frequentist elements of the bagged posterior 
which might otherwise seem challenging to interpret coherently 
(e.g., the covariance decomposition in \cref{eq:bayesbag-posterior-covariance}).

Suppose we have a model $p(x,y)$ of two variables $x$ and $y$.
In the absence of any other data or knowledge,
we would quantify our uncertainty in $x$ and $y$ via
the marginal distributions $p(x) = \int p(x|y)p(y) \dee y$ 
and $p(y) = \int p(y|x)p(x) \dee x$, respectively.
Now, suppose we are informed that the true distribution of $x$ is $p_{\optsym}(x)$,
but we are not given any samples of $x$ or $y$.
We would then quantify our uncertainty in $x$ via $p_{\optsym}(x)$,
and a natural way to quantify our uncertainty in $y$ is 
via $q(y) \defined \int p(y|x) p_{\optsym}(x) \dee x$.
The idea is that $q(x,y) \defined p(y|x) p_{\optsym}(x)$ updates the model to have the correct distribution of $x$,
while remaining as close as possible to the original model $p(x,y)$.
This is referred to as Jeffrey conditionalization~\citep{Jeffrey:1968,Jeffrey:1990,Diaconis:1982}.

Suppose $\data =(\obs{1},\ldots,\obs{\numobs})$ is the data and $y = \param$ is a parameter, so that 
$p(x,y) = p(\dataarg{\numobs},\param)$ is the joint distribution of the data and the parameter.
If we are informed that the true distribution of the data is $p_{\optsym\numobs}(\dataarg{\numobs})$, then the 
Jeffrey conditionalization approach above is to quantify our uncertainty in $\param$ by 
\[ \label{eq:jeffrey-conditionalization}
q(\param) = \int p(\param \given \dataarg{\numobs}) p_{\optsym\numobs}(\dataarg{\numobs}) \dee \dataarg{\numobs}.
\]

Now, suppose we are not informed of the true distribution exactly, but we are given data $\obsrv{1},\ldots,\obsrv{\numobs}\;\iid \dist p_{\optsym}$.
Since the empirical distribution $\empdist \defined \numobs^{-1}\sum_{n=1}^{\numobs}\delta_{\obsrv{n}}$ is a consistent estimator of $p_{\optsym}$
and $p_{\optsym\numobs}(\dataarg{\numobs}) = \prod_{n=1}^{\numobs} p_{\optsym}(\obs{n})$,
it is natural to plug in $\prod_{n=1}^{\numobs}\empdist$ for $p_{\optsym\numobs}$ in \cref{eq:jeffrey-conditionalization}.  Doing so, we arrive at the bagged posterior:
\[
q(\param)\approx\int p(\param \given \dataarg{\numobs}) \prod_{n=1}^{\numobs} \empdist(\dee \obs{n}) =
\EE\big\{p(\param \given \datarvarg{\numobs}^{\bbsym}) \given \datarvarg{\numobs}\big\}
\]
where $\obsrv{1}^{\bbsym},\ldots,\obsrv{\numobs}^{\bbsym}\;\iid \dist \empdist$ given $\datarvarg{\numobs}$.

\section{Additional experimental details} \label{app:experiment-details}

\subsection*{Hierarchical mixed effects logistic regression model}

We computed maximum likelihood estimates with the \texttt{R} package \texttt{lme4}, which uses a Laplace approximation to integrate out the random effects;
for prediction we used Monte Carlo to integrate out the random effects. 
To approximate the standard and bagged posteriors, we used Stan's implementation of dynamic Hamiltonian Monte Carlo with 4 chains (for the standard posterior) or 2 chains
(for the bagged posterior) each run for 2,000 total iterations (discarding the first half as burn-in).
For BayesBag we used $B = 100$ bootstrap datasets. 
For the MLE-based bootstrap methods we used 500 bootstrap datasets.

\subsection*{Sparse logistic regression}

\begin{table}[tp]
\caption{Real-world datasets used in binary classification experiments.}
\begin{center}
\begin{tabular}{c|c|c|c}
Name 				& $N$ 	& $D$ \\
\hline
Colon				& 62		& 2,000 \\
Leukemia			& 72		& 7,129 \\
Ovarian			& 54		& 1,536 \\
Prostate			& 102		& 5,966 \\
\end{tabular}
\end{center}
\label{tbl:datasets}
\end{table}%

\cref{tbl:datasets} summarizes the datasets used. 
We used M.~Betancourt's Stan implementation of the model from \citet{Piironen:2017}.\footnote{\url{https://betanalpha.github.io/assets/case_studies/bayes_sparse_regression.html}}
To approximate the standard and bagged posteriors, we used Stan's implementation of dynamic Hamiltonian Monte Carlo with 4 chains each run for 2,000 total iterations (discarding the first half as burn-in).
We used Stan's built-in convergence diagnostics in our preliminary experiments to confirm acceptable mixing; however, we then turned the diagnostics off because they significantly increased runtime. 

\section{Additional figures and tables} \label{app:additional-figs}

\subsection*{The mismatch index with correlated versus uncorrelated regressors} 
\begin{figure}[tbp]
\begin{center}
\begin{subfigure}[b]{.36\textwidth}
\centering
\includegraphics[width=1.6in]{{linreg-synth-uncorr-dense-linear-gaussian-c-opt-B-100-D-10-N-50-a0-2.0-reps-50-short-model-mismatch-hist}.pdf}
\caption{\textsf{default}, $N = 50$}
\end{subfigure}
\begin{subfigure}[b]{.36\textwidth}
\centering
\includegraphics[width=1.6in]{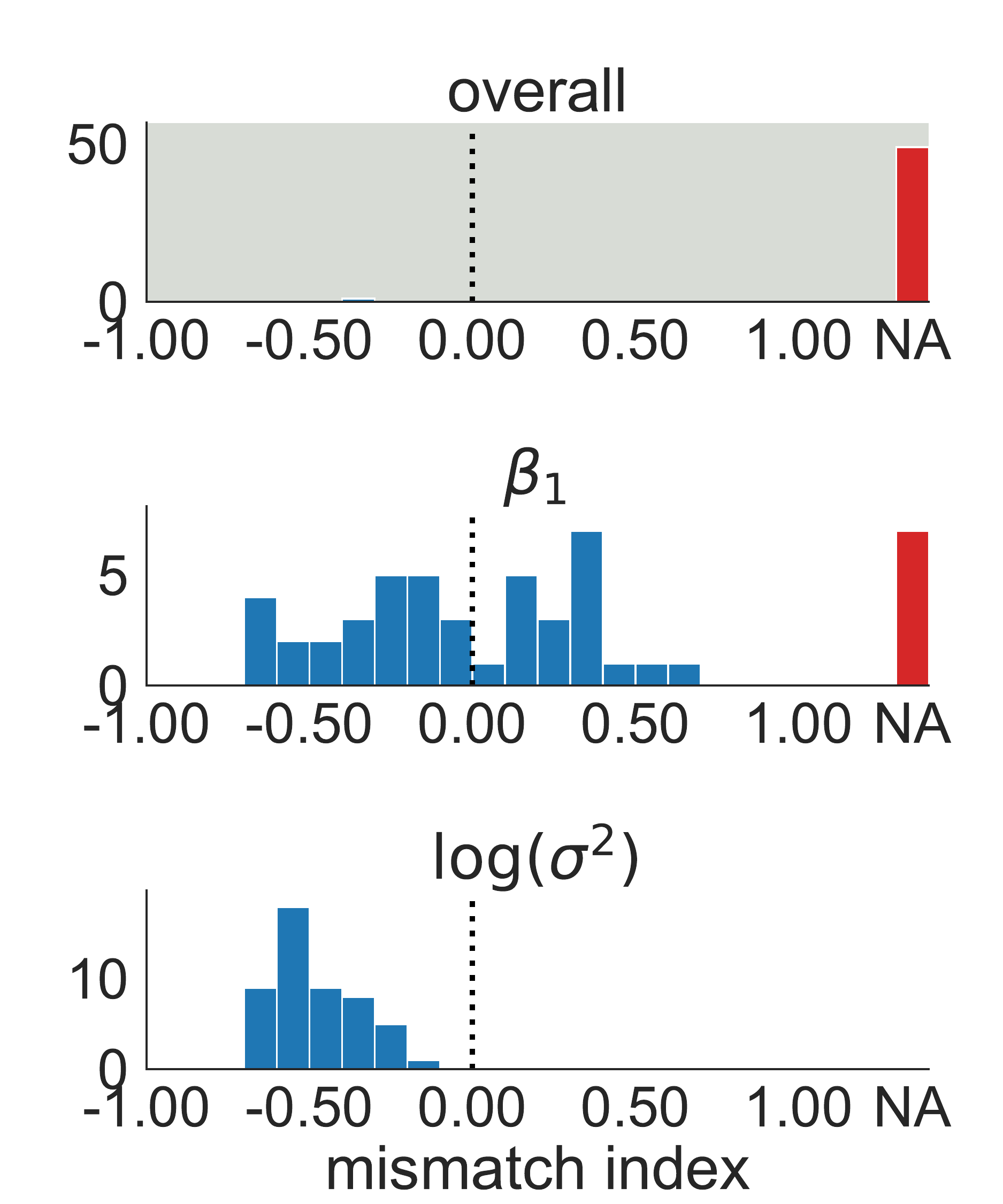}
\caption{\textsf{correlated}, $N = 50$}
\end{subfigure}  \\
\begin{subfigure}[b]{.36\textwidth}
\centering
\includegraphics[width=1.6in]{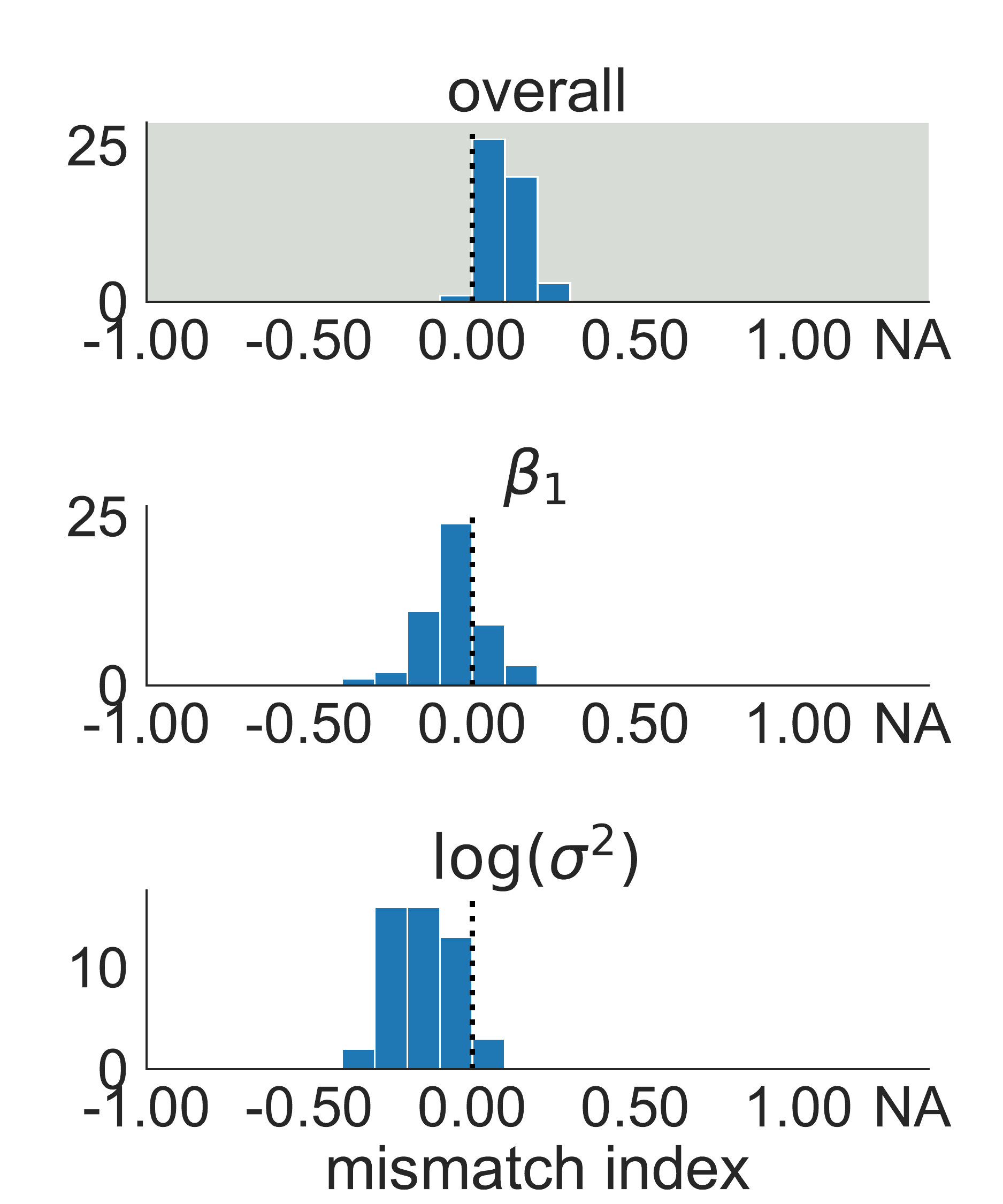}
\caption{\textsf{default}, $N = 200$}
\end{subfigure} 
\begin{subfigure}[b]{.36\textwidth}
\centering
\includegraphics[width=1.6in]{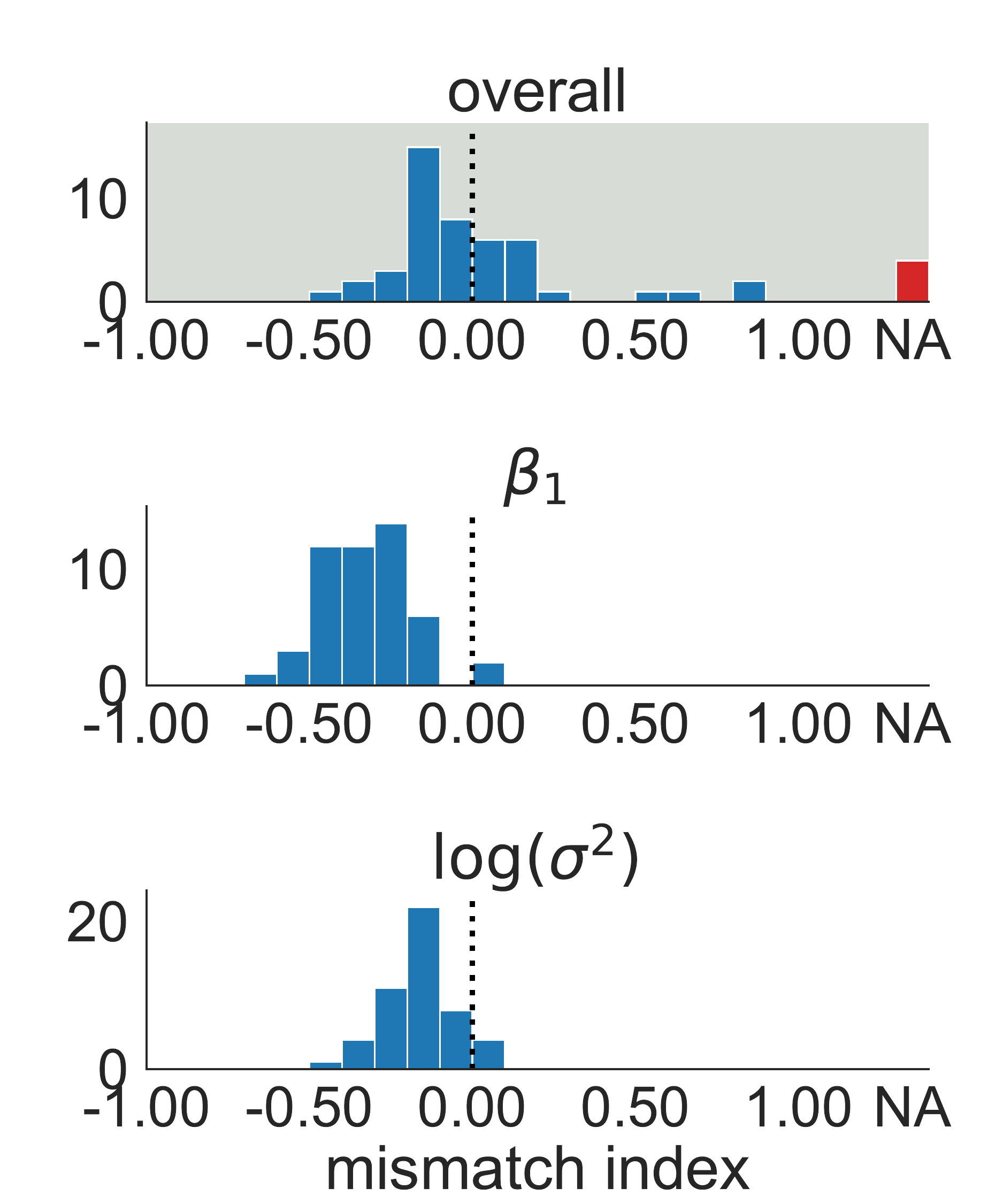}
\caption{\textsf{correlated}, $N = 200$}
\end{subfigure} 
\caption{Model--data mismatch indices $\modelmismatch$ for selected parameters as well as the overall $\modelmismatch$ value
for \textsf{default} and \textsf{correlated} data for $N \in \{50, 200\}$.
We only display one component of $\beta$ since the $\modelmismatch$ values followed very similar distributions for all components. 
}
\label{fig:linear-regression-correlated-model-mismatch}
\end{center}
\end{figure}

Continuing the discussion in \cref{sec:model-mismatch-simulations}, we consider cases involving poor identifiability in the model. 
\cref{fig:linear-regression-correlated-model-mismatch} compares the mismatch index with \textsf{default} and \textsf{correlated} data for $N \in \{50, 200\}$.
The poor identifiability of the \textsf{correlated} data was correctly detected by $\modelmismatch$.
The identifiability issue becomes less severe with more data, which is reflected in the $\modelmismatch$ values clustered around zero when $N = 200$. 
On the other hand, no identifiability issues were present for the uncorrelated \textsf{default} data, resulting in $\modelmismatch$ values that were
appropriately clustered near 0. 

\subsection*{Qualitative results for sparse logistic regression}
\begin{figure}[tbp]
\begin{center}
\begin{subfigure}[b]{\textwidth}
\centering
\includegraphics[height=2in]{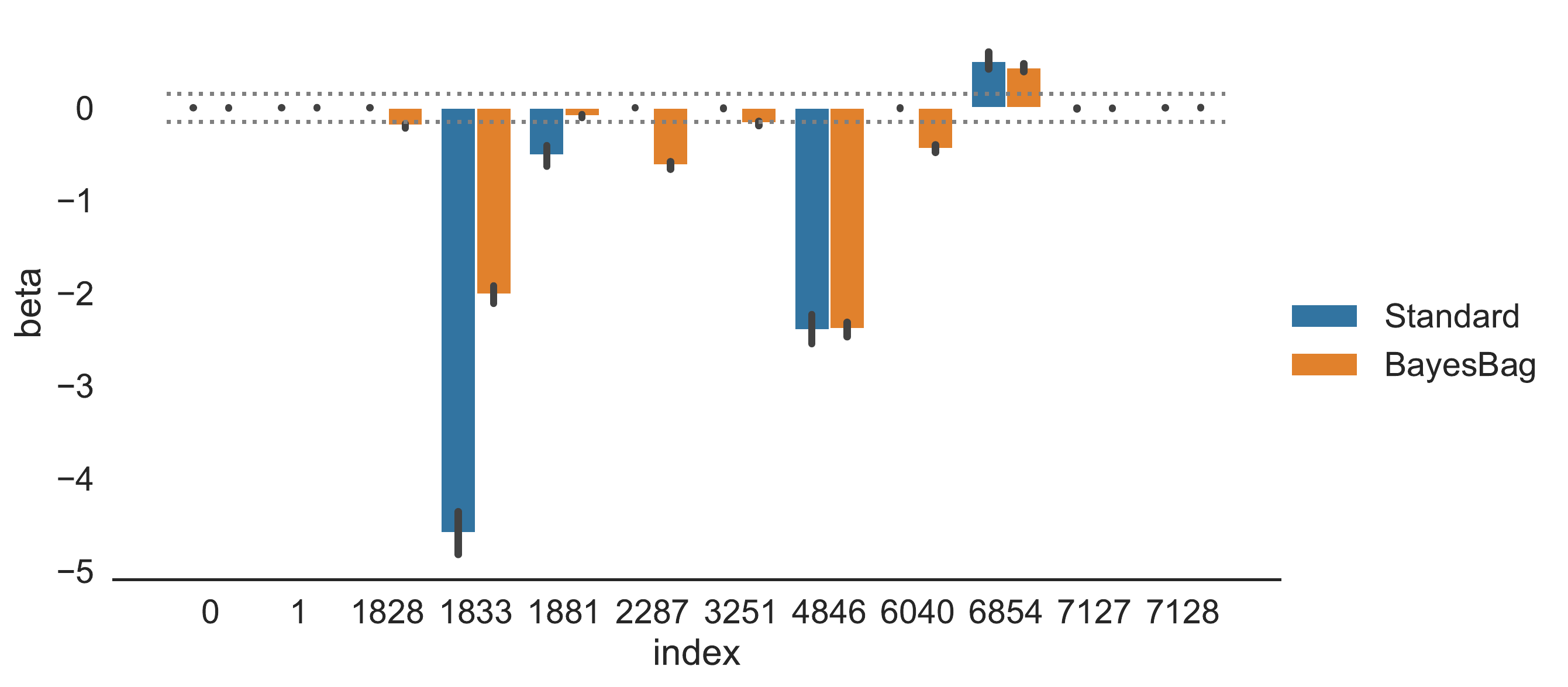}
\caption{leukemia}
\end{subfigure}
\begin{subfigure}[b]{.49\textwidth}
\centering
\includegraphics[height=2in]{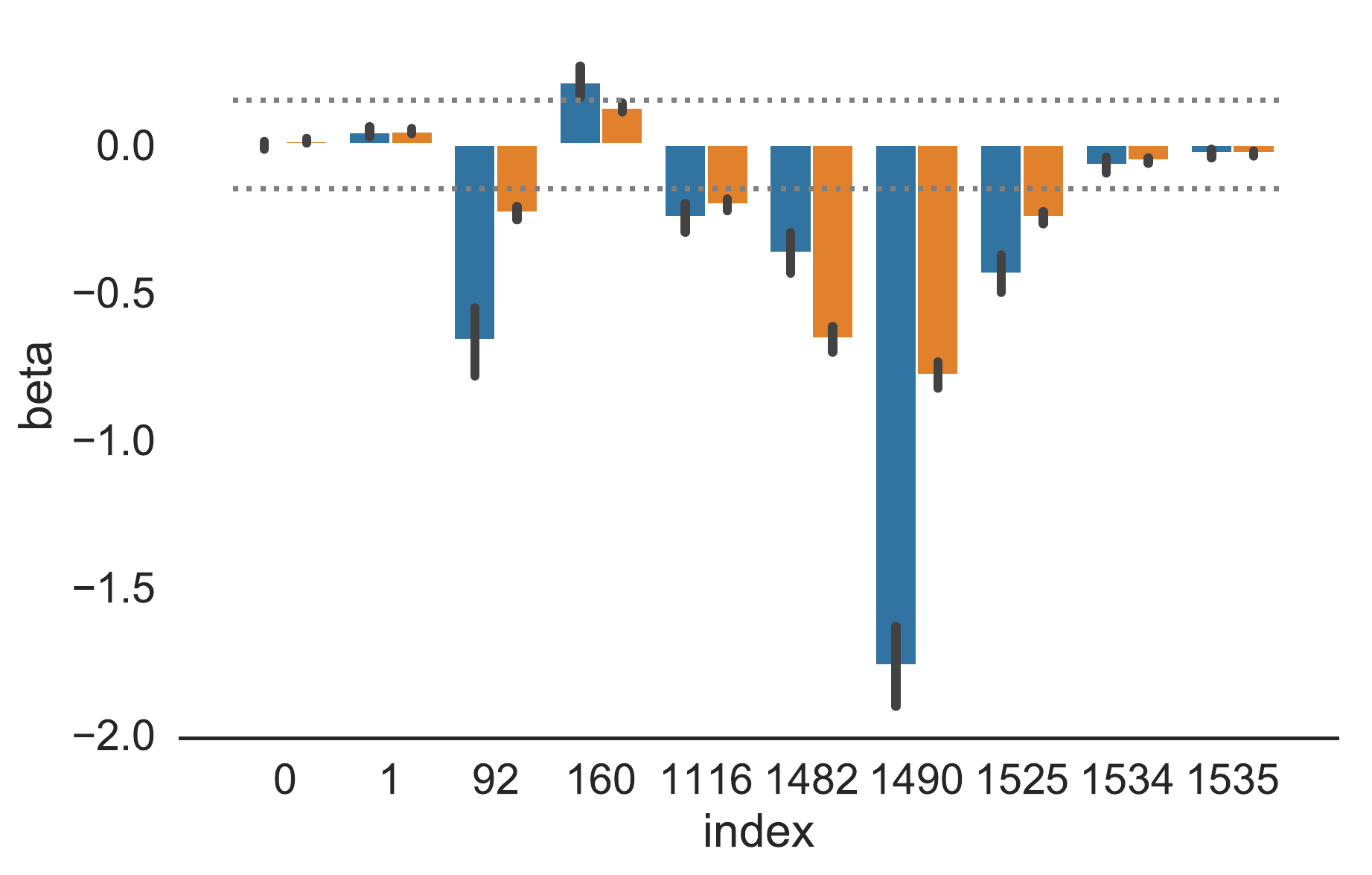}
\caption{ovarian}
\end{subfigure}
\begin{subfigure}[b]{.49\textwidth}
\centering
\includegraphics[height=2in]{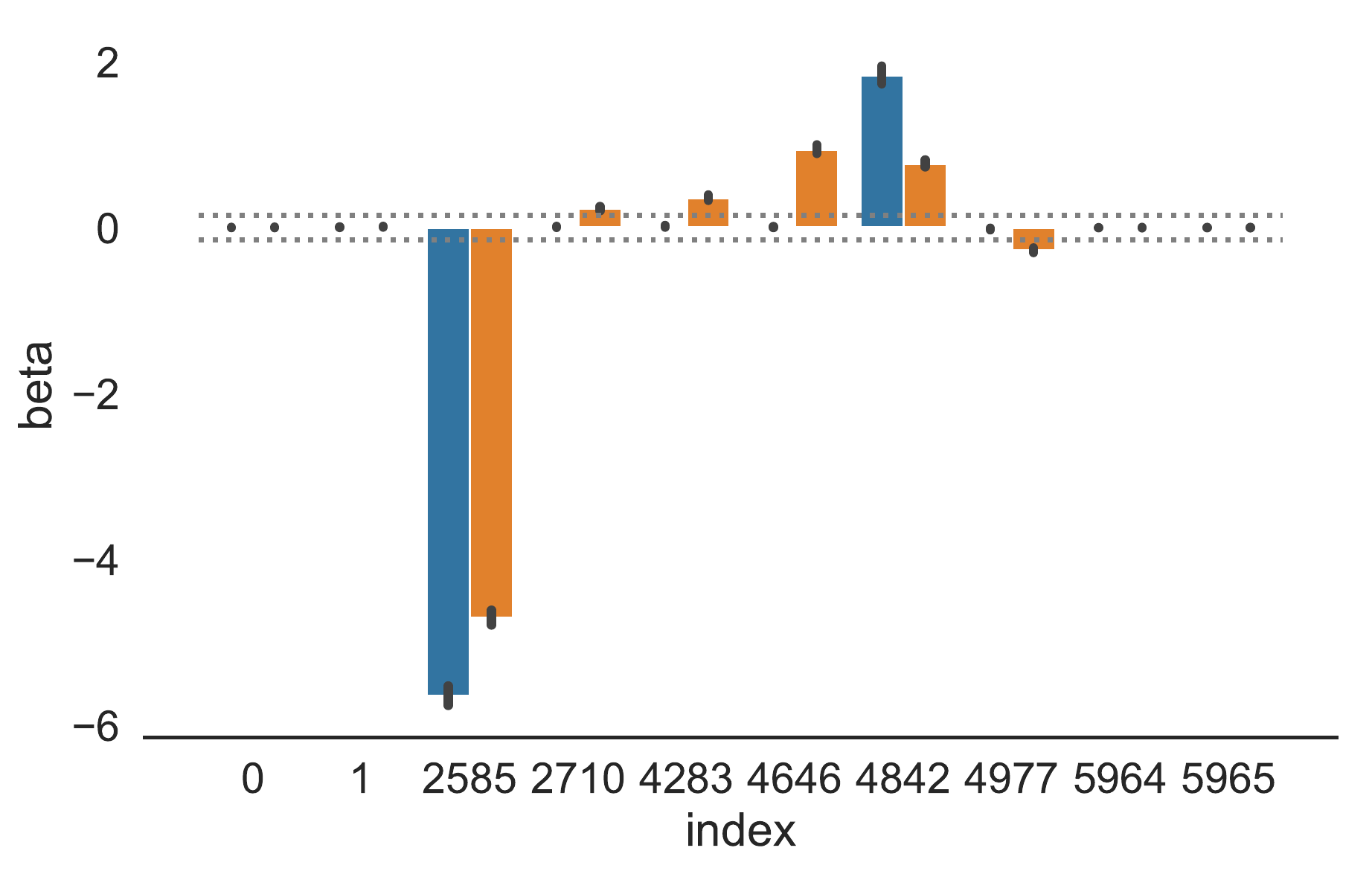}
\caption{prostate}
\end{subfigure}

\caption{For sparse logistic regression, posterior means of components with significant coefficients (absolute mean greater than 0.15) and
representative non-significant coefficients (with component indices 0, 1, $D-2$, and $D-1$).
The standard deviations are shown with black lines.  
}
\label{fig:sparse-logistic-regression-qualitative-other}
\end{center}
\end{figure}

Similarly to what \cref{fig:sparse-logistic-regression-qualitative} shows for the colon dataset, 
\cref{fig:sparse-logistic-regression-qualitative-other} shows that in the leukemia, ovarian, and prostate datasets, using BayesBag can lead to qualitatively 
different inferences compared to the standard posterior.

\clearpage

\section{Computation}\label{app:computation}

When the posterior can be computed in closed form, using BayesBag is straightforward.
If, however, approximate sampling methods such as Markov chain Monte Carlo are necessary, the computational cost could become substantial. 
In such cases we propose the basic scheme described in \cref{alg:bb-sampler}, although more advanced approaches could also be developed.
In short, the idea is to run a single long chain (or set of chains) on the standard posterior, then use the sampler hyperparameters
and posterior samples to initialize shorter chains that sample from many different bootstrap datasets.
Our proposed algorithm facilitates the use of $\optMest$ and $\modelmismatch$ since it outputs 
samples from the standard posterior as well as the bagged posterior. 

If the approximation of $\postdensityfull{\param}{\bsdatasample{b}}$ is not very accurate
(e.g., because it requires a time-consuming Markov chain Monte Carlo run), then we face a tradeoff between the error due to 
approximating each $\postdensityfull{\param}{\bsdatasample{b}}$ and the Monte Carlo error due to the BayesBag approximation given in \cref{eq:bayesbag-approximation}.
When using Markov chain Monte Carlo, we recommend assessing on how accurate different length Markov chains are likely to be 
by running long chains for the standard posterior, then using this information to decide on the best trade off between the length of the Markov chains and number of bootstrap datasets. 
Such an approach should not result in much wasted computation since we suggest obtaining a high-quality approximation to the standard posterior no matter what, 
as this permits computing quantities such as the optimal bootstrap sample size and the model--data mismatch index.

\begin{algorithm}
\caption{Basic BayesBag Sampler}\label{alg:bb-sampler}
\begin{algorithmic}[1]
\Require A Markov chain Monte Carlo procedure \Call{MCMC}{$\data$, $T$, $\param_\text{init}$, $\beta_\text{init}$} that returns adapted sampler hyperparameters 
and $T$ approximate samples from $\postdistfull{\cdot}{\data}$, with the sampler initialized at $\param_\text{init}$ with hyperparameters $\beta_\text{init}$
\Require Data $\data$, ``large'' sample number $T$, ``small'' sample number $T^{\bbsym}$, number of bootstrap datasets $B$, 
initial hyperparameters $\beta_\text{init}$

\State $\beta, \param_{1:T} \gets$ \Call{MCMC}{$\data$, $T$, $\beta_\text{init}$} %
\For{$b = 1,\dots,B$}
	\State Generate a new bootstrap dataset $\bsdatasample{b}$ of size $\bsnumobs$ from $\data$
	\State Sample $\param^{\bbsym}_{(b)\text{init}}$ uniformly from $\param_{1:T}$
	\State $\beta_{(b)}, \param^{\bbsym}_{(b)1:T^{\bbsym}} \gets$ \Call{MCMC}{$\bsdatasample{b}$, $T^{\bbsym}$, $\param^{\bbsym}_{(b)\text{init}}$, $\beta$}
\EndFor
\State $\theta^{\bbsym}_{1:BT^{\bbsym}} \gets \text{concatenate}(\param^{\bbsym}_{(1)1:T^{\bbsym}},\dots, \param^{\bbsym}_{(B)1:T^{\bbsym}})$
\State \Return posterior samples $\param_{1:T}$ and BayesBag samples $\theta^{\bbsym}_{1:BT^{\bbsym}}$
\end{algorithmic}
\end{algorithm}

\newpage

\section{Derivation of finite-sample optimal bootstrap size} \label{app:finite-sample-optimal-bootstrap-size}

We conclude with a derivation of the finite-sample optimal bootstrap sample size estimator $\optMfsest$. 
Recall that a potential shortcoming of using $\optMasymptest$ to choose the optimal bootstrap sample size is that it does not account for the influence of the prior.
If the prior remains influential, then $v^{\bbsym}_{\numobs} - v_{\numobs}$ may be deceptively small, leading $\optMasymptest$ to be too large
and the resulting bagged posterior to be overconfident.
To account for the effect of the prior, we can instead use \cref{eq:bayesbag-posterior-covariance-normal-location}.
Define $\sigma_{\optsym}^{2}$ and $s_{\optsym}^{2}$ as in \cref{sec:M-opt-selection}.
\cref{eq:bayesbag-posterior-covariance-normal-location} yields the following more refined approximation 
$v_{\bsnumobs}^{\bbsym} \approx (R_{\bsnumobs}\sigma^{2}_{\optsym} + R_{\bsnumobs}^{2}s_{\optsym}^{2})/\bsnumobs$ 
to the bagged posterior variance.
Note that since $\sigma^{2}_{\optsym}$ now plays the role that $V$ played in the Gaussian location model, $R_{\bsnumobs} = (1 + \sigma^{2}_{\optsym}v_{0}^{-1}/\bsnumobs)^{-1}$.
Since $v_{\bsnumobs}^{\bbsym}$ needs to be approximately $s^{2}_{\optsym}/\numobs$ in order to be well-calibrated, we 
set $(R_{\bsnumobs}\sigma^{2}_{\optsym} + R_{\bsnumobs}^{2}s_{\optsym}^{2})/\bsnumobs = s^{2}_{\optsym}/\numobs$ and solve for $\bsnumobs$,
which yields
\[
\optMfs
&=  \frac{\numobs}{2} + \frac{\numobs\sigma_{\optsym}^{2}}{2s_{\optsym}^{2}}  - \frac{\sigma_{\optsym}^{2}}{v_{0}}
  +   \left\{ \left(\frac{\numobs}{2} + \frac{\numobs\sigma_{\optsym}^{2}}{2s_{\optsym}^{2}}\right)^{2} - \frac{\numobs \sigma_{\optsym}^{2}}{v_{0}}\right\}^{1/2}.
\]
It remains to derive the estimators for $\sigma_{\optsym}^{2}$ and $s_{\optsym}^{2}$.
Solving $v_{\numobs} = \sigma_{\optsym}^{2}v_{0}/(\numobs v_{0} + \sigma_{\optsym}^{2})$
for $\sigma_{\optsym}^{2}$ yields the finite-sample estimator $\hat\sigma_{\optsym}^{2} = \numobs v_{0} v_{\numobs}/(v_{0} - v_{\numobs})$ 
and plugging $\hat\sigma_{\optsym}^{2}$ into the definition of $R_{\numobs}$ yields the estimator $\hat R_{\numobs} = 1 - v_{\numobs}/v_{0}$.
Combining these yields the finite-sample estimator
\[
\hs_{\optsym}^{2} 
\defined \frac{v_{0}^{2}}{(v_{0} - v_{\numobs})^{2}}(v^{\bbsym}_{\numobs} - v_{\numobs})\numobs.
\]
Observe that we recover the asymptotic versions of the variance estimators and $\optMasympt$ by taking $v_{0} \to \infty$.

\section{Proofs}

The characteristic function of a distribution $\eta$ on $\reals^K$ is denoted $\cf{\eta}(t) \defined \int \exp(\ii t^\top x) \eta(\dee x)$ for $t\in\reals^K$.
We use $\convP$ to denote convergence in probability and $\convPouter$ to denote convergence in outer probability.

\subsection{Proof of \cref{prop:bb-bbvm-gaussian-location}} \label{sec:proof-of-bb-bbvm-gaussian-location}

We use the classical characteristic function approach to proving central limit theorems. 
For $\mu \in \reals$ and $\sigma^{2} > 0$, the characteristic function of $\distNorm(\mu, \sigma^{2})$ is
\[
\cf{\distNorm(\mu, \sigma^{2})}(t) &= \exp(\ii \mu t - \sigma^{2}t^{2}/2), \qquad t \in \reals. \label{eq:gaussian-characteristic-alt}
\]
For $L \in \nats$ and $p_1,\ldots,p_K \ge 0$ with $\sum_{k=1}^{K}p_{k} = 1$, 
the characteristic function of the multinomial distribution $\distMulti(L, p)$ is
\[
\cf{\distMulti(L, p)}(t) &= \left(\sum_{k=1}^{K}p_{k}e^{\ii t_{k}}\right)^{L}, \qquad t \in \reals^{K}. \label{eq:multinomial-characteristic-alt}
\]
Let $\widetilde\Pi(\cdot \given \bsdatarvarg{\bsnumobs}) \defined \distNorm(\numobs^{1/2} R_{\bsnumobs}(\bsdatarvmeanarg{\bsnumobs} - \datarvmeanarg{\numobs}), \numobs V_{\bsnumobs})$,
noting that this is the distribution of $\numobs^{1/2}\{\bbparamsample - \EE(\bbparamsample\given\datarvarg{\numobs} )\} \given \bsdatarvarg{\bsnumobs}$.
Similarly, let $\widetilde\Pi^{\bbsym}(\cdot \given \datarvarg{\numobs})$ denote the distribution of $\numobs^{1/2}\{\bbparamsample - \EE(\bbparamsample\given\datarvarg{\numobs} )\} \given \datarvarg{\numobs}$.
Let $Y_{\numobs n} \defined \numobs^{1/2} R_{\bsnumobs}(\obsrv{n} - \datarvmeanarg{\numobs})$
and let $\bscounts \dist \distMulti(\bsnumobs, 1/\numobs)$ .
Using \cref{eq:gaussian-characteristic-alt,eq:multinomial-characteristic-alt}, we have 
\[
\cf{\widetilde\Pi^{\bbsym}(\cdot \given \datarvarg{\numobs})}(t) 
&= \EE\{\cf{\widetilde\Pi(\cdot \given \bsdatarvarg{\bsnumobs})}(t)\given \datarvarg{\numobs}\} \\
&= \EE\big[\exp\!\big\{\ii t \bsnumobs^{-1} \textstyle{\sum_{n=1}^\numobs} \bscount{n} Y_{\numobs n} - \numobs V_{\bsnumobs}t^{2}/2\big\}\given \datarvarg{\numobs}\big] \\
&= \left\{\frac{1}{\numobs}\sum_{n=1}^{\numobs} \exp(\ii t \bsnumobs^{-1} Y_{\numobs n})\right\}^{\bsnumobs} \exp(-\numobs V_{\bsnumobs}t^{2}/2).
\label{eq:bb-gaussian-location-characteristic-part-1-alt}
\]
Let $\hat{V}_{\numobs} \defined \numobs^{-1}\sum_{n=1}^{\numobs} (\obsrv{n} - \datarvmeanarg{\numobs})^{2}$. 
By Taylor's theorem, $e^{\ii s} = 1 + \ii s - s^{2}/2 + \mcR(s)$ where $\mcR(s) \le |s|^{3}/3$. 
Since $\numobs^{-1} \sum_{n=1}^\numobs Y_{\numobs n} = 0$, 
the first factor of \cref{eq:bb-gaussian-location-characteristic-part-1-alt} can be expanded as 
\[
&\left\{\frac{1}{\numobs}\sum_{n=1}^{\numobs} \bigg(1 + \ii t \bsnumobs^{-1} Y_{\numobs n} - \tfrac{1}{2} t^2 \bsnumobs^{-2} Y_{\numobs n}^2 + \mcR(t \bsnumobs^{-1} Y_{\numobs n}) \bigg)\right\}^{\bsnumobs} \\
&= \left\{1 - \tfrac{1}{2} t^2 \frac{\numobs R_{\bsnumobs}^{2}}{\bsnumobs^{2}} \hat{V}_{\numobs} + \frac{1}{\numobs}\sum_{n=1}^{\numobs}\mcR(t \bsnumobs^{-1} Y_{\numobs n})\right\}^{\bsnumobs}.
\label{eq:bb-gaussian-location-characteristic-part-2-alt}
\]
Since $\mcR(s) \le |s|^{3}/3$,
\[
\sum_{n=1}^{\numobs}\mcR(t \bsnumobs^{-1} Y_{\numobs n}) 
\le \frac{t^{3}\numobs^{3/2}R_{\bsnumobs}^{3}}{3\bsnumobs^{3}} \sum_{n=1}^{\numobs} |\obsrv{n} - \datarvmeanarg{\numobs}|^{3}.
\]
Using $|\obsrv{n} - \datarvmeanarg{\numobs}|^{3} \le |\obsrv{n}|^{3} + 3 |\obsrv{n}|^{2}|\datarvmeanarg{\numobs}| +  3 |\obsrv{n}||\datarvmeanarg{\numobs}|^{2} + |\datarvmeanarg{\numobs}|^{3}$, and applying the the strong law of large numbers to each factor, we have
\[
\limsup_{\numobs\to\infty} \frac{1}{\numobs}\sum_{n=1}^{\numobs} |\obsrv{n} - \datarvmeanarg{\numobs}|^{3} \overset{a.s.}{<} \infty.
\]
Hence, almost surely, for all $t \in \reals$, $\sum_{n=1}^{\numobs}\mcR(t \bsnumobs^{-1} Y_{\numobs n}) \to 0$ as $\numobs\to\infty$.
Further, note that $\bsnumobs/\numobs \to c$, $R_{\bsnumobs} \to 1$, and $\hat{V}_{\numobs} \convas \var(\obsrv{1})$ as $\numobs\to\infty$.
Now, we use the fact that if $a_\numobs \to a$ and $c_\numobs \to c$, then $(1 + a_\numobs/\numobs)^{\numobs c_\numobs} \to \exp(a)^c$.
Thus, almost surely, for all $t$, \cref{eq:bb-gaussian-location-characteristic-part-2-alt} converges to $\exp(-\tfrac{1}{2} t^2 \var(\obsrv{1}) / c)$.
Combining this with \cref{eq:bb-gaussian-location-characteristic-part-1-alt},
and noting that $\numobs V_{\bsnumobs} \to V/c$, we have that almost surely, for all $t\in\reals$,
\[
\cf{\widetilde\Pi^{\bbsym}(\cdot \given \datarvarg{\numobs})}(t) 
\to \exp(-\tfrac{1}{2} t^2 (\var(\obsrv{1})/c + V/c)).
\]
The result follows by L\'evy's continuity theorem~\citep[Theorem 5.3]{Kallenberg:2002}.

\subsection{Proof of \cref{thm:bb-bvm}} \label{sec:proof-of-bb-bvm}
To de-clutter the notation, we abbreviate $\Ehessloglik{\optsym} \defined \Ehessloglik{\optparam}$, 
$\Vargradloglik{\optsym} \defined \Vargradloglik{\optparam}$,
and $\gradloglikfun{\optsym} \defined \gradloglikfun{\optparam}$.
Define
\[ 
\bbempdist &\defined \bsnumobs^{-1}\sum_{n=1}^{\numobs}\bscount{n}\delta_{\obsrv{n}}, \\
\Delta^{\bbsym}_{\numobs} &\defined \numobs^{1/2}\Ehessloglik{\optsym}^{-1}(\bbempdist - \empdist)\gradloglikfun{\optparam},
\]
the empirical process $\mathbb{G}_{\numobs} = \numobs^{1/2}(\empdist - \obsdist)$, and 
the bootstrap empirical process $\mathbb{G}^{\bbsym}_{\numobs} = \bsnumobs^{1/2}(\bbempdist - \empdist)$. 
The conditions of \citet[Lemma 19.31]{vanderVaart:1998} hold by assumption, so for any sequence $h_1,h_2,\ldots\in\reals^{D}$ bounded in probability, 
\[
\mathbb{G}_{\numobs}\{\numobs^{1/2}\lambda_{\numobs} - h_{\numobs}^{\top}\gradloglikfun{\optsym}\} \convP 0,
\]
where $\lambda_{\numobs} = \loglikfun{\optparam + h_{\numobs}/\numobs^{1/2}} - \loglikfun{\optparam}$.
By \citet[Theorem 3.6.3]{vanderVaart:1996}, for almost every $\alldatarv$, conditional on $\alldatarv$, $\mathbb{G}^{\bbsym}_{\numobs}$ and $\mathbb{G}_{\numobs}$ both 
converge weakly to the same limiting process.
For the remainder of the proof we condition on $\alldatarv$, so all statements will hold for almost every $\alldatarv$. 
It follows that
\[
\mathbb{G}^{\bbsym}_{\numobs}\{\numobs^{1/2}\lambda_{\numobs} - h_{\numobs}^{\top}\gradloglikfun{\optsym}\} \convPouter 0. \label{eq:bs-emp-proc-to-zero}
\]
By the proof of \citet[Lemma 2.1]{Kleijn:2012}, 
\[
|N\empdist\lambda_{\numobs} - \mathbb{G}_{\numobs} h_{\numobs}^{\top}\gradloglikfun{\optsym} - \tfrac{1}{2}h_{\numobs}^{\top} \Ehessloglik{\optsym}h_{\numobs}| \convPouter 0
\]
and, following the same reasoning, we can expand the lefthand side of \cref{eq:bs-emp-proc-to-zero} and multiply though by $\bsscale^{1/2}$ to get 
\[
\bsscale^{1/2}(\numobs\bsnumobs)^{1/2} \bbempdist\lambda_{\numobs} - \bsscale^{1/2}\mathbb{G}^{\bbsym}_{\numobs} h_{\numobs}^{\top}\gradloglikfun{\optsym} 
 - \bsscale^{1/2}(\numobs\bsnumobs)^{1/2}\empdist\lambda_{\numobs} \convPouter 0
\]
and hence
\[
\bsnumobs \bbempdist\lambda_{\numobs} - (\bsscale^{1/2}\mathbb{G}^{\bbsym}_{\numobs} + \bsscale\mathbb{G}_{\numobs}) h_{\numobs}^{\top}\gradloglikfun{\optsym} - \tfrac{1}{2}h_{\numobs}^{\top}(\bsscale\Ehessloglik{\optsym})h_{\numobs} \convPouter 0. 
\]
Since $\bsscale\,\mathbb{G}_{\numobs} h_{\numobs}^{\top}\gradloglikfun{\optsym} = h_{\numobs}^{\top} (\bsscale \Ehessloglik{\optsym})\Delta_{\numobs}$ 
and $\bsscale^{1/2}\mathbb{G}^{\bbsym}_{\numobs} h_{\numobs}^{\top}\gradloglikfun{\optsym} (\bsscale \numobs / \bsnumobs)^{1/2} = h_{\numobs}^{\top} (\bsscale \Ehessloglik{\optsym})\Delta^{\bbsym}_{\numobs}$ by the definitions of $\Delta_{\numobs}$ and $\Delta^{\bbsym}_{\numobs}$,
it follows that for every compact $K \subset \paramspace$, 
\[
\sup_{h \in K} \Big\vert \bsnumobs \bbempdist(\loglikfun{\optparam + h/\numobs^{1/2}} - \loglikfun{\optparam}) - h^{\top} (\bsscale\Ehessloglik{\optsym}) (\Delta_{\numobs} + \Delta^{\bbsym}_{\numobs}) - \tfrac{1}{2}h^{\top}(\bsscale\Ehessloglik{\optsym})h\Big\vert \convPouter 0.
\]
We apply \citet[Theorem 2.1]{Kleijn:2012} to conclude that, letting $\bbparamsamplecopy|\bsdatarvarg{\bsnumobs} \dist \postdistfull{\cdot}{\bsdatarvarg{\bsnumobs}}$, 
the total variation distance between the distribution of $\numobs^{1/2}(\bbparamsamplecopy -  \optparam) \mid \bsdatarvarg{\bsnumobs}$
and $\distNorm(\Delta_{\numobs} + \Delta^{\bbsym}_{\numobs}, \Ehessloglik{\optsym}^{-1}/\bsscale)$ converges to zero in outer probability.
Compared to the notation of \citet[Theorem 2.1]{Kleijn:2012}, we have $\bsdatarvarg{\bsnumobs}$ in place of $X^{(n)}$, $\empdist^{\bsnumobs}$ in place of $P_{0}^{(n)}$,
$\bsscale\Ehessloglik{\optsym}$ in place of $V_{\theta^{*}}$, and $\Delta_{\numobs} + \Delta^{\bbsym}_{\numobs}$ in place of $\Delta_{n,\theta^{*}}$. 
Hence, uniformly in $t \in \reals^{D}$, the absolute difference in their characteristic functions must also converge to zero in outer probability.
Let $\eps_{\numobs}(t)$ (and similarly $\bar\eps_{\numobs}(t)$) denote a function that satisfies $\limsup_{\numobs \to \infty} \sup_{t \in \reals} \eps_{\numobs}(t) = 0$. 
We can therefore write the characteristic function of $\numobs^{1/2}(\bbparamsample - \optparam) - \Delta_{\numobs} \mid \datarvarg{\numobs}$ 
evaluated at $t \in \reals^{D}$ as
\[
\lefteqn{\EE\left[\exp\left\{\ii\Delta^{\bbsym \top}_{\numobs}t - t^{\top}\Ehessloglik{\optsym}^{-1}t/(2\bsscale)\right\}\given \datarvarg{\numobs}\right] + \eps_{\numobs}(t)} \\
\begin{split}
&= \EE\left[\exp\left\{\ii \numobs^{1/2} \bbempdist \gradloglikfun{\optsym}^{\top}\Ehessloglik{\optsym}^{-1}t \right\} \;\big|\; \datarvarg{\numobs}\right]  \exp\left\{- \ii \numobs^{1/2} \empdist\gradloglikfun{\optsym}^{\top}\Ehessloglik{\optsym}^{-1}t\right\} \\
&\phantom{=~}\times \exp\left\{ - t^{\top}\Ehessloglik{\optsym}^{-1}t/(2\bsscale)\right\} + \eps_{\numobs}(t).  \label{eq:bb-bvm-characteristic-function-initial}
\end{split}
\]
Letting $\delta\gradloglik{\obsrv{n}}{\optsym} \defined \gradloglik{\obsrv{n}}{\optsym} -\empdist\gradloglikfun{\optsym}$,
we can further expand the first line of \cref{eq:bb-bvm-characteristic-function-initial} to get 
\[
&\EE\left[\exp\left\{\ii \numobs^{1/2}\bsnumobs^{-1}\sum_{n=1}^{\numobs}\bscount{n}\gradloglik{\obsrv{n}}{\optsym}^{\top} \Ehessloglik{\optsym}^{-1}t \right\}\;\bigg|\; \datarvarg{\numobs}\right] \exp\left\{- \ii \numobs^{1/2} \empdist\gradloglikfun{\optsym}^{\top}\Ehessloglik{\optsym}^{-1}t\right\} \\
&= \left[ \frac{1}{\numobs} \sum_{n=1}^{\numobs} \exp\left\{\frac{\ii  \numobs^{1/2}\gradloglik{\obsrv{n}}{\optsym}^{\top}\Ehessloglik{\optsym}^{-1}t}{\bsnumobs}   \right\}\right]^{\bsnumobs} \exp\left\{- \ii \numobs^{1/2} \empdist\gradloglikfun{\optsym}^{\top}\Ehessloglik{\optsym}^{-1}t\right\} \\
&= \left[ \frac{1}{\numobs} \sum_{n=1}^{\numobs} \exp\left\{\frac{\ii  \numobs^{1/2}\delta\gradloglik{\obsrv{n}}{\optsym}^{\top}\Ehessloglik{\optsym}^{-1}t}{\bsnumobs}   \right\}\right]^{\bsnumobs}\\
&= \left[\frac{1}{\numobs}\sum_{n=1}^{\numobs}\left\{ 1 + \frac{\ii\numobs^{1/2}\delta\gradloglik{\obsrv{n}}{\optsym}^{\top}\Ehessloglik{\optsym}^{-1}t}{\bsnumobs} -  \frac{\numobs(\delta\gradloglik{\obsrv{n}}{\optsym}^{\top}\Ehessloglik{\optsym}^{-1}t)^{2}}{2\bsnumobs^{2}} + \mcR_{n} \right\}\right]^{\bsnumobs}  \\
&= \left\{1 -  \frac{\numobs t^{\top}\Ehessloglik{\optsym}^{-1}\empdist(\delta\gradloglikfun{\optsym}\delta\gradloglikfun{\optsym}^{\top})\Ehessloglik{\optsym}^{-1}t}{2\bsnumobs^{2}} + \mcR_{n} \right\}^{\bsnumobs},  \label{eq:bb-bvm-initial-taylor-expansion}
\]
where (recalling the notation from the proof of \cref{prop:bb-bbvm-gaussian-location})
\[
\mcR_{n} \defined \mcR\left(\frac{\ii\numobs^{1/2}\delta\gradloglik{\obsrv{n}}{\optsym}^{\top}\Ehessloglik{\optsym}^{-1}t}{\bsnumobs}\right).
\]
Arguing as in the proof of \cref{prop:bb-bbvm-gaussian-location} and using assumption (ii), we conclude that 
\[
\lim_{\numobs \to \infty} \sum_{n=1}^{\numobs}\mcR_{n} = 0.
\]

Note that $\bsnumobs/\numobs \to c$, and $\empdist(\delta\gradloglikfun{\optsym}\delta\gradloglikfun{\optsym}^{\top}) \convas \Vargradloglik{\optsym}$ as $\numobs\to\infty$.
Now, we use the fact that if $a_\numobs \to a$ and $c_\numobs \to c$, then $(1 + a_\numobs/\numobs)^{\numobs c_\numobs} \to \exp(a)^c$.
Combining all these observations with \cref{eq:bb-bvm-characteristic-function-initial,eq:bb-bvm-initial-taylor-expansion}, we have that, for all $t  \in \reals^{D}$, 
the characteristic function of $\numobs^{1/2}(\bbparamsample - \optparam) \mid \datarvarg{\numobs}$ evaluated at $t$ is
\[
\exp\left\{ \ii \Delta_{\numobs}^{\top}t - t^{\top}\Ehessloglik{\optsym}^{-1}t/(2\bsscale) -t^{\top}\Ehessloglik{\optsym}^{-1}\Vargradloglik{\optsym}\Ehessloglik{\optsym}^{-1}t/(2\bsscale) \right\} + \eps_{\numobs}(t) + \bar\eps_{\numobs}(t).
\]
The result follows from L\'evy's continuity theorem~\citep[Theorem 5.3]{Kallenberg:2002}.

\end{document}